\documentclass[11pt,a4wide]{article}
\pdfoutput=1
\usepackage{jheppub}

\usepackage[english]{babel}
\usepackage[babel]{csquotes}
\usepackage{amsfonts}
\usepackage{array, tabularx}
\usepackage{mathtools}
\usepackage{amsthm}
\usepackage{url}
\usepackage{multicol, multirow}
\usepackage{emptypage}
\usepackage{verbatim}
\usepackage{dsfont}
\usepackage{hyperref}
\usepackage{float}
\usepackage{amsmath}
\usepackage{amssymb,tikz}
\usepackage{verbatim}

\newcommand{\hb}{{\bar{h}}}
\newcommand{\zb}{{\bar{z}}}

\newcommand{\1}{{\mathds{1}}}

\newcommand{\cG}{\mathcal{G}}
\newcommand{\dd}{\mathrm{d}}
\newcommand{\ii}{\mathrm{i}}

\newcommand{\order}{O}

\newcommand{\xit}{{\xi}}
\newcommand{\xin}{{\chi}}
\newcommand{\ut}{{\tilde{u}}}

\newcommand{\INVhat}{\widehat{\mathrm{INV}}}
\DeclareMathOperator*{\INV}{INV}
\newcommand{\invhat}[1]{\widehat{\mathrm{INV}}\left[ #1 \right]}
\newcommand{\inv}[1]{\mathrm{INV}\left[ #1 \right]}

\renewcommand{\O}{{\mathcal O}}
\newcommand{\G}{{\mathcal G}}
\newcommand{\Dbar}{{\overline{D}}}
\newcommand{\dda}{{D_a}}

\DeclareMathOperator*{\dDisc}{dDisc}
\DeclareMathOperator*{\dsat}{\mathcal{D}_{\mathrm{sat}}}
\DeclareMathOperator*{\cas}{\mathcal{C}}
\DeclareMathOperator*{\res}{Res}

\theoremstyle{definition}

\theoremstyle{remark}

\theoremstyle{definition}

\allowdisplaybreaks[2]

\usepackage[font=small,format=hang,labelfont={sf,bf}]{caption}

\makeatletter
\DeclareFontFamily{OMX}{MnSymbolE}{}
\DeclareSymbolFont{MnLargeSymbols}{OMX}{MnSymbolE}{m}{n}
\SetSymbolFont{MnLargeSymbols}{bold}{OMX}{MnSymbolE}{b}{n}
\DeclareFontShape{OMX}{MnSymbolE}{m}{n}{
    <-6>  MnSymbolE5
   <6-7>  MnSymbolE6
   <7-8>  MnSymbolE7
   <8-9>  MnSymbolE8
   <9-10> MnSymbolE9
  <10-12> MnSymbolE10
  <12->   MnSymbolE12
}{}
\DeclareFontShape{OMX}{MnSymbolE}{b}{n}{
    <-6>  MnSymbolE-Bold5
   <6-7>  MnSymbolE-Bold6
   <7-8>  MnSymbolE-Bold7
   <8-9>  MnSymbolE-Bold8
   <9-10> MnSymbolE-Bold9
  <10-12> MnSymbolE-Bold10
  <12->   MnSymbolE-Bold12
}{}

\let\llangle\@undefined
\let\rrangle\@undefined
\DeclareMathDelimiter{\llangle}{\mathopen}%
                     {MnLargeSymbols}{'164}{MnLargeSymbols}{'164}
\DeclareMathDelimiter{\rrangle}{\mathclose}%
                     {MnLargeSymbols}{'171}{MnLargeSymbols}{'171}

\makeatother

\title{
An alternative to diagrams for the critical $\boldsymbol{ \mathrm O(N) }$ model: 
dimensions and structure constants to order $\boldsymbol{1/N^2}$
}

\author{Luis F. Alday,}
\author{Johan Henriksson \&}
\author{Mark van Loon}
\affiliation{Mathematical Institute, University of Oxford, Andrew Wiles Building, Radcliffe Observatory Quarter, Woodstock Road, Oxford, OX2 6GG, UK}

\emailAdd{luis.alday@maths.ox.ac.uk}
\emailAdd{johan.henriksson@maths.ox.ac.uk}
\emailAdd{mark.vanloon@maths.ox.ac.uk}

\abstract{We apply the methods of modern analytic bootstrap to the critical $\mathrm O(N)$ model in a $1/N$ expansion. At infinite $N$ the model possesses higher spin symmetry which is weakly broken as we turn on $1/N$.  By studying consistency conditions for the correlator of four fundamental fields we derive the CFT-data for all the (broken) currents to order $1/N$, and the CFT-data for the non-singlet currents to order $1/N^2$. To order $1/N$ our results are in perfect agreement with those in the literature. To order $1/N^2$ we reproduce known results for anomalous dimensions and obtain a variety of new results for structure constants, including the global symmetry central charge $C_J$ to this order. 
}

\begin{document}
\maketitle


\section{Introduction}

Conformal invariance plays a crucial role in the description of critical phenomena, where the critical exponents are directly related to the scaling dimensions $\Delta_\O$ of primary operators in the corresponding conformal field theory (CFT).
The set of scaling dimensions, together with the operator product expansion (OPE) coefficients $c_{\O_i\O_j\O_k}$ (i.e. the structure constants of the OPE), completely characterizes the dynamical information of a CFT. This is collectively referred to as the CFT-data. 
The idea of the conformal bootstrap is to use conformal symmetry and unitarity, together with associativity of the OPE, to find rigorous constraints on the CFT-data. 
Since its first application \cite{Rattazzi:2008pe} to theories in dimension $d>2$, this idea has led to a large number of successful results, see \cite{Poland:2018epd} for an extensive review. 
This idea has proven particularly useful in non-perturbative regimes, where there are very few, if any, other methods available.
However, a bootstrap approach is also useful in a perturbative regime, as it starts from the axioms of CFT and is free from much of the surplus structure of other methods. 
Consider for instance the computation of critical exponents in the $\epsilon$-expansion using Feynman diagrams. 
Evaluating them gives a general result valid away from the critical point, and only after tuning the coupling to the fixed-point value, which requires yet more diagrams, can one access the conformal scaling dimensions and thus the critical exponents.

In Minkowski space, four-point correlators develop specific singularities when two operators become null separated. The structure of these singularities together with crossing symmetry imposes constraints on the spectrum of the theory  \cite{Alday:2007mf,Fitzpatrick:2012yx,Komargodski:2012ek}. In particular, they imply the existence of infinite towers of double-twist operators, whose CFT-data reduces to that of generalized free fields (GFF) for large spin. Large spin perturbation theory (LSPT) \cite{Alday:2016njk} builds on \cite{Alday:2015eya} and allows to compute corrections to this CFT-data in inverse powers of the spin, effectively solving the crossing relations around the null limit. This method is particularly powerful in perturbative expansions, see \cite{Alday:2016njk,Alday:2016jfr}, where LSPT allows to compute the corrections to the GFF data, order by order in the coupling constant, and to all orders in an asymptotic expansion around large spin. The Lorentzian inversion formula \cite{Caron-Huot:2017vep} shows that the CFT-data that follows from this procedure can be described in terms of functions analytic in spin, and provides an elegant method of finding these functions. 

The purpose of this paper is to study the critical $\mathrm O(N)$ model in a large $N$ expansion, for $2<d<4$ and using large spin perturbation theory. The critical $\mathrm O(N)$ model is mostly physically relevant in three dimensions where it is unitary and describes critical phenomena, see \cite{Pelissetto:2000ek}. In the large $N$ limit the model possesses higher spin symmetry, and it has been conjectured \cite{Klebanov:2002ja} that the singlet sector has a holographic dual in this limit, given by type A Vasiliev theory $hs_4$, see \cite{Vasiliev:1995dn}, with specific boundary conditions. An approach to study this problem directly in $3d$ and at finite $N$ is via numerical bootstrap, which provides results for some CFT-data to very high precision \cite{Kos:2016ysd}, as well as numeric estimates for a larger set of operators \cite{Simmons-Duffin:2016wlq}. Another strategy is to study the problem analytically via a perturbative expansion: either for finite $N$ around four dimensions or in a large $N$ expansion. In the large $N$ expansion this model can be given a Lagrangian description as follows, see \cite{Fei:2014yja} for a detailed discussion. To the action of $N$ free fields $\varphi^i$, one adds the interaction terms 
\begin{equation}\label{eq:HSextraterms}
S_{\mathrm I}=\int\mathrm d^dx\left(\frac1{2\sqrt N}\sigma\varphi^i\varphi^i-\frac{1}{4\lambda N}\sigma^2\right),
\end{equation}
where $\sigma$ is an auxiliary field. Integrating out $\sigma$ gives back the quartic interaction $\lambda(\varphi^i\varphi^i)^2$. However, for large $N$, $\sigma$ gets promoted to a dynamical field and one can derive its effective action. Flowing to the IR, the second term in \eqref{eq:HSextraterms} becomes irrelevant, and the first term can be used to develop a perturbation theory with $N^{-1/2}$ as the coupling constant. The operator $\varphi^i\varphi^i$ is no longer part of the spectrum and gets replaced by $\sigma$. In this way, numerous perturbative results have been obtained: for instance, the scaling dimension of the fundamental field $\varphi^i$ was computed to order $N^{-3}$ in \cite{Vasiliev:1982dc}, while the dimension of $\sigma$ was computed to order $N^{-2}$ in \cite{Vasiliev:1981dg}. In this paper will be primarily be interested in the CFT-data of the almost conserved currents ${\mathcal J}^{(\ell)}_R \sim \varphi^i \partial^\ell \varphi^j$. Here $R$ denotes one of the irreducible representations in the product
\begin{equation}\label{eq:irreps}
\mathbf N\otimes\mathbf N=S\oplus T\oplus A,
\end{equation}
where $S=\mathbf 1$ is the singlet representation, $T=\mathbf{N(N+1)/2-1}$ is the rank two symmetric  traceless  representation, and $A=\mathbf{N(N-1)/2}$ is the rank two antisymmetric representation of $\mathrm O(N)$. The dimensions of the broken higher spin currents were computed to order $N^{-2}$ in \cite{Derkachov:1997ch} for the symmetric traceless representation and in \cite{Manashov:2017xtt} for the singlet and antisymmetric representations.

The currents ${\mathcal J}^{(2)}_S$ and ${\mathcal J}^{(1)}_A$ are conserved and correspond to the stress tensor and the global symmetry current. The corrections to the corresponding central charges $C_T$ and $C_J$ were computed to order $N^{-1}$ in \cite{Petkou:1994ad} and \cite{Lang:1992pp}, as summarized in \cite{Petkou:1995vu}. The only other known subleading corrections to any OPE coefficients are the ones for the weakly broken
currents in the $T$ and $A$ representations, which were computed to order $N^{-1}$ in \cite{Dey:2016mcs}.

Our approach does not rely on the existence of a Lagrangian description. Instead, we consider the correlator of four fundamental fields ${\cal G}_{ijkl}(u,v) \sim \langle \varphi^i\varphi^j \varphi^k \varphi^l \rangle$ and study how crossing symmetry 
\begin{equation}\label{eq:crossingijkl}
v^{\Delta_\varphi} {\mathcal G}_{ijkl}(u,v) = u^{\Delta_\varphi} {\mathcal G}_{kjil}(v,u)
\end{equation}
and the structure of null singularities, as $v \sim 0$, constrains the CFT-data for intermediate and external operators. This correlator decomposes into the three representations in $\mathbf N\otimes\mathbf N$ and crossing symmetry mixes those contributions. To leading order in a large $N$ expansion the intermediate operators are the currents ${\mathcal J}^{(\ell)}_R$, which are conserved for $N=\infty$. As we turn on $1/N$ the dimensions and OPE coefficients of these currents receive perturbative corrections. These corrections can be reconstructed from the null singularities/double discontinuities of the correlator, which upon crossing symmetry are given by
\begin{equation}
\dDisc[{\mathcal G}_{ijkl}(u,v)] = \dDisc\Big[\frac{u^{\Delta_\varphi}}{v^{\Delta_\varphi} } {\mathcal G}_{kjil}(v,u)\Big].
\end{equation}
At each order in $1/N$ only a set of operators in the dual channel contribute to the double discontinuity. Our task is then to tabulate such operators, compute their contribution to the double discontinuity and then reconstruct the OPE data from it. 
\subsection{Our assumptions}\label{sec:assumptions}

We use a minimal set of assumptions about the theory, from which all our results will follow. These regard which operators appear in the OPE and acquire anomalous dimensions at each order in \(N^{-1}\). Furthermore, we will  assume the existence of a conserved stress tensor and global current.
More precisely, our working assumptions are:
 
\begin{itemize}
\item The fundamental field $\varphi^i$ has scaling dimension $\Delta_\varphi=\mu-1+\frac1N\gamma^{(1)}_\varphi+\frac1{N^2}\gamma^{(2)}_\varphi+O(N^{-3})$, where $\mu=d/2$. 

\item The OPE of two fundamental fields in a $1/N$ expansion contains the following operators\footnote{In this equation we have written out the operators with the appropriate \emph{squared} OPE coefficients. That is, for instance $c^2_{\varphi\varphi{\mathcal J}^{(\ell)}_S}\sim N^{-1}$.}
\begin{align}\nonumber
\varphi^i\times\varphi^j\sim \delta^{ij}&\left[
\1+\frac1N {\mathcal J}^{(\ell)}_S +\frac1{N^2}[\varphi^2_S]_{n,\ell}+\frac1N \sigma+\frac1{N^2}[\sigma,\sigma]_{n,\ell}+O\left(N^{-3}\right)\right]
\\\nonumber
+
&\left[
{\mathcal J}^{(\ell)}_T+\frac1N[\varphi^2_T]_{n,\ell}+O\left(N^{-2}\right)\right]
\\+
&\left[
{\mathcal J}^{(\ell)}_A+\frac1N[\varphi^2_A]_{n,\ell}+O\left(N^{-2}\right)\right],\label{eq:OPEorders}
\end{align}
where the spin $\ell$ is even for $S$ and $T$, and odd for $A$.
Here $[\varphi^2_R]_{n,\ell}$ denote towers of double twist operators with scaling dimensions $\Delta_{R,n,\ell}=2\Delta_\varphi+2n+\ell+\gamma_{R,n,\ell}$, for $n=1,2,\ldots$.  From the explicit expressions for the OPE coefficients in GFF, it follows that they don't appear at zeroth order. This is of course consistent with the equations of motion in the free theory. 
\item
At leading twist in each representation we have a family of currents, ${\mathcal J}^{(\ell)}_R$, that are conserved at infinite $N$. They are nondegenerate and are of the following form
\begin{itemize}
\item Singlet ($S$): 
${\mathcal J}^{(\ell)}_S$ for $\ell=2,4,6,\ldots$. 
The scaling dimensions are $\Delta_{S,\ell}=2\Delta_\varphi+\ell+\frac1N\gamma^{(1)}_{S,\ell}+O(N^{-2})$, and the OPE coefficients are $a_{S,\ell}:=c^2_{\varphi\varphi\varphi^2_{S,\ell}}=\frac1Na^{(0)}_{S,\ell}+\frac1{N^2}a^{(1)}_{S,\ell}+O(N^{-3})$\footnote{We will derive the extra $N^{-1}$ suppression in the next section.}. 
For $\ell=2$ the current reduces to the stress tensor $T_{\mu\nu}$, which we assume to be protected, with $\Delta_{S,2}=d$. Its OPE coefficient is related to the central charge $C_T$ through the conformal Ward identity $ a_{S,2}=\frac{\mu^2\Delta_\varphi^2}{(2\mu-1)^2C_T}$ \cite{Petkou:1994ad}.
\item Traceless symmetric ($T$): 
${\mathcal J}^{(\ell)}_T$ for $\ell=0,2,4,\ldots$ with scaling dimensions $\Delta_{T,\ell}=2\Delta_\varphi+\ell+\frac1N\gamma^{(1)}_{T,\ell}+\frac1{N^2}\gamma^{(2)}_{T,\ell}+O(N^{-3})$ and OPE coefficients $a_{T,\ell}:=a^{(0)}_{T,\ell}+\frac1{N}a^{(1)}_{T,\ell}+\frac1{N^2}a^{(2)}_{T,\ell}+O(N^{-3})$.
\item Antisymmetric ($A$): 
${\mathcal J}^{(\ell)}_A$ for $\ell=1,3,5,\ldots$. 
The perturbative structure is the same as for $T$. The $\ell=1$ operator is the global symmetry current whose dimension is protected: $\Delta_{A,1}=d-1$, and whose OPE coefficient is related to the charge $C_J$ through the conformal Ward identity \( a_{A,1} = -\frac{1}{C_J} \) \cite{Petkou:1994ad}.
\end{itemize}
\item The scalar singlet operator $\sigma$ has dimension $\Delta_\sigma=2+\frac1N\gamma^{(1)}_\sigma+O(N^{-2})$. 
\end{itemize}

Furthermore, the existence of $\sigma$ guarantees the existence of double twist operators $[\sigma,\sigma]_{n,\ell}$ for $n=0,1,\ldots$ and $\ell=0,2,\ldots$. 
These are nondegenerate and their squared OPE coefficients \( c^2_{\varphi\varphi [\sigma,\sigma]_{n,\ell}} \) start at order $N^{-2}$. 
In section~\ref{sec:sigmasigma} we need the explicit form of their OPE coefficients, which we derive in appendix~\ref{sec:mixed} by considering the leading order inversion problem for the mixed correlator $\left\langle\varphi\varphi\sigma\sigma\right\rangle$.

\subsection{Summary of our results}
With these assumptions, we can then consider the correlator $\langle \varphi^i\varphi^j \varphi^k \varphi^l \rangle$ in a $1/N$ expansion, and to each order tabulate the class of operators, in the dual channel, that contributes to the double discontinuity. To a given order, the precise double discontinuity can be computed from CFT-data appearing at the previous order.  From this we can then compute the CFT-data of the currents in the direct channel. Our results can be summarised as follows. Regarding anomalous dimensions we reproduce all known results to order $1/N$ together with the results for the symmetric and antisymmetric representations to order $1/N^2$. Regarding the OPE coefficients, we reproduce all known results to order $1/N$ and obtain a variety of new results to order $1/N^2$. The expression for the OPE coefficients in the symmetric and antisymmetric representation is very involved and is given in terms of infinite sums, which can be exactly evaluated in different corners of the $(d,\ell)$ plane. For $\ell=1,d=3$ it can be estimated with very good precision and we can extract the value of the central charge $C_J$ to this order. We obtain  

\begin{equation}
\left.\frac{C_J}{C_{J,\text{free}}}\right|_{d=3} =1-\frac{0.720506}{N}+\frac{1.14230(2)}{N^2}+O(N^{-3}),
\end{equation}
where the coefficient to order $1/N$ is simply $-\frac{64}{9\pi^2}$. A more complete summary of our results can be found in section \ref{sec:summary}.

The rest of the paper is devoted to the details of our computation. In section~\ref{sec:LSPTforON}, we first describe our precise method for computing CFT-data in the \(\mathrm{O}(N)\) model from an analysis of the crossing symmetry equations. 
To order \(1/N\), we then analyze in turn the contributions to the CFT-data of the weakly broken currents due to the appearance in the crossed channel of the identity \(\mathds{1}\), of \(\sigma\), and of the currents \( \mathcal{J}_T^{(\ell)}\) and \( \mathcal{J}_A^{(\ell)}\), which appear in the combination \( \cG_T-\cG_A \). 
Section~\ref{sec:subsubleading} contains our computations for the \(1/N^2\) corrections to the CFT-data of \( \mathcal{J}_T^{(\ell)}\) and \( \mathcal{J}_A^{(\ell)}\).
In section~\ref{sec:H_TpA}, we find the contributions from \(1/N\) corrections to the CFT-data of the currents \( \mathcal{J}_T^{(\ell)}\) and \( \mathcal{J}_A^{(\ell)}\), now appearing in the combination \( \cG_T + \cG_A\).
We extend our results for the identity and \(\sigma\) to order \(1/N^2\) in section~\ref{sec:sigmainversionrevisited}, specifically making a connection to crossing kernels.
Finally, in section~\ref{sec:sigmasigma} we find the contribution from the operators \( [\sigma,\sigma]_{n,\ell}\), computing the double discontinuity in Mellin space.
Our results are summarized in section~\ref{sec:summary}. 
Appendices~\ref{sec:InvMethods} and \ref{app:inversionTable} give details of our inversion computations and results, appendix~\ref{sec:mixed} computes OPE coefficients in the mixed correlator, and appendices \ref{sec:unitarityboundblocks}, \ref{app:HT_pm_A} and \ref{app:MellinForSigSig} contain additional details of computations in the main text.

\section{Large spin perturbation theory for the $\mathrm O(N)$ model}

\label{sec:subleading}
\label{sec:LSPTforON}

\noindent As described in the introduction, our aim is to study the implications of crossing symmetry of the \( \langle \varphi^i(x_1) \varphi^j(x_2) \varphi^k(x_3) \varphi^l(x_4) \rangle \) correlator for the CFT-data of nearly conserved currents in the singlet, traceless symmetric and antisymmetric rank-two tensor representations.
Let us now describe our method in detail.

The correlator of four fundamental fields in the critical $\mathrm O(N)$ model can be expressed in terms of the cross-ratios as
\begin{equation}
 \cG_{ijkl}(z,\zb)=x_{12}^{2\Delta_\varphi}x_{34}^{2\Delta_\varphi}\langle \varphi^i(x_1) \varphi^j(x_2) \varphi^k(x_3) \varphi^l(x_4) \rangle .
\end{equation}
We will use $(u,v)$ and $(z,\zb)$ interchangeably for the cross-ratios, related by $u=z \zb=\frac{x_{12}^2x_{34}^2}{x_{13}^2x_{24}^2}$ and $v=(1-z)(1-\zb)=\frac{x_{14}^2x_{23}^2}{x_{13}^2x_{24}^2}$, with $x_{ij}=x_i-x_j$. 
The correlator has the following decomposition into the irreducible representations of \eqref{eq:irreps}
\begin{equation}
\cG_{ijkl}(z,\zb) = \cG_S(z,\zb) \delta_{ij}\delta_{kl} + \cG_T(z,\zb) \left(  \frac{\delta_{ik}\delta_{jl} + \delta_{il}\delta_{jk}}{2} - \frac{1}{N} \delta_{ij}\delta_{kl} \right)  + \cG_A(z,\zb) \left( \frac{\delta_{ik}\delta_{jl} - \delta_{il}\delta_{jk}}{2}  \right).
\end{equation}

The crossing equation~\eqref{eq:crossingijkl} projects onto the following equations for the different $\mathrm O(N)$ representations:
\begin{align} 
\cG_S(u,v) &= \left(\frac{u}{v}\right)^{\Delta_\varphi}\left( \frac{1}{N}\cG_S(v,u) + \frac{(N+2)(N-1)}{2N^2} \cG_T(v,u) +\frac{1-N}{2N} \cG_A(v,u)\right), \nonumber\\\label{eq:ONcrossing}
\cG_T(u,v) &=\left(\frac{u}{v}\right)^{\Delta_\varphi}\left( \cG_S(v,u)+ \frac{N-2}{2N} \cG_T(v,u) +\frac{1}{2} \cG_A(v,u)\right), \\
\cG_A(u,v) &=\left(\frac{u}{v}\right)^{\Delta_\varphi}\left( -\cG_S(v,u) + \frac{N+2}{2N} \cG_T(v,u) +\frac{1}{2} \cG_A(v,u)\right). \nonumber
\end{align}
Consider the left-hand side (direct channel) of equations \eqref{eq:ONcrossing} in a small $z$ expansion. In this limit leading twist operators, namely the conserved currents ${\mathcal J}^{(\ell)}_R$, dominate and the problem becomes essentially one-dimensional. In a $1/N$ expansion their scaling dimensions are given by
\begin{equation}
\Delta_{R,\ell} = 2 \Delta_\varphi+\ell + \frac{1}{N} \gamma_{R,\ell}^{(1)}+\ldots
\end{equation}
We can express their contribution as a sum over collinear  (i.e. $\mathrm{SL}(2,\mathbb R)$) conformal blocks $k_\hb(\zb)=\zb^\hb {_2F_1}(\hb,\hb;2\hb;\zb)$:
\begin{equation}\label{eq:LHSexpansion}
\cG_{R}(z,\zb)=z^{\Delta_\varphi}\sum_{\ell}\left(U^{(0)}_{R,\hb}+\frac12U^{(1)}_{R,\hb}\log z+\frac18U^{(2)}_{R,\hb}\log^2z+\ldots\right)r_\hb k_\hb(\zb)+\ldots, \quad\hb=\Delta_\varphi+\ell.
\end{equation}
In this expression we have organised the expansion in powers of $\log z$, which follows order by order from expanding the conformal blocks in powers of $N^{-1}$. The functions $U^{(p)}_\hb$ carry in a compact format all information about the perturbative CFT-data of the currents. 
To extract the CFT-data one uses \cite{Alday:2017vkk}
\begin{equation}\label{eq:extractAGamma}
\hat a_\hb (\gamma_\ell)^p=U^{(p)}_\hb+\frac12\partial_\hb U^{(p+1)}_\hb+\frac18\partial_\hb U^{(p+2)}_\hb+\ldots
\end{equation}
for anomalous dimensions $\gamma_\ell=\Delta_\ell-2\Delta_\varphi-\ell$ (suppressing the dependence on representation), and
\begin{equation}\label{eq:extractAfromAhat}
a_\ell=c_{\varphi\varphi{\mathcal J}^{(\ell)}}^2=r_{\tfrac{\Delta_\ell+\ell}{2}}\hat a_\hb ,\qquad r_\hb=\frac{\Gamma(\hb)^2}{\Gamma(2\hb)},
\end{equation}
for the OPE coefficients.
The functions $U^{(p)}_\hb$ have the important property that they are additive in crossed-channel contributions. Specifically this means that we will compute these functions order by order in $1/N$ by considering new contributions appearing at each order.
In the final expressions, for instance for $\gamma_\ell=\big(U^{(1)}_\hb+\frac12\partial_\hb U^{(2)}_\hb+\ldots\big)\big/\big(U^{(0)}_\hb+\frac12\partial_\hb U^{(1)}_\hb+\ldots\big)$, the depencence on crossed-channel contributions is non-linear.

By looking at a specific power of $\log z$ in \eqref{eq:LHSexpansion} we get a single sum over $\mathrm{SL}(2,\mathbb R)$ blocks:
\begin{equation}\label{eq:inversionsum}
G(\zb)=\sum_{\ell}U_\hb \, r_\hb \, k_\hb(\zb), \qquad \hb=\Delta_\varphi+\ell,
\end{equation}
where the sum over spin $\ell$ goes over even values in the case of $S$ or $T$ representations and odd values for the $A$ representation. Here $G(\zb)$ represents the part of the correlator proportional to a particular power of $z$ and $\log z$, in the given $\mathrm O(N)$ representation.

As mentioned in the introduction, in the null limit the correlator develops certain singularities, captured precisely by the double-discontinuity, to be defined below. Large spin perturbation theory \cite{Alday:2015eya} allows to reconstruct the CFT-data from these singularities, as an asymptotic expansion in inverse powers of the {\it conformal spin} $(\Delta+\ell)(\Delta+\ell-2)/4$. These expansions can be resummed by the elegant Lorentzian inversion formula \cite{Caron-Huot:2017vep}, which in this limit reduces to \footnote{The Lorentzian inversion formula was derived by Caron-Huot in \cite{Caron-Huot:2017vep} in analogy with the Froissart-Gribov formula for the $S$-matrix. The original formula contains a two-dimensional integral and results in a function with poles at the dimensions of all exchanged operators, and the (squared) OPE coefficient as the residues. In \cite{Caron-Huot:2017vep} it was further shown that the small $z$ limit gives a generating function for the contribution to each twist family. The leading twist contribution for each power of $\log z$ reduces to \eqref{eq:invIntegralDefn}, where we have chosen a normalization that agrees with \cite{Alday:2017zzv}.} 
\begin{equation} \label{eq:invIntegralDefn}
U_\hb = \inv{G(\zb)} := \frac{\Gamma(\hb)^2}{\pi^2 \Gamma(2\hb-1)} \int_0^1  \frac{\mathrm{d} \zb}{\zb^2} \,k_{\hb}(\zb)  \dDisc\left[G(\zb)\right],
\end{equation}
where the double discontinuity of a correlator is defined as the difference between the correlator and its two analytic continuations around $\zb=1$:
\begin{equation}
\dDisc[G(\zb)] := G(\zb) -\frac{1}{2} G^\circlearrowleft(\zb)-\frac{1}{2}G^\circlearrowright(\zb).
\end{equation}
We will refer to \eqref{eq:invIntegralDefn} as the \emph{inversion} of $G(\zb)$. The integral of \eqref{eq:invIntegralDefn} can be used either to generate closed-form results for $U^{(p)}_{R,\hb}$, or to generate the asymptotic series around infinite $\ell$, where contact with large spin perturbation theory is made. Indeed, the inversion of a crossed channel operator of twist $\tau=\Delta-\ell$ has an expansion
\begin{equation}\label{eq:reciprocityexpansion}
\INV\left[\left(\frac{\zb}{1-\zb}
\right)^{\Delta_\varphi}(1-\zb)^{\tau/2}\right]\sim \frac{2\hb-1}{J^{\tau+2-2\Delta_\varphi}}\left(c_0+\frac{c_1}{J^2}+\frac{c_2}{J^4}+\frac{c_3}{J^6}+\ldots\right),
\end{equation}
where $J^2=\hb(\hb-1)$ is the \emph{bare} conformal spin. This expansion is in perfect agreement with \cite{Alday:2015eya,Simmons-Duffin:2016wlq,Alday:2016njk}. Notice that only even powers of $J^{-1}$ appear in the expansion \eqref{eq:reciprocityexpansion}. This is sometimes referred to as reciprocity, and was proven in \cite{Alday:2015eya} on the level of CFT-data in an expansion in the \emph{full} conformal spin $(\Delta+\ell)(\Delta+\ell-2)/4$. The equivalence of the two statements can be shown order by order using \eqref{eq:extractAGamma}.

We then analyze the crossed channel (right-hand side) of the equations~\eqref{eq:ONcrossing} in an expansion around infinite $N$. 
At each order in $N^{-1}$, only a few (families of) operators contribute to the double-discontinuity. 
If we focus on terms that are proportional to $z^{\Delta_\varphi}\log^p z$, we get the double-discontinuity required to find $U^{(p)}_\hb$ from the inversion \eqref{eq:invIntegralDefn}. Our general strategy then is the following:
\begin{enumerate}
\item Identify the (families of) operators in the crossed channel that contribute to the double-discontinuity. 
\item Compute the double-discontinuity they produce. In general this requires infinite sums, and knowledge of the OPE coefficients with which such operators appear.
\item Invert the resulting double-discontinuity. 
\end{enumerate}
The first step of the strategy involves some rudimentary assumptions of the theory; in our case the relevant assumptions are those presented in section~\ref{sec:assumptions}. Importantly, operators with dimensions $2\Delta_\phi+n+\ell+\gamma$ for integer $n$ do not contribute with a double-discontinuity until order $\gamma^2$, as can be seen from the expansion of their crossed channel conformal blocks in the $\zb\to1$ limit. Step two requires expressions for relevant conformal blocks and techniques for computing their sums, while step three requires methods for explicitly computing the involved integrals. In the main text, we will present relevant results and techniques as we go along, while we refer details to appendices. 

A key principle in large spin perturbation theory is that we may consider the contribution from a crossed-channel operator $\O$ without knowing the precise value of its OPE coefficient $a_\O=c_{\varphi\varphi\O}^2$. Instead, we will introduce such constants as free parameters, and later derive consistency equations that fix their values, thus providing results following purely from the bootstrap approach. This principle was used in \cite{Alday:2016jfr} to derive a number of results at leading order, following from very generic assumptions. In the subsequent development \cite{Alday:2017zzv} this idea was combined with the Lorentzian inversion integral, providing a more efficient and rigorous method for the third step of the above strategy.

In both steps two and three we will make use of the quadratic Casimir operator $\cas$ of the conformal group, which has the conformal blocks as eigenfunctions \cite{Dolan:2003hv}. By introducing a constant shift it has the conformal spin $J^2$ as eigenvalue. Furthermore, in the small $z$ limit it reduces to a \emph{collinear} Casmir $\Dbar$ acting on the $\mathrm{SL}(2,\mathbb R)$ blocks with eigenvalue $J^2$. In particular, we will use this to simplify some inversion integrals, noting that
\begin{equation}
\INV\big[\Dbar G(\zb)\big]=J^2\INV\big[G(\zb)\big],
\end{equation}
where, typically, $\Dbar G(\zb)$ is easier to invert than $G(\zb)$.

\subsection{Contribution from the identity operator}

Let us start by considering the large $N$ expansion of the crossing equation for the traceless symmetric ($T$) representation:
\begin{equation}
\G_T(u,v)=\left(\frac{u}{v}\right)^{\Delta_\varphi}\left(\G_S(v,u)+\frac{\G_T(v,u)+\G_A(v,u)}2-\frac1N\G_T(v,u)\right).
\end{equation}
The only contributions up to order $N^{-1}$ come from the identity operator and from $\sigma$, both appearing in the $S$ representation.\footnote{The \( \mathcal{J}_R^{(\ell)} \) are already present in the correlator at this order, but they do not contribute to the double-discontinuity since the factor of \(v^{\Delta_\varphi}\) cancels.
They do contribute to the double-discontinuity at higher order with a factor of \(\log^2 v\), which arises from non-zero squared anomalous dimensions: see section~\ref{sec:S_at_N-1} for details.}
The contribution from these operators takes the form
\begin{equation}\label{eq:firstopsinS}
\G_S(v,u)=1+\frac{a_\sigma^{(0)}}{N}v^{\Delta_\sigma/2}g^{(d)}_{\Delta_\sigma,0}(v,u)+O(N^{-2}),
\end{equation}
where $G^{(d)}_{\tau,\ell}(u,v)=u^{\tau/2}g^{(d)}_{\tau,\ell}(u,v)$ is the conformal block in $d$ dimensions.

The inversion of the identity contribution is straightforward, and involves computing the integral
\begin{equation}\label{eq:Adef}
A[p](\hb):=\frac{\Gamma(\hb)^2}{\pi^2\Gamma(2\hb-1)}\int_0^1\frac{d \zb}{\zb^2}k_\hb(\zb)\dDisc\left[\left(\frac{\zb}{1-\zb}\right)^{p}\right]=\frac{2(2\hb-1)\Gamma(\hb+p-1)}{\Gamma(p)^2\Gamma(\hb-p+1)}
\end{equation}
for $p=\Delta_\varphi$. 
The result $A[p](\hb)$ will be heavily used in the following and can easily be found by noting that $\dDisc[(1-\zb)^{-p}]=2\sin^2(\pi p)(1-\zb)^{-p}$ and by using an integral representation for the hypergeometric function in $k_\hb(\zb)$.

Using the different crossing prefactors for the $S$ and $A$ representation we readily get\footnote{We have chosen a normalization for the conformal blocks that differs by a factor $(-2)^\ell$ from e.g. \cite{Dolan:2000ut}.
In our convention, this means that the squared OPE coefficients of the intermediate operators in the antisymmetric representation are negative, since they have odd spin.
}
\begin{equation}
NU^{(0)}_{S,\hb}=U^{(0)}_{T,\hb}=-U^{(0)}_{A,\hb}=A[\Delta_\varphi](\hb)+O(N^{-1}).
\end{equation}
Extracting the OPE coefficients from this using \eqref{eq:extractAfromAhat}, we find that to this order they agree with the result from a conformal block decomposition of the generalized free theory correlator
\begin{equation}\label{eq:GstaGFF}
\mathcal G_S(u,v)=1+\frac1Nu^{\Delta_\varphi}+\frac1N\left(\frac uv\right)^{\Delta_\varphi}\hspace*{-5pt},\qquad \mathcal G_{T/A}(u,v)=u^{\Delta_\varphi}\pm \left(\frac uv\right)^{\Delta_\varphi}\hspace*{-5pt}.
\end{equation}
Here, as in the rest of the paper, the upper sign refers to the $T$ representation, and the bottom sign to the $A$ representation.

\subsection{Contribution from the scalar $\sigma$}
\label{sec:sigmatree}

Let us now turn to the second term in \eqref{eq:firstopsinS}.
Although at this order in $N$ we have $\Delta_\sigma=2$, it is worth considering the general problem of inverting a scalar operator.
To be precise, we would like to compute the contributions to $U^{(0)}_\hb$ and $U^{(1)}_\hb$ due to a scalar operator with dimension $\Delta$ appearing with OPE coefficient $a_\Delta=c_{\varphi\varphi\O_\Delta}^2$ in the crossed channel of the $\varphi$ four-point function, i.e.
\begin{equation}\label{eq:inversionproblemblock}
U^{(0)}_\hb|_\Delta+\frac12U^{(1)}_\hb|_\Delta\log z=\frac{a_\Delta\Gamma(\hb)^2}{\pi^2\Gamma(2\hb-1)}\int_0^1\frac{d \zb}{\zb^2}k_\hb(\zb)\dDisc\left[\zb^{\Delta_\varphi}(1-\zb)^{\frac\Delta2-\Delta_\varphi}\left.g^{(d)}_{\Delta,0}(v,u)\right|_{\text{small $z$}}\right]\!.
\end{equation}
The scalar conformal block for identical external scalars in $d=2\mu$ dimensions was given in \cite{Dolan:2000ut}:
\begin{equation}
\label{eq:ScalarBlockAnyD}
g^{(d)}_{\Delta,0}(v,u) = \sum_{m,n=0}^\infty \frac{\left(\Delta/2\right)^2_m\left(\Delta/2\right)^2_{m+n}}{m! n! \left( \Delta+1-\mu\right)_m \left(\Delta\right)_{2m+n}} v^m(1-u)^n,
\end{equation}
where $(a)_n=\frac{\Gamma(a+n)}{\Gamma(a)}$ is the Pochhammer symbol. 
Performing the sum over $n$ and expanding for small $z$ gives
\begin{equation}\label{eq:scalarblocksmallzterm}
\left.g^{(d)}_{\Delta,0}(v,u)\right|_{\text{small $z$}}=\sum_{m=0}^\infty -\frac{ \Gamma (2 m+\Delta ) \left(\frac{\Delta }{2}\right)_m^2 \left(2
   S_1(m+\frac{\Delta }{2}-1)+\log (z \zb )\right)}{\Gamma \left(\frac{\Delta }{2}\right)^2
   (\Delta )_{2 m} (\Delta -\mu +1)_mm!}(1-\zb)^m,
\end{equation}
where $S_1(n)$ denotes the analytic continuation of the harmonic numbers. 
If we perform the sum over $m$ and use some identities for hypergeometric functions, we arrive at the closed-form expression
\begin{align}\nonumber
\left.g^{(d)}_{\Delta,0}(v,u)\right|_{\text{small $z$}}=-\frac{\Gamma(\Delta)}{\Gamma(\frac\Delta2)^2}\Big[
&\left(2S_1(\tfrac\Delta2-1)+\log(z\zb)\right){_2F_1}\left(\tfrac\Delta2,\tfrac\Delta2;\Delta+1-\mu;1-\zb\right)
\\
&+\dda\,{_2F_1}\left(\tfrac\Delta2+a,\tfrac\Delta2;\Delta+1-\mu;1-\zb\right)
\Big],\label{eq:genericcrossedblock}
\end{align}
where $\dda f=\left.\frac{\partial f}{\partial a}\right|_{a=0}$.
We defer the solution of the inversion problem for general $\Delta$ until section~\ref{sec:sigmainversionrevisited}, and focus now on the contribution from the operator $\sigma$. 
At leading order in $N^{-1}$ we can take $a_\Delta=c_{\varphi\varphi\sigma}^2=\frac{a^{(0)}_\sigma}N$ and $\Delta=\Delta_\sigma^{(0)}=2$. 
Furthermore, we may put $\Delta_\varphi=\mu-1$. The expression inside the double-discontinuity in \eqref{eq:inversionproblemblock} becomes
\begin{equation}\label{eq:toinvertsigmatree}
-\left(\frac{\zb}{1-\zb}\right)^{\mu-1}(1-\zb)\left[\log(z\zb)\,{_2F_1}\left(1,1;3-\mu;1-\zb\right)+2\dda  \,{_2F_1}\left(1+a,1;3-\mu;1-\zb\right)\right].
\end{equation}
The integral in \eqref{eq:inversionproblemblock} now involves products of hypergeometric functions. 
In the remaining part of this paper we will encounter a variety of similar functions that enter the inversion integral. 
We develop different methods for computing these inversion integrals. 
In the body of the paper we will describe briefly these methods as we use them, but we refer further details to the comprehensive appendix~\ref{sec:InvMethods}.

One method of finding the inversion of \eqref{eq:toinvertsigmatree}, which will also be useful in the inversion integrals that appear in the subsequent sections, is to make use of the (shifted) quadratic Casimir \(\cas\), which has the conformal block \( G_{\Delta,\ell}(z,\zb)\) as an eigenfunction, with the conformal spin as the associated eigenvalue.
It is defined by \cite{Alday:2016jfr}
\begin{equation}\label{eq:CasimirDef}
\cas=D_z+D_{\zb}+(2\mu-2)\frac{z \zb}{z-\zb}\left((1-z)\partial_z-(1-\zb) \partial_\zb\right)-\frac{\tau(\tau+2-4\mu)}4,
\end{equation}
with $D_x=x^2(1-x)\partial _x^2-x^2\partial_x$ and $\tau=\Delta-\ell$.
In the collinear (small \(z\)) limit, it reduces to
\begin{equation} \label{eq:dbardefn} 
\Dbar=(1-\zb)\zb^2\partial_\zb^2-\zb^2\partial_\zb,
\end{equation}
with the collinear conformal blocks as eigenfunctions: \(\Dbar k_{\hb}(\zb) = \hb(\hb-1) k_{\hb}(\zb)\).
In the inversion integral \eqref{eq:invIntegralDefn}, \(\Dbar\) is self-adjoint \cite{Hogervorst:2017sfd}, so that we may act with it on \(k_{\hb}(\zb)\), to produce a factor of \(J^2=\hb(\hb-1)\) .
Therefore, if we know the inversion of $\Dbar g(\zb)$, the inversion of $g(\zb)$ follows from simply dividing by $J^2$.\footnote{This holds up to the addition of functions in the kernel of $\Dbar$. 
For a detailed discussion see appendix \ref{sec:collCas}.}
For instance, the $\log z$ piece of the above inversion satisfies
\begin{equation}\label{eq:Dbarof2F1intoconst}
\Dbar \left(\frac{\zb}{1-\zb}\right)^{\mu-1}(\zb-1){_2F_1}\left(1,1;3-\mu;1-\zb\right)=-(\mu-2)^2\left(\frac{\zb}{1-\zb}\right)^{\mu-1},
\end{equation}
from which we immediately conclude that
\begin{equation}\label{eq:U1TAatN1}
U^{(1)}_{T/A,\hb}=\mp2(\mu-2)^2\frac{a^{(0)}_\sigma}{N}\frac{A[\mu-1](\hb)}{J^2}.
\end{equation}
Here, as in \eqref{eq:GstaGFF}, the upper sign refers to the $T$ representation, and we used that the exact form of the crossing equation guarantees that the results in the $A$ representation are equal up to a sign to this order in $N^{-1}$. 
 Similarly for the non-$\log z$ piece we get
\begin{align}
&\Dbar \left(\frac{\zb}{1-\zb}\right)^{\mu-1}(\zb-1)\left[\log\zb\,{_2F_1}\left(1,1;3-\mu;1-\zb\right)+2\dda \,{_2F_1}\left(1+a,1;3-\mu;1-\zb\right)\right]
\\
&\quad=-\left(\frac{\zb}{1-\zb}\right)^{\mu-1}\left((\mu-2)^2\log\zb+(1-\zb){_2F_1}\left(1,1;3-\mu;1-\zb\right)\right).
\end{align}
The first term can now be integrated directly, for instance by replacing the $k_\hb(\zb)$ by its defining infinite sum and integrating term by term. This gives $A[\mu-1](\hb)\mathbf{S_1}[\mu-1](\hb)$, where 
\begin{equation}\label{eq:Qdefalpha}
\mathbf{S_1}[\alpha](\hb)=2S_1(\hb-1)-S_1(\hb+\alpha-2)-S_1(\hb-\alpha)
\end{equation}
is a combination of harmonic numbers with a large $J$ expansion that contains only even powers of $J$ and is free from $\log J$ terms.
The second term is again of the form \eqref{eq:Dbarof2F1intoconst} and will thus contribute with a term proportional to $A[\mu-1](\hb)/J^4$. In summary, this gives 
\begin{equation}\label{eq:U0TAatN1}
U^{(0)}_{T/A,\hb}=\pm A[\Delta_\varphi](\hb)
\pm(\mu-2)^2\frac{a^{(0)}_\sigma}{N}\frac{A[\mu-1](\hb)}{J^2}\left(\mathbf{S_1}[\mu-1](\hb)-\frac1{J^2}
\right).
\end{equation}

\subsection{CFT-data for $T$ and $A$ at $N^{-1}$}
\label{sec:TA_Nmin1}
As discussed above, at order $N^{-1}$, the only operators that contribute to the CFT-data in the $T$ and $A$ representations are the identity $\1$ and the auxiliary field $\sigma$. 
The CFT-data at this order can therefore be computed from the functions $U^{(p)}_{T/A}$, as given in \eqref{eq:U1TAatN1} and \eqref{eq:U0TAatN1}, using the general relation \eqref{eq:extractAGamma}.

For the anomalous dimensions we only need the $N^0$ term in \eqref{eq:U0TAatN1}, which can be evaluated at $\Delta_\varphi=\mu-1$. This directly gives
\begin{equation} \label{eq:gamma_TA_Nmin1}
\gamma_{T,\ell}=\gamma_{A,\ell}=-\frac{2(\mu-2)^2}{J^2}\frac{a^{(0)}_{\sigma}}{N}+O(N^{-2}),
\end{equation}
where $J^2=(\ell+\mu-1)(\ell+\mu-2)+O(N^{-1})$. The results for $T$ ($A$) take even (odd) values for $\ell$. Specifically, the $\ell=1$ operator is the conserved global symmetry current $\mathcal J^{(1)}_A$, which has $\Delta_{A,1}=d-1$. Since $\Delta_{T/A,\ell}= 2\Delta_\varphi + \ell + \gamma_{T/A,\ell} = d - 2 +\ell +2\gamma_\varphi^{(1)}/N+\gamma^{(1)}_{T/A,\ell}/N$, this imposes the relation
\begin{equation}\label{eq:asigma0res}
a^{(0)}_{\sigma}=\frac{\mu(\mu-1)}{(\mu-2)^2}\gamma^{(1)}_\varphi,
\end{equation}
which we will use in the following.

The OPE coefficients are extracted using \eqref{eq:extractAGamma} and \eqref{eq:extractAfromAhat},
\begin{align}\nonumber
a_{T/A,\ell}&=\pm
\frac{2\Gamma(\ell+\mu-1)^2\Gamma(\ell+2\mu-3)}{\Gamma(\mu-1)^2\Gamma(\ell+1)\Gamma(2\ell+2\mu-3)}
\Big[1+
\frac{2\gamma_\varphi^{(1)}}{N}\Big(
S_1(\ell+\mu-2)-S_1(\mu-2)
\\\label{eq:OPETAfirstorder}
& \hspace*{-12pt} +\frac{\mu(\mu-1)}{J^2(\ell+\mu-1)}+\frac{(\ell-1)(\ell+2\mu-2)}{J^2}\left(S_1(\ell+2\mu-4)-S_1(2\ell+2\mu-4)\right)\Big)
+O(N^{-2})
\Big],
\end{align}
where, as always, the upper sign refers to \( T\) and the lower sign to \(A\).
The OPE coefficient for the global symmetry current is related to the charge $C_J$ through the conformal Ward identity \( a_{A,1} = -\frac{1}{ C_J} \) \cite{Petkou:1994ad}, from which we get
\begin{equation}\label{eq:cJfirstorder}
\frac{C_J}{C_{J,\mathrm{free}}}=1-\frac{2(2\mu-1)}{\mu(\mu-1)}\frac{\gamma^{(1)}_\varphi}{N}+O(N^{-2}).
\end{equation}

\subsection{CFT-data for $S$ at $N^{-1}$} \label{sec:S_at_N-1}
Let us consider the crossing equation \eqref{eq:ONcrossing} for the singlet representation in large $N$:
\begin{equation}\label{eq:crossingS}
\mathcal G_S(u,v)\! =\!\left(\frac{u}{v}\right)^{\Delta_\varphi}\!\!\left(\frac1N \mathcal G_S(v,u) + \!\frac12\left(\mathcal G_T(v,u)\!-\!\mathcal G_A(v,u) \right) + \!\frac1{2N}\left(\mathcal G_T(v,u)\!+\!\mathcal G_A(v,u) \right)\!+\! \order(N^{-2})\!\right).
\end{equation}
Now recall that the singlet tree-level OPE coefficients are of order \(N^{-1}\), so that we study corrections to their CFT-data by analyzing equation \eqref{eq:crossingS} at order \(N^{-2}\).

On the RHS of this equation, the contributions to the double discontinuity of \(\G_S(v,u)\) are from the identity and from \(\sigma\), yielding the same contributions as in the previous section.
The new term is part of the combination \(\frac{1}{2} \left( \G_T - \G_A\right) \): let us call this contribution  \(I_{T-A}\).
It arises from the fact that the leading twist operators in the \(T\) and \(A\) have order \( N^{-1}\) corrections to their scaling dimensions, as found in section~\ref{sec:TA_Nmin1}.
These then generate a contribution with double discontinuity at order \(N^{-2}\) from expanding \( (1-\zb)^{\tau_\ell/2} = (1-\zb)^{\Delta_\varphi}\left( 1+ \frac{1}{2}\gamma_\ell \log(1-\zb) + \frac18 \gamma_\ell^2 \log^2(1-\zb) + \dots \right)\) and noting that \(\log^2(1-\zb)\) has a non-trivial double discontinuity.
Explicitly, the relevant contribution is
\begin{align}
I_{T-A}=\left(\frac{\zb}{1-\zb}\right)^{\Delta_\varphi}\frac12\Bigg(
&\sum_{\ell=0,2,\ldots}a_{T,\ell} \gamma_{T,\ell}^2 G^{(d)}_{2\mu-2+\ell,\ell}(1-\zb,1-z)\\&-
\sum_{\ell=1,3,\ldots}a_{A,\ell} \gamma_{A,\ell}^2 G^{(d)}_{2\mu-2+\ell,\ell}(1-\zb,1-z)
\Bigg)\frac{\log^2(1-\zb)}{8} + \order(N^{-3}).\nonumber
\end{align}
Ultimately we are interested in the small $z$ limit of $I_{T-A}$. Since the conformal blocks have a regular expansion in the opposite limit, $z\to1$, we cannot swap the order of limit and summation. To directly compute the sum defining $I_{T-A}$ is a formidable task, as it involves an infinite sum over conformal blocks in generic dimension. Our strategy will be to employ the technology of \emph{twist conformal blocks}, introduced in \cite{Alday:2016njk,Alday:2016jfr}, to write a differential equation that relates $I_{T-A}$ to the same sum without the factors of $\gamma_{R,\ell}$. This sum is explicitly known, it is just the $\frac12(T-A)$ part of the free field correlator!

First we make the following observation \cite{Alday:2016jfr}: the anomalous dimensions in the representations \(T\) and \(A\) take the same functional form:
\begin{equation}
\gamma^{(1)}_{T,\ell}=\gamma^{(1)}_{A,\ell}=-\frac{2 (\mu -1) \mu}{J^2}  \gamma_\varphi^{(1)}.
\end{equation}
Furthermore, since \( \left( \gamma_{T/A ,\ell} \right)^2 \sim \frac{1}{N^2} \), all other parts can be evaluated at tree level.
Specifically we can use the tree level values of $a_{T,\ell}$ and $a_{A,\ell}$, which also have the same functional form \(a_{T,\ell} = - a_{A,\ell}\), so that 
\begin{align}
I_{T-A} & = \frac{1}{N^2} \frac{\log^2(1-\zb)}{8} \left(\frac{\zb}{1-\zb}\right)^{\mu-1}    \sum_{\ell=0,1,2,\dots} \frac{ a_{T,\ell}^{(0)} }{2} \frac{4(\mu -1)^2 \mu ^2}{J^4} \left( \gamma_\varphi^{(1)}\right)^2  G^{(d)}_{2\mu-2+\ell,\ell}(1-\zb,1-z)\nonumber\\
&=\frac{1}{N^2} \frac{\log^2(1-\zb)}{8}~ \left(\frac{\zb}{1-\zb}\right)^{\mu-1}  \kappa ~  H^{(2)}_{T-A}(1-\zb,1-z),
\end{align}
where \( \kappa=4 (\mu -1)^2 \mu ^2\left( \gamma_\varphi^{(1)}\right)^2\), and where
\begin{equation}
H^{(m)}_{T-A}(z,\zb) := \sum_{\ell=0,1,2,\dots} \frac{a_{T,\ell}^{(0)}}{2 J^{2m}} G^{(d)}_{2\mu-2+\ell,\ell}(z,\zb)=\frac12\left(H^{(m)}_T(z,\zb)-H^{(m)}_A(z,\zb)\right)
\end{equation}
are the level $m$ twist conformal blocks. They are defined as a sum of conformal blocks over a family of operators with identical twist, in our case $\tau=2\mu-2$, modulated by a power $J^{-2m}$ of the conformal spin. At level $0$, the twist conformal block is simply the contribution of that twist family to a four-point function, and for twist at the unitarity bound $\tau=2\mu-2$ it reduces to the free theory correlator. 

Acting with the quadratic Casimir $\cas$, defined in \eqref{eq:CasimirDef}, with $\tau=2\mu-2$, we generate a recursion relation for the level $m$ twist conformal blocks:
\begin{equation}\label{eq:Hrecursionrelation}
{\cas}^m H^{(m)}_{T-A}(z,\zb)=H^{(0)}_{T-A}(z,\zb)=\left(\frac uv\right)^{\mu-1}.
\end{equation}
We then employ a combination of two methods to find explicitly the small $z$ limit of $H_{T-A}^{(2)}(1-\zb,1-z)$. Solving \eqref{eq:Hrecursionrelation} for $m=2$ in the (crossed channel) collinear limit gives a solution with a free parameter. On the other hand, we can use that $H_{T-A}^{(2)}(z,\zb)$ is a sum over conformal blocks at the unitarity bound. As we show in appendix~\ref{sec:unitarityboundblocks}, the expressions for these blocks and their sums simplify dramatically. This results in an explicit sum that fixes the final coefficient. We give the full details of the computation in appendix~\ref{app:HT_pm_A}, and the result is
\begin{equation}
\left. H_{T-A}^{(2)} (1-\zb, 1-z) \right|_{z^0} = \xit^{\mu-1} \left( g_0^{(-)}(\xit) \log z + f_0^{(-)}(\xit) \right),
\end{equation}
with \( \xit \equiv (1-\zb)/\zb\), where 
\begin{align}
g_0^{(-)}(\xit) = -\frac{\pi  \csc (\pi  \mu )}{\mu -2},
\end{align}
and 
\begin{align} \label{eq:H_TminA_nonLogRes}
f^{(-)}_0 (\xit ) = \pi  \csc (\pi  \mu )  & \bigg[\frac{\pi  \cot (\pi  \mu )}{\mu -2} -\frac{1}{(\mu -2)^2} +(\mu -1)\xit  \,_4F_3\!\left(\!\left. {1,1,3-\mu,\mu}~\atop~{\!\!\!2,2,2} \right| -\xit\right) \nonumber \\
& \qquad-\frac{\, _2F_1(2-\mu ,\mu -1;1;-\xit )-1}{(\mu -2)^2}\bigg].
\end{align}

It is now trivial to solve for the contribution to \(U^{(1)}_S\): 
\begin{align}
U^{(1)}_{S,\hb} 
&= \frac{1}{N} U^{(1)}_{T,\hb} + \frac{2}{N^2}  \, \inv{\frac{\kappa}{8}\log^2(1-\zb) g_0^{(-)}(\xit)} \nonumber\\
&= \frac{1}{N} U^{(1)}_{T,\hb} -(2 \hb- 1) \frac{4 \pi \csc (\pi  \mu )  (\mu -1)^2 \mu ^2}{(\mu -2)J^2N^2 }\left( \gamma_\varphi^{(1)}\right)^2.
\end{align}
From this we can extract the first order correction to the anomalous dimensions 
\begin{equation}\label{eq:gammaSres}
\gamma_{S,\ell}^{(1)} = -\frac{2\gamma_{\varphi}^{(1)}}{J^2}\left(   (\mu -1) \mu +  \gamma_{\varphi}^{(1)} \frac{\pi  \csc (\pi  \mu ) \Gamma (\mu +1)^2 \Gamma (\ell+1)}{(\mu -2) \Gamma (\ell+2 \mu -3)}   \right),
\end{equation}
where \( J^2= (\ell+\mu-1)(\ell+\mu-2)+ O\left(N^{-1}\right) \). 
Conservation of the stress tensor implies that \( \Delta_{S,2}=d\), i.e. $2\gamma^{(1)}_{\varphi } + \gamma^{(1)}_{S,2}=0$. 
This leads to a quadratic equation for $\gamma^{(1)}_{\varphi }$, whose solutions are
\begin{equation}\label{eq:gammaphisols}
\gamma^{(1)}_{\varphi }=0,\quad \gamma^{(1)}_{\varphi }= \frac{(\mu-2)\Gamma(2\mu-1)}{\Gamma(\mu+1)\Gamma(\mu)^2\Gamma(1-\mu)}.
\end{equation}
We recognize this as the known results for the free theory and the interacting critical \(\mathrm{O}(N)\) theory.
This followed entirely from conformal symmetry and the assumptions given in section~\ref{sec:assumptions}.
Using the non-trivial solution for $\gamma^{(1)}_{\varphi }$ from \eqref{eq:gammaphisols}, we can check that all results computed above agree with their literature values. In particular the singlet anomalous dimensions \eqref{eq:gammaSres} agree with the values given in \cite{Manashov:2017xtt} (computed first in \cite{Lang:1992zw}) and the non-singlet OPE coefficients \eqref{eq:OPETAfirstorder} agree with the results of \cite{Dey:2016mcs}\footnote{Note that there is a typo in equation~(4.21) of \cite{Dey:2016mcs}. We thank the authors for sharing with us a file with the correct result.}.

It is interesting to note that if we assume that the shadow relation $\Delta_\sigma=d-\Delta_{S,0}$ holds also at order $N^{-1}$, we can compute
\begin{equation}\label{eq:gammasigmares}
\gamma_\sigma^{(1)}=\frac{4(\mu-1)(2\mu-1)}{\mu-2}\gamma_\varphi^{(1)},
\end{equation}
in agreement with the literature value \cite{Vasiliev:1981yc,Manashov:2017xtt}.

For the OPE coefficients, we invert the expression \eqref{eq:H_TminA_nonLogRes}.
Using the inversion methods from appendix \ref{sec:InvMethods}, we find \(U^{(0)}_S\):
\begin{align}
U^{(0)}_{S,\hb} 
&=\frac1N U^{(0)}_{T,\hb}  +  \frac{1}{N^2} \inv{\frac{\kappa}{8}\log^2(1-\zb) f_0^{(-)}(\xit)} \nonumber\\
&=\frac1N  U^{(0)}_{T,\hb}  + \frac{2(2 \hb-1)\mu ^2 (\mu -1)^2  \pi  \csc (\pi  \mu ) }{(\mu -2)^2N^2} \left(\gamma_\varphi^{(1)}\right)^2 \Bigg[   \frac{(\mu -2)\pi \cot (\pi  \mu )}{J^2}-\frac{1}{J^2}   + \nonumber\\
& \qquad\qquad (\mu -2) \frac{ \mathbf{S_1}[\mu-1](\hb) - 1/J^2}{J^2} \Bigg].
\end{align}
Hence we find the full OPE coefficients:
\begin{align}
a_{S,\ell }&= \frac{1}{N} a^{\text{free}}_{\mu,\ell}  + \frac{1}{N^2} \Bigg[ 2\, a^{\text{free}}_{\mu,\ell}   \gamma _{\varphi }^{(1)} \bigg(S_1(\ell+\mu -2)-S_1(\mu -2)  + \frac{\ell^2+2 \mu ^2+2 (\mu -1) \ell-3 \mu +1}{(\ell+\mu -1)^2 (2 \ell+2 \mu -3)}   \nonumber \\
& \quad  + \frac{(\ell-1) (2 \mu +\ell-2) \left(S_1(\ell+2 \mu -4)-S_1(2 \ell+2 \mu-3)\right)}{(\ell+\mu -1) (\ell+\mu -2)} \bigg) \nonumber \\ 
& \quad + \frac{2 \pi  (\mu -1)^2 \mu ^2\csc (\pi  \mu ) \Gamma (\ell+\mu -1)^2}{(\mu  -2) (\ell+\mu -2) (\ell+\mu -1) \Gamma (2 \ell+2 \mu -3)}  \left(\gamma _{\varphi }^{(1)}\right)^2  \bigg(  \frac{2 (\ell+\mu -2)}{(\ell+\mu -1) (2 \ell+2 \mu -3)}\nonumber\\
&\quad+ \frac{1}{2-\mu } +\pi  \cot (\pi  \mu ) +2 S_1(2 \ell+2 \mu -3)-S_1(\ell+2 \mu -4)-S_1(\ell) \bigg) \Bigg] + O(N^{-3}),\label{eq:OPESres}
\end{align}
where \(a^{\text{free}}_{\mu,\ell} =\frac{2 \Gamma (\ell+\mu -1)^2 \Gamma (\ell+2 \mu -3)}{\Gamma (\ell+1) \Gamma (\mu -1)^2 \Gamma (2 \ell+2 \mu -3)} \) are the OPE coefficients in the theory of a single free scalar field.

From this result we can also compute the order \(1/N\) correction to the central charge:
\begin{equation}
\frac{C_T}{C_{T,\mathrm{free}}}=1 +   \frac{\gamma_\varphi^{(1)}}{\mu(\mu+1)}  \left(\frac4{\mu-2}-\mu - 2\pi \mu \cot(\pi\mu)-2\mu\, S_1(2\mu-2)\right)\frac{1}{N}+\order(N^{-2}),
\end{equation}
matching the known result \cite{Petkou:1994ad}.

\section{Non-singlet CFT-data at order $N^{-2}$}
\label{sec:subsubleading}

In this section we move on to computing the CFT-data at order $N^{-2}$. 
We will consider only the $T$ and $A$ representations, whose crossing equations read
\begin{equation}\label{eq:crossingTANm2}
\G_{T/A}(u,v)=\left(\frac{u}{v}\right)^{\Delta_\varphi}\left(\pm\G_S(v,u)+\frac{\G_T(v,u)+\G_A(v,u)}2\mp\frac1N\G_T(v,u)\right).
\end{equation}
On the right-hand side, the singlet ($S$) representation now contains three different contributions. 
Firstly the identity operator gives $A[\Delta_\varphi](\hb)$ in $U_\hb^{(0)}$ to any order. 
Secondly, there are subleading corrections to the contribution from $\sigma$: we denote this contribution $I_\sigma$ and compute it in section~\ref{sec:sigmainversionrevisited}. 
Thirdly the operators $[\sigma,\sigma]_{n,\ell}$ will contribute at this order. 
To find their contribution $I_{[\sigma,\sigma]_{n,\ell}}$ we need the compute a double sum over $n=0,1,2,\ldots$ and $\ell=0,2,4,\ldots$. 
As discussed in section~\ref{sec:sigmasigma}, this sum can be given in a closed form as an integral in Mellin space, but the inversion is highly non-trivial.

The operators contributing in the $T$ and $A$ representations on the right-hand side of \eqref{eq:crossingTANm2} are the leading twist operators themselves, whose contribution is suppressed by $\gamma_{T/A}^2\sim N^{-2}$. 
We can therefore ignore the last term in \eqref{eq:crossingTANm2}, and thus note that the contribution from the $T$ and $A$ representation takes a special form $I_{T+A}$, similar to what we saw in section~\ref{sec:S_at_N-1}. 
This means that we will be able to determine this contribution in a similar way to $I_{T-A}$, which we do first. 

\subsection{Inverting $H_{T + A}^{(2)}$} \label{sec:H_TpA}
Consider the contributions from the $T$ and $A$ representations in \eqref{eq:crossingTANm2}, which at order \(N^{-2}\) will arise from the squares of anomalous dimensions, exactly as for \(I_{T-A}\) in section \ref{sec:S_at_N-1}.
The computation of the double discontinuity is almost identical to section~\ref{sec:S_at_N-1}, with a contribution
\begin{equation}
I_{T+A}= \frac{1}{N^2} \frac{\log^2(1-\zb)}{8}~ \left(\frac{\zb}{1-\zb}\right)^{\mu-1}  \kappa ~  H^{(2)}_{T+A}(1-\zb,1-z),
\end{equation}
where the twist conformal blocks \(H_{T+A}^{(m)}\) are defined exactly analogously to the \(H_{T-A}^{(m)}\):
\begin{equation}
H^{(m)}_{T+A}(z,\zb) := \sum_{\ell=0,1,2,\dots} \frac{1}{2}  \left( a_{T,\ell}^{(0)}\delta_{\ell,\text{even}}  +  a_{A,\ell}^{(0)}\delta_{\ell,\text{odd}}  \right)\frac{1}{ J^{2m}} G^{(d)}_{2\mu-2+\ell,\ell}(z,\zb),
\end{equation}
which satisfy the equations
\begin{equation}
{\cas}^m H^{(m)}_{T+ A}(z,\zb)=H^{(0)}_{T+A}(z,\zb)= u^{\mu-1}
\end{equation}
for $\cas$ defined in \eqref{eq:CasimirDef}.

The twist conformal block \(H_{T+A}^{(2)}(1-\zb,1-z) \) is found in a very similar way to the twist conformal block \(H_{T-A}^{(2)}(1-\zb,1-z) \), and details of the computation, which relies crucially on the fact that the intermediate operators saturate the unitarity bound, can be found in appendix \ref{app:HT_pm_A}.
The result is that 
\begin{equation}
\left. H_{T+A}^{(2)} (1-\zb, 1-z) \right|_{z^0} = \xit^{\mu-1} \left( g_0^{(+)}(\xit) \log z + f_0^{(+)}(\xit) \right),
\end{equation}
where
\begin{align} \label{eq:H_TplusA_logRes}
g_0^{(+)}(\xit) = -\frac{1}{(\mu-2)^2},
\end{align}
and\footnote{Here \( \left(\Dbar-K^2\right)^{-1} \left[h(\xit)\right]\) is defined to be the solution \(q\) to the linear ODE \(\Dbar\, q(\xit) - K^2 q(\xit) = h(\xit) \) with a boundary condition that is described in appendix \ref{sec:HTpmADiffEqn}. }
\begin{align} \label{eq:H_TplusA_nonLogRes}
 f^{(+)}_0 (\xit ) = \frac{1}{(\mu -2)^2}& \bigg[ \frac{(\mu -2)^2 (S_2(\mu -2)-\zeta_2)+\mu -3}{\mu -2}\, _2F_1(2-\mu ,\mu-1;1;-\xit ) \nonumber \\
& \quad  + \frac{1}{2} (\mu -2) (\mu -1) \mu \left(\Dbar-K^2\right)^{-1} \left[\xit  \,_3F_2\!\left(\! \left. {1,1,\mu +1}~\atop~{\!\!\!\!2,3} \right| -\xit\right)\right] \nonumber\\
  &  \quad -\frac{\mu -3}{\mu -2} - S_1(\mu -2) \bigg],
\end{align}
with \(K^2 = (\mu-1)(\mu-2)\), and where \(\Dbar\) was defined in \eqref{eq:dbardefn}.
The contribution to \(U^{(1)}_{T/A,\hb}\) is straightforwardly found as the inversion of \(g_0^{(+)} \): 
\begin{align} \label{eq:U1_TA_fromTplusA_Nmin2}
U^{(1),(T+A)}_{T/A,\hb} &= 2  \, \inv{\frac{\kappa}{8}\log^2(1-\zb) g_0^{(+)}(\xit)} \nonumber\\
&= -(2 \hb-1) \frac{4  (\mu -1)^2 \mu ^2}{J^2 (\mu -2)^2} \left( \gamma _{\varphi }^{(1)} \right)^2 .
\end{align}
The contribution to \(U^{(0)}_{T/A,\hb}\) is more complicated, since it involves the inversion of a complicated hypergeometric \(\,_3F_2\).
The result is 
\begin{align} 
U^{(0),(T+A)}_{T/A,\hb} &= \frac{1}{N^2} \inv{\frac{\kappa}{8}\log^2(1-\zb) f_0^{(+)}(\xit)} \nonumber\\
&= \frac{2 \hb-1}{N^2}   \frac{\mu ^2 (\mu -1)^2}{(\mu -2)^3}  \left( \gamma _{\varphi }^{(1)} \right)^2 \Bigg( 
 \frac{2 \left(3-\mu -(\mu -2) S_1(\mu -2)\right)}{J^2}  \nonumber\\
& \qquad + \frac{2 \left(\mu-3 +(\mu -2)^2 \left(S_2(\mu -2)-\zeta _2\right) \right)}{J^2-K^2} + \frac{ \mu  (\mu -1) (\mu -2)^2 B(\hb,\mu)}{J^2-K^2}      \Bigg),\label{eq:U0_TA_fromTplusA_Nmin2}
\end{align}
where we defined
\begin{equation} \label{eq:Bdefn}
(2\hb-1)B(\hb,\mu) = \inv{\frac{1}{4} \log^2(1-\zb)\, \xit  \,_3F_2\!\left(\! \left. {1,1,\mu +1}~\atop~{\!\!\!\!2,3} \right| -\xit\right)  }. 
\end{equation}
The inversion of this can be found via various methods, e.g. via the Mellin space method of appendix \ref{sec:Mellinmethods}, or by hitting it once with the Casimir \(\Dbar\), collecting powers of \(\zb \) and \( (1-\zb) \), and inverting term-by-term as in appendix \ref{sec:wSeriesInv}.
Using the Mellin space method, we find that:
\begin{align} \label{eq:BResExact}
B(\hb,\mu)   = &   \frac{1}{J^2 \left(J^2-2\right)} \, _4F_3\!\left(\!\left. {1,1,2,\mu+1}~\atop~{\!\!3,3-\hb,\hb+2} \right|1\right)       \nonumber\\
&   -\frac{2 \pi  \Gamma \left(\hb\right) \Gamma \left(\mu +\hb-1\right)}{J^2 \Gamma (\mu +1) \sin \left(\pi  \hb\right)  \Gamma \left(2 \hb\right)} \, _3F_2\!\left(\!\left. {\hb-1,\hb,\hb+\mu -1}~\atop~{\!\!2 \hb,\hb+1} \right|1\right).
\end{align}
It has a large \(J\) expansion of the form:
\begin{equation} \label{eq:BLargeJSer}
B(\hb,\mu) \sim \frac{1}{J^4} -\frac{2 (\mu -2)}{3 J^6} + \frac{\mu ^2-\frac{7 \mu }{3}+\frac{2}{3}}{J^8} + \dots.
\end{equation}
%

\subsection{Contribution from scalar block revisited}\label{sec:sigmainversionrevisited}

The contribution from $\sigma$ at order $N^{-2}$ will contain subleading corrections for both $\Delta_\varphi$ and $\Delta_\sigma$, compared with what we saw at tree-level in section~\ref{sec:sigmatree}.

To compute the CFT-data at order $N^{-2}$ for the $T$ and $A$ representations, we need to evaluate the $U^{(p)}_\hb|_{\Delta}$ for $\Delta_\varphi=\mu-1+\gamma^{(1)}_\varphi/N$ and $\Delta=2+\gamma^{(1)}_\sigma/N$. 
We can truncate these expansions to this order since $a_\sigma$ is already of order $N^{-1}$.
Inserting these expansions in \eqref{eq:genericcrossedblock} gives a well-defined inversion problem at $O(N^{-2})$.

The $\log z$ computation can be done explicitly, for example by identifying the asymptotic $J^2$ series.
The result is 
\begin{align} \label{eq:U1fromsigmaN-2}
U^{(1)}_\hb|_{\Delta_\sigma}=2(\mu-2)^2\left(\frac{a_{\sigma}^{(0)}}{N}+\frac{a_{\sigma}^{(1)}}{N^2}\right)&\frac{A[\Delta_\varphi](\hb)}{J^2}
\bigg[
-1+\frac{\gamma^{(1)}_\varphi}N\left(\frac1{J^2}-\frac{2}{\mu-2}\right)
\\
&
+\frac{\gamma_\sigma^{(1)}}{N}\left(
\frac{3-\mu}{\mu-2}+S_1(\hb-1)-S_1(\mu-2)
\right)
\bigg]+O(N^{-3}),\nonumber
\end{align}
where $a_{\sigma}^{(0)}$ is given by \eqref{eq:asigma0res} and $\gamma_\sigma^{(1)}$ and $a_{\sigma}^{(1)}$ are yet to be determined. 
Here we have multiplied out the factor $A[\Delta_\varphi](\hb)$ rather than its tree-level counterpart  $A[\mu-1](\hb)$, which simplifies the term proportional to $\gamma^{(1)}_\varphi$.

For $U^{(0)}_\hb|_{\Delta_\sigma}$, we expand the double-discontinuity in $\xit$ to generate a large $J^2$ series, which we again recognize:
\begin{align}\nonumber
U^{(0)}_\hb|_{\Delta_\sigma}&=(\mu-2)^2\left(\frac{a_{\sigma}^{(0)}}{N}+\frac{a_{\sigma}^{(1)}}{N^2}\right)\frac{A[\Delta_\varphi](\hb)}{J^2}\bigg[
\mathbf{S_1}[\Delta_\varphi](\hb)-\frac1{J^2}
\\&+
\frac{\gamma_\varphi^{(1)}}{N}\left(
\frac{2\mathbf{S_1}[\Delta_\varphi](\hb)}{\mu-2}-\frac{\mathbf{S_1}[\Delta_\varphi](\hb)}{J^2}+\frac2{J^4}+\frac{2(\mu-3)}{(\mu-2)J^2}
\right)
\label{eq:U0_TA_fromsigmaNm2}
\\&-
\frac{\gamma_\sigma^{(1)}}{N}\left(
\zeta_2+
\left(\mathbf{S_1}[\Delta_\varphi](\hb)-\frac1{J^2}\right)\left(\frac{3-\mu}{\mu-2}+S_1(\hb-1)-S_1(\mu-2)\right)
\right)
\bigg]+O(N^{-3}),\nonumber
\end{align}
where $\mathbf{S_1}$ was defined in \eqref{eq:Qdefalpha}.

\subsubsection{Relation to crossing kernels} \label{sec:crossingKernels}

Let us be a bit more general and consider the problem of inverting the contribution from a single scalar operator with generic dimension $\Delta$ appearing in the crossed channel of the correlator of identical scalars with dimension $\Delta_\varphi$. 

\begin{quotation}
\noindent \emph{What contribution to the CFT-data of leading twist operators follows from the appearance of a single scalar operator in the crossed channel?}
\end{quotation}

Recall that this problem amounts to computing \eqref{eq:inversionproblemblock}, where the scalar block in the small $z$ limit is given by \eqref{eq:genericcrossedblock}. In the case of $U^{(1)}_\hb|_\Delta$ one can compute the integral in terms of two ${_4F_3}$ functions using the Mellin space methods in appendix~\ref{sec:Mellinmethods}. 
Here we will proceed with an alternative approach that also generalizes to $U^{(0)}_\hb|_\Delta$. 
Replacing the hypergeometric functions by their defining infinite sums in both the inversion kernel $ k_\hb(\zb)/\zb^2$ and in the $\log z$ piece of \eqref{eq:scalarblocksmallzterm}, we get the following expression for $U^{(1)}_\hb|_\Delta$ after exchanging the order of summation and integration:
\begin{align}\nonumber
U^{(1)}_\hb|_\Delta&=-4a_\Delta\sin^2\left(\pi(\Delta_\varphi-\tfrac\Delta2)\right)\frac{\Gamma(\hb)^2}{\pi^2\Gamma(2\hb-1)}\frac{\Gamma(\Delta)}{\Gamma(\frac\Delta2)^2}
\\
&\quad\times\sum_{m,n=0}^\infty\frac{(\hb)_n^2}{(2\hb)_n(1)_n}\frac{(\frac\Delta2)_m^2}{(1+\Delta-\mu)_m(1)_m}\int\limits_0^1 \mathrm{d} \zb~\zb^{\Delta_\varphi+\hb-2+n}(1-\zb)^{\frac\Delta2-\Delta_\varphi+m}.
\end{align}
The integral mixes the $m$ and $n$ dependence, which means that the result is a double infinite sum that defines a \emph{Kamp\'e de F\'eriet function}:
\begin{align}
U^{(1)}_\hb|_\Delta&=-4a_\Delta\sin^2\left(\pi(\Delta_\varphi-\tfrac\Delta2)\right)\frac{\Gamma(\hb)^2}{\pi^2\Gamma(2\hb-1)}
\frac{\Gamma(\Delta)}{\Gamma(\frac\Delta2)^2}\frac{\Gamma(1+\frac\Delta2-\Delta_\varphi)\Gamma(\Delta_\varphi+\hb-1)}{\Gamma(\frac\Delta2+\hb)}\nonumber
\\\label{eq:crossingkernellog}
&\quad \times F^{03}_{11}\left(\begin{matrix}\\\frac\Delta2+\hb\end{matrix}\middle|\begin{matrix}\left\{\frac\Delta2,\frac\Delta2,\frac\Delta2-\Delta_\varphi+1\right\},\left\{\hb,\hb,\hb+\Delta_\varphi-1\right\}\\\left\{\Delta-\mu+1\right\},\left\{2\hb\right\}\end{matrix}\middle|1,1\right).
\end{align}
The Kamp\'e de F\'eriet function is a hypergeometric function of two variables defined by \cite{exton1978handbook}
\begin{align}\nonumber
& F^{pq}_{rs}\left(
\begin{matrix}a_1,\ldots, a_p\\c_1,\ldots c_r\end{matrix}
\middle|
\begin{matrix}\{b_1,\ldots, b_q\},\{b'_1,\ldots, b'_q\}\\\{d_1,\ldots, d_s\},\{d'_1,\ldots,d'_s\}\end{matrix}
\middle|x,y\right)
\\&\qquad\qquad\qquad\qquad=\sum_{m,n=0}^\infty\frac{(a_1)_{m+n}\cdots (a_p)_{m+n}}{(c_1)_{m+n}\cdots (c_r)_{m+n}}\frac{(b_1)_{m}(b'_1)_{n}\cdots (b_q)_{m}(b'_q)_{n}}{(d_1)_{m}(d'_1)_{n}\cdots (d_s)_{m}(d'_s)_{n}}\frac{x^my^n}{m!n!}.
\label{eq:KdFdef}
\end{align}
The problem for $U^{(0)}_\hb|_\Delta$ is analogous. The terms $S_1(m+\tfrac\Delta2-1)$ and $\log \zb$ can be rewritten in terms of derivatives of a Gamma function and $\zb$ power respectively, and after some simplifications we get
\begin{align}\nonumber
U^{(0)}_\hb|_\Delta&=-2a_\Delta\sin^2\left(\pi(\Delta_\varphi-\tfrac\Delta2)\right)\frac{\Gamma(\hb)^2}{\pi^2\Gamma(2\hb-1)}\frac{\Gamma(\Delta)}{\Gamma(\frac\Delta2)^2}\frac{\Gamma(1+\frac \Delta2-\Delta_\varphi)\Gamma(\Delta_\varphi+\hb-1)}{\Gamma(\frac\Delta2+\hb)}
\\\nonumber
&\times\bigg\{\left[2S_1(\tfrac\Delta2-1)-S_1(\tfrac\Delta2+\hb-1)+S_1(\hb+\Delta_\varphi-2)\right]
\\&\quad\quad\times
F^{03}_{11}\left(\begin{matrix}\\\frac\Delta2+\hb\end{matrix}\middle|\begin{matrix}\left\{\frac\Delta2,\frac\Delta2,\frac\Delta2-\Delta_\varphi+1\right\},\left\{\hb,\hb,\hb+\Delta_\varphi-1\right\}\\\left\{\Delta-\mu+1\right\},\left\{2\hb\right\}\end{matrix}\middle|1,1\right)
\label{eq:U0fromsingleblock}
\\&\quad\quad+\dda
F^{03}_{11}\left(\begin{matrix}\\\frac\Delta2+\hb+a\end{matrix}\middle|\begin{matrix}\left\{\frac\Delta2,\frac\Delta2+2a,\frac\Delta2-\Delta_\varphi+1\right\},\left\{\hb,\hb,\hb+\Delta_\varphi-1+a\right\}\\\left\{\Delta-\mu+1\right\},\left\{2\hb\right\}\end{matrix}\middle|1,1\right)
\bigg\},
\nonumber
\end{align}
where, as above, $\dda f=\left.\frac{\partial f}{\partial a}\right|_{a=0}$. Let us emphasize that the expressions \eqref{eq:crossingkernellog},\eqref{eq:U0fromsingleblock} for $U^{(0)}_\hb|_\Delta$ are completely general, valid within any four-point function of identical scalar operators, and without reference to any perturbative parameters. 

The problem above is closely related to crossing kernels and $6j$ symbols \cite{Sleight:2018epi, Cardona:2018dov, Sleight:2018ryu, Liu:2018jhs, Cardona:2018qrt, Albayrak:2019gnz}.
In \cite{Sleight:2018ryu} expressions were given for the leading order anomalous dimension of a double-field operator from a generic operator in the crossed channel. 
In order to compare with these results we have to make some further assumptions.

If we assume that the $\varphi\times\varphi$ OPE contains the identity and the scalar of dimensions $\Delta$ and OPE coefficient $a_\Delta$, as well as the GFF operators $[\varphi,\varphi]_{n,\ell}$, but no further operators until order $a_\Delta^2$, then at order $a_\Delta$ the anomalous dimensions are given by
\begin{equation}\label{eq:crossingkernelres}
\gamma_\ell|_\Delta=\frac{U^{(1)}_\hb|_\Delta}{A[\Delta_\varphi](\hb)}+O\big(a_\Delta^2\big).    
\end{equation}
Furthermore, up to this order, the contributions from several scalar operators are additive provided their OPE coefficients are of the same order in expansion. 
One can check that the expression \eqref{eq:crossingkernelres} agrees with the result for leading order anomalous dimensions of the leading twist operators, given in (1.41) of \cite{Sleight:2018ryu}.

Notice, however, that the result \eqref{eq:crossingkernelres} is only valid for the anomalous dimensions to leading orders in $a_\Delta$. 
If, for instance $\Delta=\Delta^{(0)}+\Delta^{(1)}a_\Delta$, the anomalous dimensions at order $O\big(a_\Delta^2\big)$ will not be given by just expanding the parameter $\Delta$ in \eqref{eq:crossingkernelres}. 
The reason is that even with the most minimal assumptions further contributions need to be considered. 
Firstly, the denominator would have to include the terms $U^{(0)}_\hb|_\Delta+\frac12\partial_\hb U^{(1)}_\hb|_\Delta$. 
Secondly, more operators will now necessarily contribute to the numerator. 
These are the double-twist operators themselves which contribute with a non-zero double-discontinuity proportional to $\gamma_\ell\sim a_\Delta^2$ in $U^{(1)}_\hb$, and operators $[\O,\O]_{n,\ell}$ which contribute to both $U^{(1)}_\hb$ and $U^{(2)}_\hb$ \footnote{It is clear from the left-hand side that we must have a non-zero $U^{(2)}_\hb$ at order $a_\Delta^2$.
Our minimal assumptions guarantee the existence of $[\O,\O]_{n,\ell}$ \cite{Fitzpatrick:2012yx,Komargodski:2012ek}, and furthermore their inversion gives the correct large $J$ expansion $U^{(2)}_\hb\sim a_\Delta^2/J^{2\Delta}$. 
In a more generic theory the origin of $U^{(2)}_\hb$ could be more complicated.
}. 

\subsection{Contribution from $[\sigma,\sigma]_{n,\ell}$}\label{sec:sigmasigma}
The final contribution to the CFT-data at this order is the contribution \(I_{[\sigma,\sigma]_{n,\ell}} \) from the double field operators \( [\sigma,\sigma]_{n,\ell} \), which we can think of as the operators \( \sigma \square^n \partial^{\mu_1}\dots\partial^{\mu_\ell} \sigma\).
This contribution takes the form
\begin{equation}
I_{[\sigma,\sigma]_{n,\ell}}  = \xit^{1-\mu} \left.S(v,u) \right|_{z^0} ,\qquad S(v,u)=\sum_{n,\ell}c^2_{\varphi^i\varphi^i[\sigma,\sigma]_{n,\ell}}G_{4+2n,\ell}^{(d)}(1-\zb,1-z).
\end{equation}
The fact that the OPE coefficient \(c_{\varphi\varphi\sigma} \sim N^{-1/2}\) implies that the double field operators couple to \( \varphi\) with OPE coefficients \(c_{\varphi\varphi[\sigma,\sigma]_{n,\ell}} \sim N^{-1} \), which means that they contribute to our correlator \( \langle \varphi \varphi \varphi \varphi \rangle\) at order \(N^{-2}\).
These OPE coefficients can be found by studying crossing for the mixed correlator \( \langle \varphi\varphi\sigma\sigma \rangle \), in which these operators appear with OPE coefficients \(c_{\varphi\varphi[\sigma,\sigma]_{n,\ell}} c_{\sigma\sigma [\sigma,\sigma]_{n,\ell}} \sim N^{-1}\). In appendix~\ref{sec:mixed} we show that they can be computed from just a single crossed-channel operator, namely $\varphi^i$, resulting in the expression
\begin{equation} \label{eq:sigsigOPECoeffs}
c^2_{\varphi^i\varphi^i[\sigma,\sigma]_{n,\ell}}=\frac{1}{N^2} K(\mu)^2  \frac{(\mu-1-n)_n^2}{(\mu-3-n)_n^2}  \frac{1}{J^4_{2n+4,\ell}} \left. a^{\mathrm{GFF}}_{n,\ell}\! \right|_{\Delta=2} \, +\order(N^{-3}).
\end{equation}
Here the \( a^{\mathrm{GFF}}_{n,\ell}|_{\Delta=2}\) are the generalized free field OPE coefficients (normalized as in eq.~(2.10) of \cite{Henriksson:2018myn}) of a free scalar of dimension 2, \(J^2_{2n+4,\ell}=(n+\ell+2)(n+\ell+1)\) is the conformal spin of the operator \( [\sigma,\sigma]_{n,\ell}\), and \(K(\mu) =\frac{\mu-2}{\mu-3}a_\sigma^{(0)}= \frac{\mu(\mu-1)}{(\mu-2)(\mu-3)} \gamma_{\varphi}^{(1)} \) is a constant introduced to simplify the discussion below.

\subsubsection*{Mellin amplitude}
From \eqref{eq:sigsigOPECoeffs}, and the known expressions for the subleading corrections to the conformal blocks, we can find the function \(S(v,u)\) order by order in \(v \). 
The first few terms read
\begin{equation}\label{eq:Svuser}
S(v,u)=K(\mu)^2\left(\frac{\log ^2u}{2} v^2 +\frac{ (7-2 \mu ) \log^2 u+(16-5 \mu)\log u-2 (\mu
   -2)\zeta _2 }{(\mu -4)^2}v^3+\ldots
   \right)\!.
\end{equation}
Unfortunately, we have not been able to find a closed form expression for this series in position space. However, we have been able to identify a closed form for its Mellin transform.
We follow the conventions for holographic correlators given in \cite{Rastelli:2017udc}, and define the Mellin amplitude \(\mathcal{M}(s,t)\) for identical external operators of dimension $\Delta_\varphi$ by

\begin{equation}\label{eq:MellinDef}
\mathcal G(u,v)=\frac14\int_{-\ii\infty}^{\ii \infty}\frac{dsdt}{(2\pi \ii)^2}\Gamma\left(\Delta_\varphi-\tfrac s2\right)^2\Gamma\left(\Delta_\varphi-\tfrac t2\right)^2\Gamma\left(\Delta_\varphi-\tfrac{\ut}{2}\right)^2\, u^{\frac s2}v^{\frac t2-\Delta_\varphi}\, \mathcal M(s,t),
\end{equation}
where $s+t+\ut=4\Delta_\varphi$ are the ``Mandelstam variables".
Our strategy will be to construct explicitly the Mellin amplitude $\mathcal{M}(s,t)$ such that it generates the series \eqref{eq:Svuser} for $S(v,u)$ order by order.
Notice first that the integrand has double poles at the dimensions $s,t=2\Delta_\varphi+2n+\ell$.
To shift the poles to the dimensions of the double field operators \( [\sigma,\sigma]_{n,\ell}\), we therefore define
\begin{equation}\label{eq:Mellinmiddlestep}
\mathcal M_{[\sigma,\sigma]_{n,\ell}}(s,t)=\frac{\Gamma\left(\Delta_\sigma-\frac s2\right)}{\Gamma\left(\Delta_\varphi-\tfrac s2\right)^2}\mathsf M_{[\sigma,\sigma]_{n,\ell}}(s,t).
\end{equation}
By demanding that we reproduce the series expansion of \(S(v,u)\) computed previously, we can fix the poles of \(\mathsf M_{[\sigma,\sigma]_{n,\ell}}(s,t)\) completely.
We first find the appropriate terms with triple poles in $t$ that generate the $\log^2u$ part of the series \eqref{eq:Svuser}. We then add a term that makes the expression symmetric under $t\leftrightarrow \ut$. In fact, it turns out that the resulting expression correctly reproduces the entire series for $S(v,u)$, including the terms with no $\log u$. The result is
\begin{equation}\label{eq:reducedMsigmasigma}
\mathsf M_{[\sigma,\sigma]_{n,\ell}}(s,t)    =  K(\mu)^2   \sum_{ k=2}^\infty     \left( \frac1{t+4-2 k-2\Delta_\varphi}
+\frac1{\ut+4-2 k-2\Delta_\varphi} \right)
\tilde{R}( k,s), 
\end{equation}
where \( \ut=4\Delta_\varphi-s-t  \), and where
\begin{equation}\label{eq:RsumforS}
\tilde{R}( k,s)= \frac{ 2\pi\csc(\pi\mu)\Gamma (\mu -2) }{\Gamma (\mu -3)^2}  \frac{ (-1)^{\frac s2} \Gamma \left(\frac{s}{2}-1\right)}{ \Gamma\left( \frac s2+2-\mu\right)^2}  \sum_{i=0}^{\frac s2-2}\frac{
\Gamma(i+3-\mu)   }{\Gamma (i+1) (3+i- k-\frac{s}{2}) }.
\end{equation}
Note that the integrand has only single poles in \(s\), but triple poles in \(t\).
This implies that \(S(v,u)\) has no terms proportional to \(\log v\), but has terms of the form \(\log^2 u, \log^1 u, \log^0 u\), corresponding roughly to the CFT-data \(U_\hb^{(2)}\), \(U_\hb^{(1)}\) and \(U_\hb^{(0)}\) respectively.

By carefully computing the residues, an infinite sum expression for \(S(v,u) \) can be found -- for details see appendix \ref{app:MellinForSigSig}.
In particular, we are able to compute the \(u^0\log^2 u\) piece explicitly.
After changing to the coordinates \( (z,\zb) \), we find that
\begin{align}
S(v,u)|_{z^0 \log^2 z} &=  K(\mu)^2 (1-\zb)^2 \bigg[  \,_2F_1\!\left(\!\left. {\,\,1,2}~\atop~{\!\!4-\mu}\right| 1-\zb\right) +     (\mu -3) ~ \dda \,_2F_1\!\left(\!\left. {\,\,1,2}~\atop~{\!\!4-\mu+a}\right| 1-\zb\right) \!    \bigg] \nonumber\\
&= \frac{1}{8} 4 K(\mu)^2 \xit\,^2   \,\,_3F_2\!\left(\!\left. {\,\,2,3-\mu,3-\mu}~\atop~{\!\!4-\mu,4-\mu}\right| -\xit\right),   
\end{align}
where, as before, \(\xit=(1-\zb)/\zb\).
This has a particularly simple inversion, and gives
\begin{equation} \label{eq:fullU2_TA_Nmin2}
U^{(2)}_{T/A,\hb} = \pm \frac{ 4\mu^2(\mu-1)^2}{N^2}   \left( \gamma_{\varphi}^{(1)}\right)^2     \frac{A[\mu-1](\hb) }{J^4} \,\, + \order(N^{-3}),
\end{equation}
in perfect agreement with the result \eqref{eq:gamma_TA_Nmin1} for the squared anomalous dimension \(\gamma^2_{T/A,\ell}\).

We have also found the \(z^0 \log z \) piece, which contributes to \(U^{(1)}_{T/A,\hb}\), by expanding it to very high order in \( \xit\) and recognizing the function as
\begin{align}
S(v,u)|_{z^0 \log z} = \frac12 2K(\mu)^2 \xit\,^2 \dda \Bigg[&  (\mu-2)   \,_3F_2\!\left(\!\!\!\left. {\,\,1+a,3-\mu-a,3-\mu\!}~\atop~{\!\!4-\mu,4-\mu\!} \right| - \xit\right)         \nonumber\\
& \quad +  (\mu-3)   \,_2F_1\!\left(\!\!\!\left. {\,\,1-a,3-\mu+2a\!}~\atop~{\!\!4-\mu\!} \right| - \xit\right)        \Bigg] \!.
\end{align}
This gives the following contribution to \(U^{(1)}_{T/A,\hb}\):
\begin{align}  
\hspace{-0.2cm}U^{(1),[\sigma,\sigma]_{n,\ell}}_{T/A,\hb} &=2  \INV\left[\xit^{-\Delta_\varphi} S(v,u)|_{z^0 \log z} \right] \nonumber \\
&= \frac{1}{N^2} 2 \mu^2 (\mu-1)^2 \left( \gamma_\varphi ^{(1)} \right)^2 A[\Delta_\varphi](\hb) \left( \frac{1}{J^4}  - \frac{R(\hb,\mu)}{J^2} - \frac{\mathbf{S_1}[\Delta_\varphi](\hb) - 1/J^2}{J^4}  \right),
\label{eq:U1_TA_fromSigSig_Nmin2}
\end{align}
where \( \mathbf{S_1} \) was defined in \eqref{eq:Qdefalpha} and where we defined the function\footnote{This function, up to an additional term, appeared first in \cite{Derkachov:1997ch} as a two-dimensional integral. In \cite{Alday:2015ewa} the explicit form used here was given.}
\begin{equation} \label{eq:Rdefn}
R(\hb,\mu) = \frac{1}{\hb(\hb+2-\mu)}  \,_{3}F_{2}\left(\left. {1,\hb,3-\mu}~\atop~{ \!\!\!\hb+1,\hb+3-\mu}\right|1\right).
\end{equation}

Finally, we have not been able to find the full \( z^0 \log^0 z \) piece of \( S(v,u)\) in a closed form; however in appendix~\ref{app:MellinForSigSig} we find a closed form expression for the piece proportional to $\zeta_2$:
\begin{align}
 S(v,u)|_{z^0\zeta_2} = K(\mu)^2 (1-\zb)^2\Bigg[&\frac{\,_2F_1\!\left(\!\left. {\,\,1,2}~\atop~{\!\!4-\mu}\right| 1-\zb\right)}{\mu -2}-\frac{\,_3F_2\!\left(\!\left. {\,1,2,2}~\atop~{4-\mu ,4-\mu}\right| 1-\zb\right)}{\mu -2}   \nonumber\\
 &\quad  +(\mu -3) \dda \,_2F_1\!\left(\!\left. {\,\,1,2}~\atop~{\!\!4-\mu+a}\right| 1-\zb\right)\!\Bigg]  ,\label{eq:Svuz2part}
\end{align}
which we find to have a straightforward inversion.
The other piece of \( S(v,u)\), not proportional to \(\zeta_2\), can be evaluated to arbitrarily high order in \(\xit\), which yields a large \(J\) expansion to arbitrarily high order.

The contribution to \(U^{(0)}_{T/A}\) then takes the form
\begin{equation}\label{eq:U0fromsigmasigma}
U^{(0),[\sigma,\sigma]_{n,\ell}}_{T/A,\hb}=K(\mu)^2\frac{A[\Delta_\varphi]}{N^2}\left(
\rho_1(\hb,\mu)+\zeta_2\rho_2(\hb,\mu)
\right),
\end{equation}
where 
\begin{equation}\label{eq:rho2threed}
\rho_2(\hb,\mu)=(\mu-2)^3(\mu-3)^2\left(
\frac{R(\hb,\mu)}{J^2-K^2}-\frac1{J^2(J^2-K^2)}+\frac{1-\mu}{J^4(J^2-K^2)}
\right),
\end{equation}
and where \(\rho_1(\hb,\mu)\) can be computed to arbitrarily high order in \(1/J^2\).
For example, the first few terms read
\begin{align}\nonumber
\rho_1(\hb,\mu)&=(\mu -1)(\mu -2)^3 (\mu -3)^2 \Bigg(
\frac{\mu ^2-5 \mu +12}{2 J^8}
+\frac{\mu ^4-10 \mu ^3+39 \mu ^2-34 \mu -48}{2 J^{10}}
\\
&\quad+\frac{11 \mu ^6-159 \mu ^5+943 \mu ^4-2817 \mu ^3+6686 \mu ^2-16648 \mu +20448}{24
J^{12}}
+O\big(J^{-14}\big)
\Bigg).
\label{eq:rho1series}
\end{align}
In section~\ref{sec:OPE_TA_Nmin2} we will use this series to provide numeric estimates for OPE coefficients.

\subsection{CFT-data at order $N^{-2}$}

By combining the contributions \(I_{\sigma}\), \(I_{T+A}\) and \(I_{[\sigma,\sigma]_{n,\ell}}\), we can now give the full CFT-data \(U^{(p)}_{T/A,\hb}\) up to order \(N^{-2}\).

Firstly, \(U^{(2)}_{T/A,\hb}\) was already given in \eqref{eq:fullU2_TA_Nmin2}, since it receives a contribution only from \( [\sigma,\sigma]_{n,\ell}\).
To this order \(U^{(2)}_{T/A,\hb} = \hat a_\ell \gamma_\ell ^2\) gives the squares of anomalous dimensions, so that it is fully fixed by \(U^{(1)}_{T/A,\hb}\) and \(U^{(0)}_{T/A,\hb}\) at previous orders, a non-trivial consistency condition that it satisfies.

For \(U^{(p)}_{T/A,\hb}\), \(p=0,1\) we have the following expressions:
\begin{align}
U^{(p)}_{T,\hb}&= U^{(p),\sigma}_{T/A,\hb} + U^{(p),[\sigma,\sigma]_{n,\ell}}_{T/A,\hb} + U^{(p),(T+A)}_{T/A,\hb}  & \text{for }p=0,1, \label{eq:fullU1_T_Nmin2} \\
U^{(p)}_{A,\hb}&= - U^{(p),\sigma}_{T/A,\hb} - U^{(p),[\sigma,\sigma]_{n,\ell}}_{T/A,\hb} + U^{(p),(T+A)}_{T/A,\hb}  & \text{for }p=0,1, \label{eq:fullU1_A_Nmin2}
\end{align}
where the various contributions were given in equations \eqref{eq:U1_TA_fromTplusA_Nmin2}, \eqref{eq:U0_TA_fromTplusA_Nmin2}, \eqref{eq:U1fromsigmaN-2}, \eqref{eq:U0_TA_fromsigmaNm2}, \eqref{eq:U1_TA_fromSigSig_Nmin2} and \eqref{eq:U0fromsigmasigma}.\footnote{The order \(1/N\) results have been absorbed into \(  U^{(p),\sigma}_{T/A,\hb} \).}

Using this, the anomalous dimensions \(\gamma_{T/A,\ell}\) can be computed to order \(N^{-2}\).
They are most easily expressed in terms of \(\hb=\mu-1+\ell+\gamma_\varphi\):
\begin{align} \label{eq:gamma_TA_Nmin2}
\gamma_{T/A}(\hb) &=  -2\gamma_\varphi^{(1)} \frac{\mu(\mu-1)}{J^2} \frac{1}{N} + 2 \left(\gamma_\varphi^{(1)}\right)^2 \frac{1}{N^2}\Bigg[ \mu^2(\mu-1)^2 \Bigg(-\frac{\left(2 \bar{h}-1\right)}{J^6}   +   \frac{\left(\mu ^2-\mu +1\right)}{J^4 (\mu -1) \mu }     \nonumber\\
& \qquad  -\frac{2 \left(4 \mu ^3-18 \mu ^2+21 \mu -8\right)}{J^2 (\mu -2)^2 (\mu -1) \mu }    -\frac{ R\left(\bar{h},\mu \right)}{J^2}   +  \frac{4 (2 \mu -1) \left(S_1\left(\bar{h}-1\right)-S_1(\mu -2)\right)}{J^2 (\mu -2) \mu }   \Bigg)   \nonumber\\
&\qquad   -\frac{ (\mu -2)^2 a^{(1)}_{\sigma }\Big/ \left(\gamma_\varphi^{(1)}\right)^2 }{ J^2}     \mp   \frac{\Gamma (\mu +1)^2 \Gamma \left(\bar{h}+2-\mu\right)}{J^2 (\mu -2)^2 \Gamma \left(\bar{h}-2+\mu\right)}\Bigg],
\end{align}
with the upper sign for \(T\) and the  lower sign for \(A\). 
The conservation of the global symmetry current implies that \( 2\Delta_\varphi + \gamma_{A,\ell=1} \overset{!}{=} d -1 \), which fixes the OPE coefficient \(a^{(1)}_\sigma\) in terms of \(\gamma_{\varphi}^{(2)}\)
\begin{equation}
a_\sigma^{(1)} = \frac{(\mu -1) \mu}{(\mu -2)^2}  \left(\gamma_\varphi^{(2)} +  \left(\gamma_\varphi^{(1)}\right)^2 \left(-8 \mu +\frac{17}{\mu -2}+\frac{14}{(\mu -2)^2}+\frac{1}{\mu -1}-\frac{1}{\mu }+12\right)\right)\!.
\end{equation}
We expect that \(\gamma_\varphi^{(2)}\) can be fixed by demanding conservation of the stress-energy tensor, similar to how \(\gamma_\varphi^{(1)}\) was fixed by this conservation in section \ref{sec:S_at_N-1}.
Since this is a computation at next order, we will instead use the literature value of \(\gamma_\varphi^{(2)}\) \cite{Vasiliev:1981yc,Manashov:2017xtt} to find that
\begin{align}\label{eq:asigma1res}
a_\sigma^{(1)} &= 2 \left(\gamma_\varphi^{(1)}\right)^2 \frac{ \mu}{(\mu -2)^4}  \bigg(\!\! -4 \mu ^4+23 \mu ^3-43 \mu ^2+31 \mu -6 \nonumber\\
& \qquad \qquad -(\mu -2) (\mu -1)  \left(2 \mu ^2-3 \mu +2\right) \left(\pi  \cot (\pi  \mu )+S_1(2 \mu -3)\right) \bigg).
\end{align}
With this value, which agrees with the literature value computed in \cite{Lang:1993ct}, we find that our result \eqref{eq:gamma_TA_Nmin2} for \(\gamma_{T,A}(\hb)\) matches that found in \cite{Derkachov:1997ch,Manashov:2017xtt}.

\subsubsection{OPE coefficients and $C_J$ correction} \label{sec:OPE_TA_Nmin2}

The final piece of CFT-data that we can compute is the OPE coefficients in the $T$ and $A$ representations to order $N^{-2}$. They are computed from the $U^{(p)}_{T/A,\hb}$ using 
\eqref{eq:extractAGamma}
and \eqref{eq:extractAfromAhat}.
The resulting expressions are lengthy and we give the explicit form in appendix~\ref{app:completeresults}. Notice that they depend on the unknown function $\rho_1(\hb,\mu)$ defined in \eqref{eq:U0fromsigmasigma}. 
If one is interested in a particular OPE coefficient in a given dimension, one can use the series expansion \eqref{eq:rho1series} for $\rho_1(\hb,\mu)$ to generate numeric estimates.
In particular, we are interested in computing the corrections to the central charge $C_J$, which follows from the OPE coefficient in the antisymmetric representation, with $\ell=1$ and $\mu=3/2$. 
Thus we need a numeric estimate for $\rho_1(\mu,\mu)$.

Define 
\begin{equation}
s_\mu(\zb):=\left.S(v,u)\right|_{z^0\log^0z}=\sum_{n=3}^\infty c_n(\mu)w^n, \qquad w=1-\zb.
\end{equation}
We are able to compute the first $140$ coefficients $c_n(\mu)$.
In any specific dimension we can quickly compute more: in three dimensions we have computed the first $4300$ terms. For reference we give the first orders here:
\begin{equation}
s_{\mu=3/2}(\zb)=K\!\left(\tfrac32\right)^2\left(
\frac{4\zeta _2}{25} w^3+\left(\frac{48 \zeta _2}{175}-\frac{27}{2450}\right)w^4+\left(\frac{3904
   \zeta _2}{11025}-\frac{79}{3150}\right) w^5
+\ldots\right)\!.
\end{equation}
However the convergence is very slow. 
In order to get better control of the tail $n\rightarrow\infty$, we study the large \(n\) behaviour of the coefficients \(c_n(\mu)\) and find that it takes the following form:
\begin{align}\nonumber
c_n(\mu)&=\frac{\Gamma(4-\mu)(3-\mu)\Gamma(n)}{\Gamma(n+2-\mu)}\left[
a_2(\mu)S_1(n)^2+a_1(\mu)S_1(n)+a_0(\mu)+O\left(\frac{1}n\right)
\right]
\\&\quad+\frac1n\left[
\tilde a_{0,1}(\mu)S_1(n)+\tilde a_{0,0}(\mu)+\frac{\tilde a_{1,1}(\mu)S_1(n)}n+ \frac{\tilde a_{1,0}(\mu)}n+O\left(\frac{1}{n^2}\right)
\right]\!. \label{eq:smuzbcoefs}
\end{align}
This form was derived by finding recursion relations for the $c_n(\mu)$ in various integer $\mu$, and studying the solutions around large $n$. 
Notice that the first term scales as $n^{\mu-2}\log^2n$ and thus dominates at large positive $\mu$, whereas the second term scales as $n^{-1}\log n$ and dominates at negative $\mu$. 
At negative integer $\mu$ the terms $\tilde a_{k,1}(\mu)\log n+\tilde a_{k,0}(\mu)$ develop poles that exactly cancel with the corresponding poles induced by the $a_2(\mu)S_1(n)^2+a_1(\mu)S_1(n)+a_0(\mu)$ order by order. 
This leads to a determination of $a_2(\mu)$, $a_1(\mu)$ and $a_0(\mu)$.\footnote{
We get $a_2(\mu)=\frac12\left(S_1(2-\mu)+S_1(\mu-2)\right)$ and some lengthy expressions for $a_1(\mu)$ and $a_0(\mu)$, which are given in appendix~\ref{sec:taylorseriessmu}. 
We have determined many of the $\tilde a_{k,p}$, the first ones are $\tilde a_{0,1}=\frac{(\mu-3)^2}{(\mu-1)^2}$ and $\tilde a_{0,0}=\frac{(\mu-3)^2(5\mu-8)}{(\mu-2)(\mu-1)^3}$.
}

\begin{figure}
  \centering
\includegraphics[width=92.5mm]{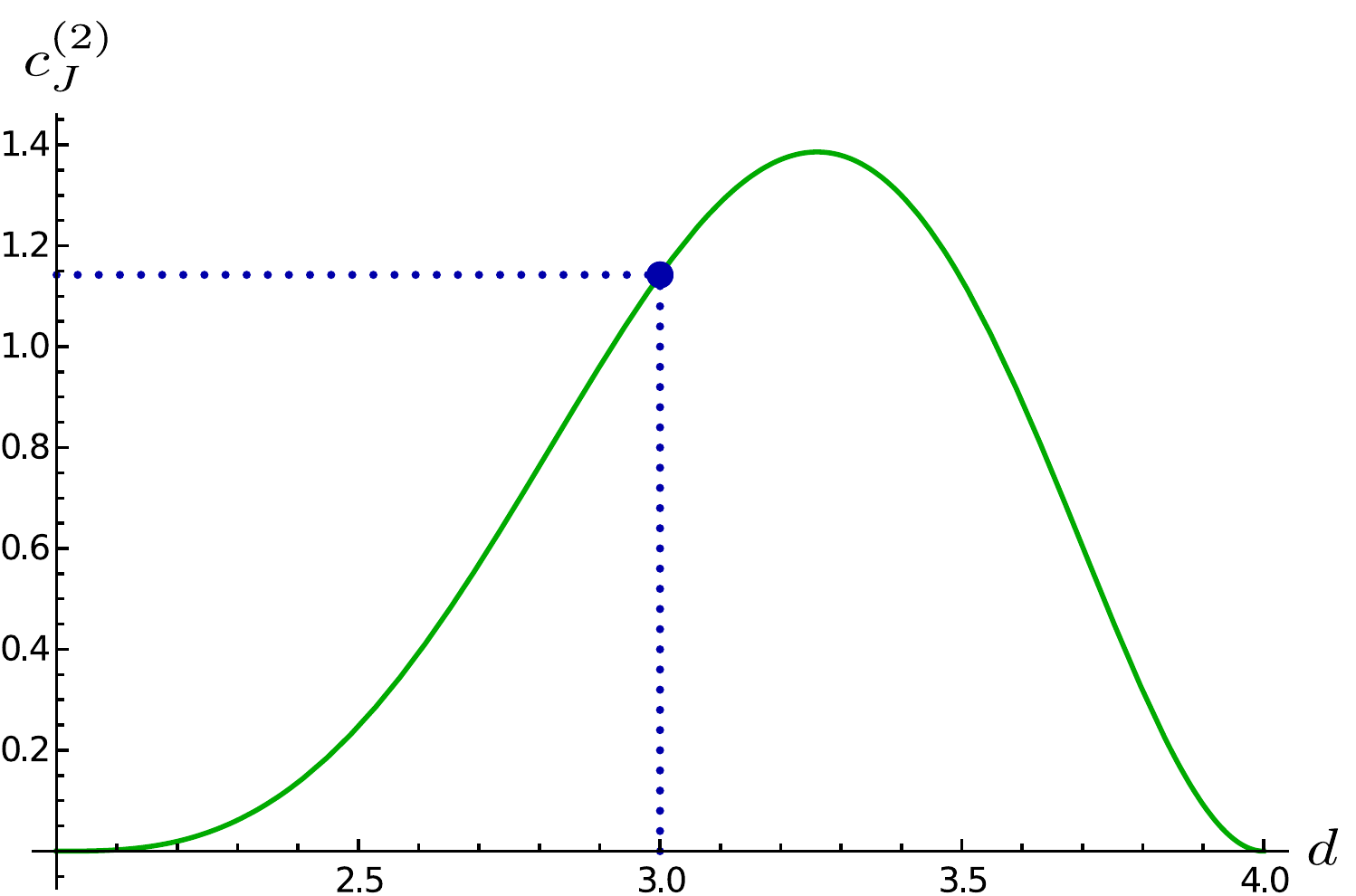}
  \caption{The numeric estimate for the \(1/N^2\) correction \(c_J^{(2)}\) in \eqref{eq:cJsecondorder} to the ratio of the global current charge \(C_J/C_{J,\mathrm{free}
}\) in the interacting and in the free theory.
  }\label{fig:cJ2plot}
\end{figure}

The analytic continuation of the coefficients $a_p(\mu)$ away from integer $\mu$ now means that we can derive the leading large $n$ dependence in any dimension. 
This will map to the small $\zb$ expansion 
\begin{align}\nonumber
s_{\mu}(\zb)\sim\frac{\pi (\mu-2)(\mu-3)^2\zb^{1-\mu}}{\sin(\pi\mu)}\Big[&
\left((S_1(\mu-2)-\log \zb)^2-S_2(\mu-2)+\zeta_2\right)a_2(\mu)
\\&+\left(S_1(\mu-2)-\log \zb\right)a_1(\mu)+a_0(\mu)
\Big]+\ldots.\label{eq:smusmallzb}
\end{align}
Under the inversion integral~\eqref{eq:invIntegralDefn} this expansion elucidates the structure of the poles in \(\hb\): a term of the form $\zb^\alpha\log^\beta \zb$ maps to a pole as follows:
\begin{equation}
\INV\left[\left(\tfrac{\zb}{1-\zb}\right)^{\Delta_\varphi}\zb^\alpha \log^\beta\zb\right]=2\sin^2(\pi \Delta_\varphi)\frac{(-1)^\beta \beta!}{(\hb-(1-\Delta_\varphi-\alpha))^{\beta+1}}+\ldots.
\end{equation}
In the case at hand, with $\Delta_\varphi=\mu-1$, we get a triple pole at $\hb=1$, or equivalently at $J^2=0$. 
From the expressions \eqref{eq:smusmallzb} we can thus work out the expansion near $J^2=0$ of the inversion of $s_{\mu}(\zb)$. In a similar manner, the second line of \eqref{eq:smuzbcoefs} will give poles as $J^2=K^2$. Thus we have full control of the pole structure of this inversion:
\begin{align}
\rho_1(\hb,\mu)+\zeta_2\rho_2(\hb,\mu) &\sim\frac{r_3(\mu)}{J^6}+\frac{r_2(\mu)}{J^4}+\frac{r_1(\mu)}{J^2}  	& \text{near } J^2&=0, \label{eq:rhofuncsaround0}\\
&\sim \frac{\tilde r_2(\mu)}{(J^2-K^2)^2}+\frac{\tilde r_1(\mu)}{J^2-K^2}									    & \text{near } J^2&=K^2, \label{eq:rhofuncsaroundK2}
\end{align}
where
\begin{align}
r_3(\mu)&=2(\mu-2)^2(\mu-3)^2a_{2}(\mu),
\\
r_2(\mu)&=(\mu-2)^2(\mu-3)^2\big(2(S_2(2-\mu)+1)a_2(\mu)+a_1(\mu)\big),
\\
r_1(\mu)&=(\mu-2)^2(\mu-3)^2\big(
(S_1(2-\mu)^2-S_2(2-\mu)-\zeta_2)a_2(\mu)+S_1(2-\mu)a_1(\mu)+a_0(\mu)
\big),
\end{align}
and
\begin{align}
\tilde r_2(\mu)&=(2\mu-3)^2\tilde a_{0,1}(\mu),
\\
\tilde r_1(\mu)&=\big(1+(2\mu-3)S_1(2-2\mu)-2(2\mu-3)S_1(1-\mu)\big)\tilde a_{0,1}(\mu)-(2\mu-3)\tilde a_{0,0}(\mu).
\end{align}
In order to interpolate between these poles at finite $J^2$ and the asymptotic expansion around infinite $J^2$, we contruct Pad\'e approximants of the form
\begin{equation}\label{eq:padeansatz}
\text{Pad\'e}_n[\rho_1+\zeta_2\rho_2](J^2)=\frac{\sum_{k=0}^n\alpha_kJ^{2k}}{J^6(J^2-K^2)^2\left(1+\sum_{k=1}^{n-2}\beta_kJ^{2k}\right)}.
\end{equation}
After determining the constants $\alpha_k,\beta_k$ we get estimates for $\rho_1(\mu,\mu)$. These can be converted to estimates for the $C_J$ corrections:
\begin{equation}\label{eq:cJsecondorder}
\frac{C_J}{C_{J,\mathrm{free}}}=1+\frac{c_J^{(1)}}N+\frac{c_J^{(2)}}{N^2}+O(N^{-3}),
\end{equation}
where $c_J^{(1)}$ matches \eqref{eq:cJfirstorder} computed before, and $c_J^{(2)}$ is extracted from \eqref{eq:aTANm2Final}. 
In figure \ref{fig:cJ2plot} we present graphically the numeric results for the second order correction $c_J^{(2)}$ in different dimensions based on Pad\'e approximants of order $n=30$.
In particular, note that the behaviour of the graph is consistent with the known expansions in \(d=4-\epsilon\) and \(d=2+\epsilon\) where both the value and the slope of \(c_J^{(2)}\) vanish.
In \(d=3\) we can compute the large \(J\) series to order 800, to provide the more precise estimate:
\begin{equation}\label{eq:rho1estimate}
\rho_1\left(\tfrac32,\tfrac32\right)=-0.3422(1) .
\end{equation}
This was derived from computing Pad\'e approximants \eqref{eq:padeansatz} in three dimensions for $n$ of increasing order, and matching the asymptote. 
We can now compute the correction to the central charge $C_J$ in three dimensions.
\begin{align}\nonumber
\left.\frac{C_J}{C_{J,\text{free}}}\right|_{\mu=3/2}&=1-\frac1N\frac{64}{9\pi^2}
+\frac{1}{N^2}\frac{16}{9\pi^4}\Big(
-\rho_1\left(\tfrac32,\tfrac32\right) 
+8\log(2)^2+\tfrac{100}3\log 2+6\pi\log2
\\\nonumber&
\hspace{124pt} -7\pi+\tfrac{11}2\pi^2+\tfrac3{16}\pi^3-24\beta(2)+\tfrac{55}9
\Big)+O(N^{-3})
\\&
=1-\frac{0.720506}{N}+\frac{1.14230(2)}{N^2}+O(N^{-3}).\label{eq:cJ3d}
\end{align}
where $\beta(2)=0.91597\ldots$ is Catalan's constant. 
In this expression we used that
$\partial_\hb R\left(\tfrac32,\tfrac32\right) =16+ 4 \pi-32\beta(2)$. Notice that the only unknown quantity here is $\rho_1\left(\tfrac32,\tfrac32\right)$ for which we used the estimate in \eqref{eq:rho1estimate}. In principle, one can write down a closed form expression for $\rho(\hb,\mu)$ in terms of the inversion integral of a nested sum, or as the inversion integral of a limit of an inverse Mellin transform, and therefore compute $\rho_1\left(\tfrac32,\tfrac32\right)$ to arbitrary precision.


\section{Summary of results and matching with other regimes}
\label{sec:summary}

In this paper we have used large spin perturbation theory to compute scaling dimensions and OPE coefficients in the critical $\mathrm O(N)$ model in the large $N$ limit. In section~\ref{sec:LSPTforON} we studied the corrections at order $N^{-1}$ for currents $\mathcal J^{(\ell)}_R$ in all representations, and found that they follow purely from assuming the existence of the identity operator and the operator $\sigma$ with scaling dimension $\Delta_\sigma=2+O(N^{-1})$. All CFT-data at this order was determined in terms of the leading anomalous dimension $\gamma_\varphi^{(1)}$ of $\varphi^i$, which was fixed in \eqref{eq:gammaphisols} by demanding conservation of the stress tensor and global symmetry current. Let us collect our results from section \ref{sec:LSPTforON}:
\begin{itemize} 
\item The leading order anomalous dimension $\gamma_\varphi^{(1)}$ \eqref{eq:gammaphisols}.
\item The scaling dimensions \eqref{eq:gamma_TA_Nmin1} and OPE coefficients \eqref{eq:OPETAfirstorder} of non-singlet currents $\mathcal J^{(\ell)}_R$, $R=T,A$, at order $N^{-1}$.
\item The OPE coefficient $a_\sigma=c^2_{\varphi\varphi\sigma}$ at leading order \eqref{eq:asigma0res}.
\item The scaling dimensions \eqref{eq:gammaSres} and OPE coefficients \eqref{eq:OPESres} of singlet currents $\mathcal J^{(\ell)}_S$ at order $N^{-1}$. Apart from $\ell=2$, these OPE coefficients are new results.
\item By assuming the shadow relation $\Delta_\sigma+\Delta_{S,0}=d$, we computed the leading anomalous dimension of $\sigma$ in \eqref{eq:gammasigmares}.
\end{itemize}
In section~\ref{sec:subsubleading} we focused on the computation at order $N^{-2}$, with the goal of deriving new results. Here we used our results at previous orders, as well as some input from the literature. Specifically we made use of the second order anomalous dimension $\gamma_\varphi^{(2)}$ of $\varphi^i$, and the $N^{-1}$ part of the correlator $\langle\varphi^i\varphi^j\sigma\sigma\rangle$, \eqref{eq:phiphisigmasigmacorr}. The results of section~\ref{sec:subsubleading} are:
\begin{itemize}
\item The scaling dimensions \eqref{eq:gamma_TA_Nmin2} of non-singlet currents $\mathcal J^{(\ell)}_R$, $R=T,A$, at order $N^{-2}$.
\item The OPE coefficient $a_\sigma=c^2_{\varphi\varphi\sigma}$ at subleading order \eqref{eq:asigma1res}.
\item The OPE coefficients of non-singlet currents $\mathcal J^{(\ell)}_R$, $R=T,A$, at order $N^{-2}$, given explicitly in \eqref{eq:aTANm2Final}, but more conveniently expressed in terms of the functions $U^{(p)}_{T/A,\hb}$ given in \eqref{eq:fullU2_TA_Nmin2} and \eqref{eq:fullU1_T_Nmin2}--\eqref{eq:fullU1_A_Nmin2}. These OPE coefficients are new results.
\item Of particular interest is the new result \eqref{eq:cJsecondorder} for the second order correction to the central charge $C_J$, plotted in figure~\ref{fig:cJ2plot} for $2<d<4$ and evaluated in \eqref{eq:cJ3d} for $d=3$. 

\end{itemize}
\begin{figure}
  \centering
\includegraphics[width=100mm]{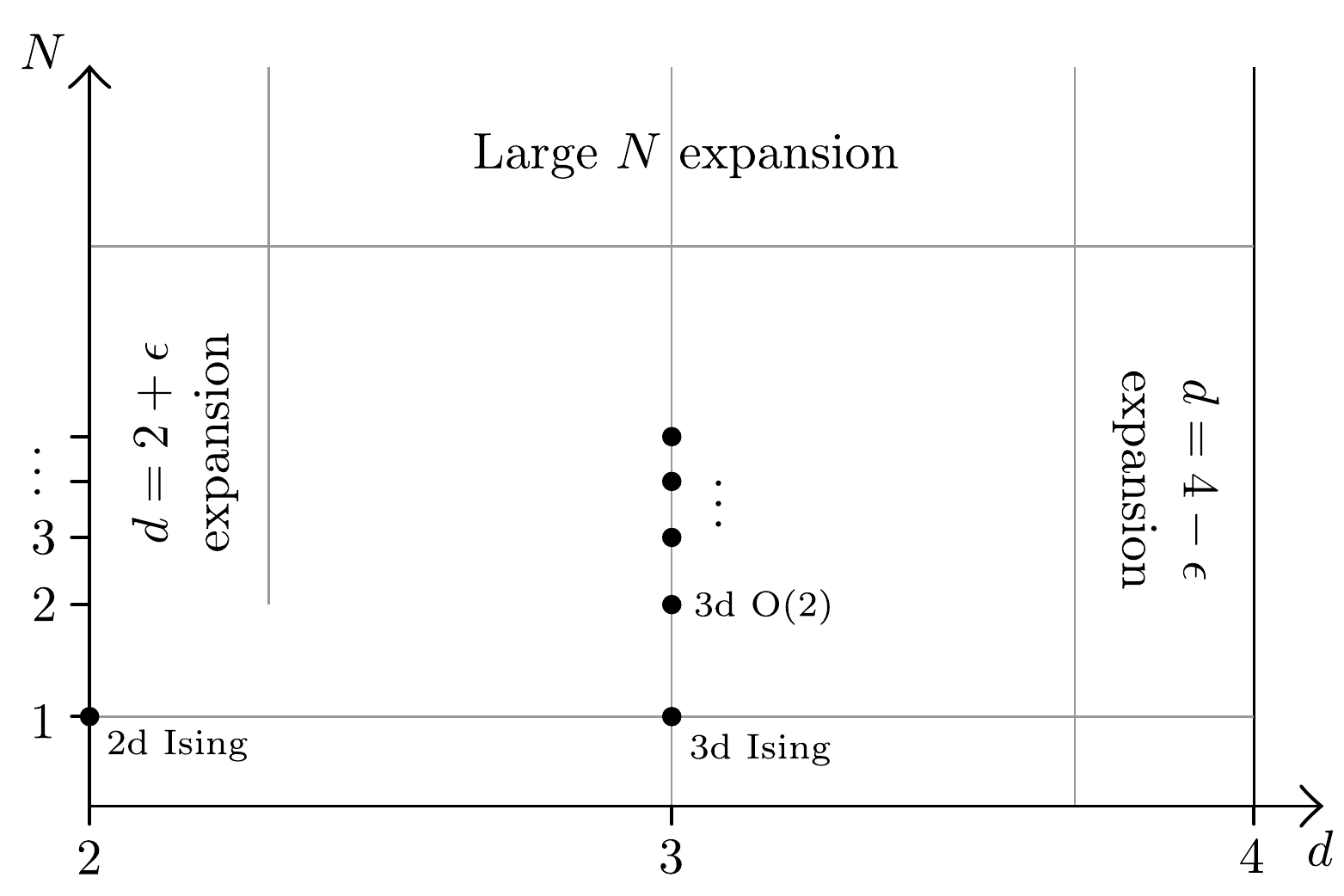}
  \caption{The critical $\mathrm O(N)$ model is defined for $2<d<4$ and $N$ positive integer.
  }\label{fig:Nd-diagram}
\end{figure}
The large $N$ expansion is only one of the perturbative expansions that can be used to study the critical $\mathrm O(N)$ model, and it is interesting to see how our results compare with existing results in other regimes, as pictured in figure~\ref{fig:Nd-diagram}.
In each space-time dimension $2<d<4$ \footnote{It was proposed in \cite{Fei:2014yja} that the critical $\mathrm O(N)$ model, for sufficiently large $N$, extends also to the range $4<d<6$. Some anomalous dimensions have been computed in a $d=6-\epsilon$ expansion \cite{Fei:2014yja,Giombi:2016hkj}, which match the corresponding expressions at large $N$. It would be interesting to compute corrections to OPE coefficients the $d=6-\epsilon$ expansion to compare to our new results at large $N$.
} and at each $N=1,2,\ldots$ the critical $\mathrm O(N)$ model exists as an interacting conformal field theory with global $\mathrm O(N)$ symmetry and a field $\varphi^i$ transforming in the fundamental ($\mathbf N$) representation\footnote{Notice that at any $d$ there is also the theory of $N$ free scalars, and that below the multicritical dimensions $d=3,\frac83,\frac52,\ldots$ there are also multicritical $\mathrm O(N)$ symmetric CFTs. It is believed that the multicritical theories cannot be reached from the large $N$ expansion \cite{Yabunaka:2017uox}.}.
All these theories are continuously connected when varying the parameters $d$ and $N$\footnote{For generic, non-integer values of $N$ and $d$, the theories are not expected to be unitary \cite{Hogervorst:2015akt,Maldacena:2011jn}. This manifests itself by the appearance of evanescent operators that may acquire complex dimensions and OPE coefficients. These operators do not appear at the order to which we are working in the $1/N$ expansion.
}.
In table~\ref{tab:operatornames} we give a partial dictionary covering some operators discussed in this paper with their standard names in the various regimes. 
The case $N=1$ corresponds to the Ising model, where the limit $d\to2$ gives the 2d Ising model. The upper limit $d=4$ is the critical dimension for $\phi^4$ theory, whereas in the lower limit the $d=2$ twist gap between the stress-tensor and the identity operator vanishes.

\begin{table}
\centering
\begin{tabular}{cclll}
Irrep & Spin & $\epsilon$-expansion & Large $N$ & Ising ($N=1$)
\\\hline\hline
$S$&$0$&$\1$&$\1$&$\1$
\\\hline
$\mathbf N$&$0$&$\varphi^i$&$\varphi^i$&$\sigma$
\\\hline\hline
$S$&$0$&$\varphi^2=\varphi^i\varphi^i$&$\sigma$&$\epsilon$
\\
$S$&$2$&$\mathcal J^{(2)}_S=T^{\mu\nu}$&$\mathcal J^{(2)}_S=T^{\mu\nu}$&$T^{\mu\nu}$
\\
$S$&$4$&$\mathcal J^{(4)}_S=\varphi^i\partial^4\varphi^i$&$\mathcal J^{(4)}_S=\varphi^i\partial^4\varphi^i$&$C^{\mu\nu\rho\sigma}$
\\
$S$&$\ell$ even&$\mathcal J^{(\ell)}_S=\varphi^i\partial^\ell\varphi^i$&$\mathcal J^{(\ell)}_S=\varphi^i\partial^\ell\varphi^i$&$\O^{\{\mu_1\cdots\mu_\ell\}}$
\\\hline
$T$&$\ell$ even&$\mathcal J^{(\ell)}_T=\varphi^{\{i}\partial^\ell\varphi^{j\}}$&$\mathcal J^{(\ell)}_T=\varphi^{\{i}\partial^\ell\varphi^{j\}}$& ---
\\\hline
$A$&$\ell$ odd&$\mathcal J^{(\ell)}_A=\varphi^{[i}\partial^\ell\varphi^{j]}$&$\mathcal J^{(\ell)}_A=\varphi^{[i}\partial^\ell\varphi^{j]}$& ---
\\\hline\hline
$S$ & $0$ & $\varphi^4=(\varphi^i\varphi^i)^2$ & $[\sigma,\sigma]_{0,0}=\sigma^2$ & $\epsilon'$
\\\hline
\end{tabular}
\caption{Some operators with low twist and their names in the different regimes.}\label{tab:operatornames}
\end{table}

At the edges of the diagram in figure~\ref{fig:Nd-diagram}, one can study the critical $\mathrm O(N)$ model using various perturbative expansions, whereas in the centre there is no Lagrangian description and no infinitesimal expansion parameter. 
In the $d=4-\epsilon$ expansion, the theory can be described by a Lagrangian with coupling $\lambda(\varphi^i\varphi^i)^2$, where $\lambda$ is computed perturbatively in $\epsilon$ such that the $\beta$-function vanishes. 
This gives the $\mathrm O(N)$ symmetric Wilson--Fisher fixed-point. There the scaling dimensions for the scalars $\varphi^i$, $\varphi^2$ and $\varphi^4$ have been computed to order $\epsilon^7$ using diagrammatic methods \cite{Schnetz:2016fhy}\footnote{The previous highest order results, at order $\epsilon^6$, were computed in \cite{Kompaniets:2017yct}. We thank Erik Panzer for making us aware of \cite{Schnetz:2016fhy}.}, and similarly for singlet currents $\mathcal J^{(\ell)}_S$ to order $\epsilon^4$ \cite{Derkachov:1997pf,Manashov:2017xtt}. 
For currents $\mathcal J^{(\ell)}_R$ in all three $\mathrm O(N)$ representations, both scaling dimensions and OPE coefficients were computed to order $\epsilon^4$ using large spin perturbation theory \cite{Henriksson:2018myn}\footnote{See also \cite{Dey:2016mcs} where this CFT-data, including scalar operators, were computed to order $\epsilon^3$ using bootstrap in Mellin space.}.

It can be explicitly checked that the CFT-data of the operators in table~\ref{tab:operatornames} agree in the overlap of the available expansions. Consider for instance the operator $\sigma$ in the large $N$ expansion. To order $\epsilon^7/N^2$ the expansion $\Delta_\sigma=2+\gamma_\sigma^{(1)}/N+\ldots$, given to order $N^{-2}$ in \cite{Vasiliev:1981dg}\footnote{Note that there
is a typo in the definition of the parameter $\alpha$ in \cite{Vasiliev:1981dg}; as pointed out in \cite{Fei:2014yja} it should properly read $\alpha=\mu-2+\eta/2$.}, agrees with $\Delta_{\varphi^2}=d-2+\frac{N+2}{N+8}\epsilon+\ldots$, given to order $\epsilon^7$ in the $4-\epsilon$ expansion in \cite{Schnetz:2016fhy}.
Likewise the OPE coefficient $c^2_{\varphi\varphi\sigma}=\frac{a_\sigma^{(0)}}N+\frac{a_\sigma^{(1)}}{N^2}+\ldots$, with $a_\sigma^{(0)}$ and $a_\sigma^{(1)}$ are given in \eqref{eq:asigma0res} and \eqref{eq:asigma1res} respectively, agrees with the expansion of $C_{S,0}$ of \cite{Dey:2016mcs} to order $\epsilon^3/N^2$.
The new results for OPE coefficients in this paper makes it possible to perform similar consistency checks for the currents $\mathcal J_R^{(\ell)}$ to order $\epsilon^4/N^2$ in all representations and for all spins. Using the $4-\epsilon$ expansions computed in \cite{Henriksson:2018myn} we find perfect agreement. Unfortunately this does not give any constraint on the function $\rho_1(\hb,\mu)$, since it does not appear until order $\epsilon^5/N^2$, which can be seen from the explicit prefactor $(\gamma^{(1)}_\varphi)^2/(\mu-2)^2\sim \epsilon^2$ in \eqref{eq:aTANm2Final} and the additional factor $(\mu-2)^3\sim\epsilon^3$ in the series \eqref{eq:rho1series}.

\begin{table}
\centering
\begin{tabular}{c|llll}
\multirow{2}{*}{$N$}&\multicolumn{2}{l}{Truncated values}&Numeric  &Pad\'e 
\\
&$1/N$&$1/N^2$&bootstrap &approximant 
\\\hline\hline
$2$&$0.6397469$&$0.9253216(5)$& $0.9050(16)$ &
\\
$3$&$0.7598313$&$0.8867534(2)$ & $0.9065(27)$ & 0.896922
\\
$4$&$0.8198735$&$0.8912671(1)$ & & 0.906422
\\
$5$&$0.8558988$&$0.9015907(1)$ & & 0.915022
\\
$6$&$0.8799156$&$0.9116462(1)$ &  & 0.922546
\\
$10$&$0.9279494$&$0.9393724$ &  & 0.943830
\\
$20$&$0.9639747$&$0.9668304 $&$ 0.9674(8)$ & 0.967555
\\\hline
\end{tabular}
\caption{Evaluation of the truncated series in $d=3$ for $C_J$ given in \eqref{eq:cJsecondorder}. The column $1/N^2$ contains the new results of this paper and show great improvement when compared to results from numeric bootstrap \cite{Kos:2015mba} and Pad\'e approximants (we use the $\text{Pad\'e}_{[3,3]}$ of \cite{Henriksson:2018myn} which uses both the $4-\epsilon$ and the $2+\epsilon$ expansions). For $N=2$, $C_J/C_{J,\mathrm{free}}$ is related to the universal conductivity $\sigma_\infty$ of the 3d $\mathrm O(2)$ model \cite{Katz:2014rla} and has been estimated using Monte Carlo simulations to $0.914(10)$ \cite{Chen:2013ppa}.}\label{tab:numericmatch}
\end{table}
For $N\geqslant3$ one can develop a perturbative expansion around $d=2+\epsilon$, starting from a non-linear sigma model, where some scaling dimensions are known to order $\epsilon^2$ \cite{Giombi:2016hkj}. The fact that the graph in figure~\ref{fig:cJ2plot} shows a horizontal tangent at $d=2$ agrees with the expectation $\frac{C_J}{C_{J,\mathrm{free}}}=1-\frac{2}N+\frac\epsilon{N}+O(\epsilon^2)$ in the non-linear sigma model \cite{Diab:2016spb}.

In three dimensions and at finite $N$, numeric conformal bootstrap provides results for some CFT-data to very high precision \cite{Kos:2016ysd}, as well as numeric estimates for a larger set of operators in $N=1$ \cite{Simmons-Duffin:2016wlq}. 
Furthermore, in \cite{Cappelli:2018vir} the operator spectrum was numerically traced along the $N=1$ line from $d=3$ towards $d=2$, where it was shown how the spectrum reorganizes into appropriate Virasoro primaries as one approaches the 2d Ising model.
In table~\ref{tab:numericmatch} we compare our large $N$ result for the central charge $C_J$ with some numeric values in the unitary theories in three dimensions at finite $N$ and can see that the inclusion of the new $N^{-2}$ term significantly improves the truncated expansion results when comparing with the numeric results in the literature. 

In conclusion, this paper provides a self-contained computation of leading $N^{-1}$ corrections CFT-data in the critical $\mathrm O(N)$ model without any reference to diagrammatic methods, and derives new results for OPE coefficients at order $N^{-2}$. All results agree with other perturbative results and improve predictions at finite $N$.
In the computations of this paper, several technical methods were developed which we would like to draw some attention towards. Appendix~\ref{sec:InvMethods} is a summary of the toolbox required for the computations of the inversion integrals. In appendix~\ref{sec:unitarityboundblocks} we notice that expressions involving conformal blocks of spinning operators at the unitarity bound take a particularly simple form. Appendix~\ref{sec:mixed} shows that the leading order direct-channel OPE coefficients in the mixed correlator follows purely from the scalar $\varphi^i$ in the crossed channel. Finally, in section~\ref{sec:sigmainversionrevisited} we wrote down some closed form expressions for the inversion of an arbitrary scalar operator in the crossed channel.

\section*{Acknowledgements}

We would like to thank Parijat Dey, Simone Giombi, Tassos Petkou and Eugene Skvortsov for enlightening discussions, and Simone Giombi and Eugene Skvortsov for comments on the draft. This project has received funding from the European Research Council (ERC) under the European Union's Horizon 2020 research and innovation programme (grant agreement No 787185). M. van Loon was supported by an EPSRC studentship, award reference 1657727.

\appendix

\section{Inversion methods} \label{sec:InvMethods}
In this appendix we elaborate on some of the methods used to evaluate the inversion integral
\begin{equation} \label{eq:invHatDefn}
\invhat{g(\zb)} := \frac{\Gamma(\hb)^2}{\pi^2\Gamma(2\hb-1)}\int_0^1\frac{d \zb}{\zb^2}k_\hb(\zb) g(\zb) = \frac{\Gamma(\hb)^2}{\pi^2\Gamma(2\hb-1)}\int_0^1\frac{d \zb}{\zb^2} \zb^\hb \, _2F_1(\hb,\hb;2\hb;\zb) g(\zb),
\end{equation}
where the \( \INVhat\) indicates that we are not taking the double discontinuity, i.e. we have the defining relation:
\begin{equation}
\mathrm{INV}\left[ G \right] = \invhat{\mathrm{dDisc}[G]}.
\end{equation}

\subsection{Action of the collinear Casimir} \label{sec:collCas}
The inversion integral above solves the following problem:
\begin{equation} \label{eq:essenceOfInversion}
\sum_{\ell} a_{\ell} f_{2\Delta_\varphi+\ell,\ell}(\zb) = G(\zb) \quad \Rightarrow \quad a_{\ell} = \mathrm{INV}\left[ G(\zb) \right],
\end{equation}
where the \(f_{\Delta,\ell}(\zb) = r_{\frac{\Delta+\ell}{2}}k_{\frac{\Delta+\ell}{2}}(\zb) \) are collinear conformal blocks in a convenient normalization.
We now apply the collinear Casimir \(\Dbar\), defined in \eqref{eq:dbardefn}, to both sides of equation \eqref{eq:essenceOfInversion}, assume it is appropriate to interchange \(\Dbar\) and the infinite sum over spins, and use the fact that the \(f\) are eigenfunctions under the collinear Casimir \(\Dbar\) with eigenvalue \(J^2\), to see that we should expect
\begin{equation}
a_{\ell} = \mathrm{INV}\left[ G(\zb) \right] \qquad \Rightarrow \qquad (J^2+\alpha) a_{\ell} = \mathrm{INV}\left[ (\Dbar+\alpha) G(\zb) \right], \quad \alpha\in\mathbb{R}.
\end{equation}
This can be very useful since \((\Dbar+\alpha) G(\zb)\) is often easier to invert than \(G(\zb)\) itself.

At the level of the double discontinuities this can be made more precise since the inversion integral \eqref{eq:invHatDefn} can be seen as an inner product on an appropriate function space \(V\) \cite{Hogervorst:2017sfd}:
\begin{equation}
\invhat{g(\zb)} = \langle k_\hb(\zb), g(\zb) \rangle, \qquad \langle f(\zb), h(\zb) \rangle := \int_0^1 \frac{\dd \zb}{\zb^2}\, f(\zb) h(\zb).
\end{equation}
Furthermore, \(\Dbar\) is a self-adjoint operator in this space \cite{Hogervorst:2017sfd}, so that it easily follows that 
\begin{equation}
\invhat{\Dbar g(\zb)} = \langle k_\hb(\zb), \Dbar g(\zb) \rangle = \langle \Dbar k_\hb(\zb), g(\zb) \rangle = J^2 \invhat{g(\zb)},
\end{equation}
assuming that both \(g(\zb)\) and \(\Dbar g(\zb)\) lie within \(V\).
In practice this means that the procedure works up to a contribution \(\frac{2\hb-1}{J^2}\), which is the inversion of \(\log^2(1-\zb)\).
To see this explicitly:
\begin{equation}
\inv{\log^2(1-\zb)} = \invhat{4\pi^2} = \frac{4 (2 \hb-1)}{J^2},
\end{equation}
so that
\begin{equation}
\inv{\frac{2 \zb^2}{1-\zb}} = \inv{ \Dbar \log^2(1-\zb)} = 4 (2 \hb-1) \neq 0= \invhat{\Dbar(4\pi^2)}.
\end{equation}
This does not pose any real problem, since for any inversion we can always check whether a term of the form \(\frac{2\hb-1}{J^2}\) needs to be added by comparing to the asymptotic large \(J\) expansion from appendix \ref{sec:xitSeriesInv}, or to a numeric evaluation of the integral.

\subsection{Asymptotic expansions for large \(J\)} \label{sec:xitSeriesInv}
The basic idea behind this method is as follows: in this paper the functions \( g(\zb) \) in equation \eqref{eq:invHatDefn} can always be expanded, up to an overall power, as a power series in \((1-\zb) \), or equivalently a power series in \(\xit = (1-\zb)/\zb\), i.e. \(g(\zb) = \xit^\alpha \sum_n g_n \xit^n\).
Let us make a change of variables in the integral to \(\xit\), using the identity
\begin{equation} \label{eq:2F1zbtoxit}
_2 F_1\left(a,c-b;c;\frac{\zb}{\zb-1}\right) = (1-\zb)^a\, _2 F_1(a,b;c;\zb),
\end{equation}
to rewrite it as 
\begin{equation} \label{eq:invHatXit}
\invhat{g(\zb)} = -\frac{\Gamma(\hb)^2}{\pi^2\Gamma(2\hb-1)}  \int_0^\infty \mathrm{d}\xit  ~  \xit^{-\hb} \,_2F_1(\hb,\hb;2\hb;-\xit^{-1} )g(\xit),
\end{equation}
where by a small abuse of notation, we have written \(g(\xit)\) to mean \(g(\zb(\xit))\).
For large \(\hb\), this integral is dominated by the contribution from the region \( \xit \ll 1\).
Therefore, while the power series and integral may not generally be interchanged outside the radius of convergence, a term-by-term integration of the power series in \(\xit\) will give an expansion of the inversion that is asymptotic in large \(\hb\), or equivalently in large \(J^2 = \hb(\hb-1) \):
\begin{equation}
\invhat{g(\zb)} \sim \sum_{n=0}^\infty g_n ~ \invhat{\xit^{\alpha+n}}.
\end{equation}
The inversion of a single power \(\invhat{\xit^p} \) occurs when discussing the generalized free theory: by definition \( A[-p](\hb) := \inv{\xit^p} = 2\sin^2(\pi p)~\invhat{\xit^p} \).
This inversion can be found in many various ways, e.g. via an integral representation for the hypergeometric function, or using either of the methods in sections \ref{sec:wSeriesInv} and \ref{sec:Mellinmethods}, and gives
\begin{equation}
\frac{\invhat{\xit^{p}}}{(2\hb-1)} = \frac{\Gamma(1+p)^2\Gamma(\hb-p-1)}{\pi^2\Gamma(\hb+p+1)}\quad \Rightarrow \quad \frac{ \inv{\xit^{\alpha+n}}}{ \inv{\xit^{\alpha}}} = \frac{(-1)^n (\alpha +1)_n^2}{\left(\alpha -\hb+2\right)_n \left(\alpha +\hb+1\right)_n}
\end{equation}
so that
\begin{equation}
\inv{g(\zb)} \sim A[-\alpha](\hb)\sum_{n=0}^\infty g_n \, \frac{(-1)^n (\alpha +1)_n^2}{\left(\alpha -\hb+2\right)_n \left(\alpha +\hb+1\right)_n}
\end{equation}
provides an asymptotic expansion in \(1/\hb\) for \(\inv{g(\zb)} \).

In simple cases this asymptotic expansion may be recognized as that of a known function, such as rational functions of \(J^2\) and simple combinations of harmonic numbers \(S_1(\hb-1)\).
More generally an ansatz of appropriate functions can be fitted to a sufficiently high order asymptotic expansion, and the final result can then be checked for finite \(\hb\) through a numeric evaluation of the integral.

\subsection{Power series in \(\zb\) and \(1-\zb\)} \label{sec:wSeriesInv}
The approach in section \ref{sec:xitSeriesInv} is always able to provide an asymptotic expansion around \(J=\infty\), but unfortunately misses out on one of the main advantages of the inversion formula: the fact that it converges for finite spins down to \(\ell=2\).
However, a simpler approach would be to not make the change of variables to \(\xit\) but expand both the integration kernel and the double discontinuity \(g(\zb)\) as power series in \(\zb\) and \( 1-\zb \) respectively.
The kernel has radius of convergence \(1\), as do many of the double discontinuities encountered in this paper.
Since it is fine to integrate a power series term-by-term inside its radius of convergence, we expect that the resulting infinite sum gives the full non-asymptotic answer for the inversion integral.\footnote{There may be an issue involving convergence at the radius of convergence, which is not guaranteed. 
If the final result is convergent though, one expects that this interchange is valid by Abel's theorem.}

To be more explicit, we expand the kernel and double discontinuity as power series:
\begin{align}
\zb^{\hb-2} \, _2F_1(\hb,\hb;2\hb;\zb) = \sum_{n=0}^\infty \frac{(\hb)_n^2}{n! (2\hb)_n} \zb^{\hb-2+n}, \qquad g(\zb) = \left(\frac{\zb}{1-\zb}\right)^\alpha \sum_{k=0}^\infty g_k (1-\zb)^k.
\end{align}
Then we interchange the orders of summation and integration, and perform the simple beta-integral to find
\begin{equation} \label{eq:wInversionSeries}
\frac{\inv{g(\zb)}}{\inv{\xit^{-\alpha}}} = \frac{\Gamma (\hb)\Gamma (\hb-\alpha +1) }{\Gamma (1-\alpha ) \Gamma (2 \hb) }  \sum_{n,k=0}^\infty \frac{  (\hb)_n^2 (\hb+\alpha -1)_n (1-\alpha )_k  \,g_k}{ n! \, (2 \hb)_n (\hb)_{k+n}}.
\end{equation}

Depending on the form of the \(g_k\), this sum may be used to give a nice expression for the inversion. 
For example, if \(g(\zb)\) is a hypergeometric function of \((1-\zb)\), equation \ref{eq:wInversionSeries} becomes a Kamp{\'e} de Feriet function, which we use to derive the expressions for the crossing kernels in section \ref{sec:crossingKernels}.

In some special cases, the sum may even be evaluated to give an elementary function.
An example is the case \(P_0(\zb) = \left(\frac{\zb}{1-\zb} \right)^{\mu-1} (1-\zb) \), i.e. \(\alpha=\mu-1\) and \(g_k = \delta_{k,1}\).
The sum over \(k\) is trivial and the sum over \(n\) yields a \(\,_3F_2\) hypergeometric function at 1, to which we apply an identity \cite{hypAtUnitArgument} to find
\begin{equation}
\inv{P_0(\zb)} = (\mu-2)^2 \inv{\xit^{1-\mu}(1-\zb)} = \frac{A[\mu-1](\hb)}{\hb(\hb+2-\mu)}~ _3F_2\!\left(\!\left. {1,\hb,3-\mu} ~\atop~{\!\!\!\hb+1,\hb+3-\mu} \right| 1\right).
\end{equation}
We recognize this as a multiple of the function \(R(\hb,\mu)\), defined in equation \eqref{eq:Rdefn}.

Similarly one can incorporate logarithms like \(\log(\zb)\) in the power series of \(g(\zb)\).
A function of particular interest to us is the function \(Q_{\alpha}(\zb) =  -\left(\frac{\zb}{1-\zb} \right)^{\alpha} \log(\zb) \), which again trivializes the sum over \(k\).
The sum over \(n\) evaluates to derivatives of hypergeometric \(_2F_1\) functions at 1.
Since these hypergeometric functions are simply a ratio of gamma functions \cite{hypAtUnitArgument}, we can explicitly evaluate the result to find that
\begin{equation}
\inv{Q_\alpha(\zb)} = A[\alpha](\hb)~\mathbf{S_1}[\Delta](\hb),
\end{equation}
where \(\mathbf{S_1}[\alpha](\hb) = 2S_1(\hb-1)-S_1(\hb+\alpha-2)-S_1(\hb-\alpha)\), as defined in \eqref{eq:Qdefalpha}.

\subsection{Mellin space methods} \label{sec:Mellinmethods}
The idea of this method is to replace both the kernel \(\zb^\hb \,_2F_1(\hb,\hb;2\hb;\zb)\) and the double discontinuity \( g(\zb)\) by their Mellin representation \cite{Hogervorst:2017sfd, Liu:2018jhs}.
In many cases, such as when \(g\) is a hypergeometric function, this makes the \(\zb\) integral into a simple Beta-function integral, which can be performed with some suitable regularization.
What remains are two inverse Mellin transforms: for a suitable class of functions, discussed below, one of these can be performed using Barnes' second lemma \cite{Hogervorst:2017sfd}.
For simple enough functions, the final inverse Mellin transform may then be performed again using known lemmas, or a sum over poles may be performed to give an infinite sum representation of the inversion.

We will actually deviate from the method above slightly in a way that avoids both the need for regularization and for using Barnes' second lemma, instead directly giving the inversion integral as a single inverse Mellin transform.
First we define the Mellin transform of a function \(f(x)\) as
\begin{equation} \label{eq:MellinTransformDefn}
F(s) = \int_0^\infty \dd x~ x^{s-1} f(x).
\end{equation}
Defining \( F(s) = \int_0^\infty \dd x ~ x^{s-1} f(x)\) and \( G(s) = \int_0^\infty\dd x ~ x^{s-1} g(x)  \) to be the Mellin transforms of \(f\) and \(g\) respectively, we get that
\begin{equation} \label{eq:MellinConv}
\int_0^\infty  \dd x ~  x^{s-1} f(x)g(x)= \frac{1}{2\pi\ii}\int_{-\ii\infty}^{\ii\infty}\dd t~  F(t) G(s-t).
\end{equation}
To see the relevance to the inversion problem, we use the substitution \(\zb\rightarrow \xin = \xit^{-1} =\frac{\zb}{1-\zb}\) and some identities of hypergeometric functions to recast the inversion integral as
\begin{equation}
\invhat{g} =   \frac{\Gamma(\hb)^2}{\pi^2\Gamma(2\hb-1)} \int_0^\infty \frac{\mathrm{d}\xin}{\xin^2} \, \xin^\hb \,_2F_1(\hb,\hb;2\hb;-\xin) g(\xin),
\end{equation}
which is now of the form \eqref{eq:MellinConv}.

There are two free parameters in this method: we can shift around the powers of \(\xin\) between \(f\) and \(g\), and we can evaluate the Mellin transform at different \(s\). 
The latter does not affect the integrand, but the former does: however the poles and residues do not change, so that the inversion is insensitive to this.\footnote{The contour depends slightly on this choice: it needs to be chosen such that certain series of poles lie to its right, and others to its left.}
For most inversions, it has been useful to pick \(f(\xin) =  \frac{\Gamma(\hb)^2}{\pi^2\Gamma(2\hb-1)} \,_2F_1(\hb,\hb;2\hb;-\xin)\), and absorb the rest into \(g(\xin)\). 
For example, to invert \(\xit^p = \xin^{-p}\), we use \(g(\xin)= \xin^{-p+\hb-1}\), so that
\begin{equation}
F(s) = \frac{2\hb-1}{\pi^2} \frac{\Gamma(s)\Gamma(\hb-s)^2}{\Gamma(2h-s)}, \qquad G(s) = 2\pi \delta(\ii (-p+\hb-1+s)).
\end{equation}
Therefore we find that
\begin{equation}
\INVhat[\xit^p] =  \frac{1}{\ii} \int_{-\ii\infty}^ {\ii\infty} \mathrm{d}t ~ \delta(\ii (-p+\hb-1+t)) \frac{\Gamma(-t)\Gamma(\hb+t)^2}{\Gamma(2\hb+t)} = \frac{2\hb-1}{\pi^2} \frac{\Gamma(1+p)^2\Gamma(\hb-p-1)}{\Gamma(\hb+p+1)}.
\end{equation}

A general class of functions that can be inverted in this way, are the hypergeometric functions of \(-\xit\), i.e. functions of the form
\begin{equation} \label{eq:generalp+1Fp}
g(\xit) = \xit^{-k} \, _{p+1}F_p\left(\left. {a_1,\dots,a_{p+1}}~\atop~{\!\!\!\!\! b_1,\dots,b_p}\right|-\xit\right).
\end{equation}
For instance, a function in this class appeared in \cite{Li:2019dix} when considering the inversion of the stress tensor. 
We find that
\begin{equation}
\frac{\pi^2}{2\hb-1} \INVhat[g(\xit)] = \frac{1}{2\pi \ii} \int_{-\ii\infty}^{\ii\infty} \dd t ~ \frac{\Gamma (-t) \Gamma (1-k+t)^2 \Gamma (\hb+k-t-1)}{\Gamma (\hb-k+t+1)}\frac{\prod_{j=1}^{p+1}\Gamma(a_j+t)}{\prod_{j=1}^p\Gamma(b_j+t)}
\end{equation}
Closing the contour to the right, we find poles at \( t\in \mathbb{N}\) and \(t \in \hb+k-1 + \mathbb{N}\).
Computing the residues carefully, we find two sums corresponding to hypergeometric functions, leading to the final result
\begin{align}
\frac{\pi^2}{2\hb-1}  \INVhat[g(\xit)] &= \frac{\Gamma(2\hb)\Gamma(1-k)^2\Gamma(\hb+k-1)}{\Gamma(\hb)^2\Gamma(1+\hb-k)} \,_{p+3}F_{p+2}\left(\left. {1-k,1-k,a_1\,\dots,\,a_{p+1}}~\atop~{ \!\!\!2-\hb-k,1+\hb-k,b_1,\dots,b_p}\right|1\right)
\nonumber\\
&\quad + \Gamma(1-\hb-k)\frac{\prod_{j=1}^{p}\Gamma(b_j)}{\prod_{j=1}^{p+1}\Gamma(a_j)}      \frac{\prod_{j=1}^{p+1}\Gamma(\tilde{a}_j)}{\prod_{j=1}^p\Gamma(\tilde{b}_j)} 
\,_{p+3}F_{p+2}\left(\left. {\hb,\hb,\tilde{a}_1\,\dots,\,\tilde{a}_{p+1}}~\atop~{ \!\!\!2\hb,\hb+k,\tilde{b}_1,\dots,\tilde{b}_p}\right|1\right), \label{eq:p+1Fpinv}
\end{align}
where we defined \(\tilde{a}_j = a_j + \hb+k-1\), and similarly \(\tilde{b}_j = b_j + \hb+k-1\).

This is a non-asymptotic result that is valid for finite \(\hb\).
The analytic structure of this result is quite interesting.
The first hypergeometric function in equation \eqref{eq:p+1Fpinv} can be expanded in \(1/\hb\) from the definition of the hypergeometric function as an infinite sum; in fact, it is precisely the hypergeometric function that one would naively obtain if one expanded \(g(\xit)\) as a power series in \(\xit\) and integrated term-by-term, as in appendix \ref{sec:xitSeriesInv}.
This single hypergeometric function indeed gives the correct asymptotic expansion in large \(\hb\), but is incorrect for finite \(\hb\): it even has an infinite series of poles in \(\hb\), which are precisely cancelled by the other hypergeometric function.

An example of such a function is the special function \(B(\hb,\mu)\), defined in equation \eqref{eq:Bdefn}, for which the expression \eqref{eq:BResExact} was found using the method above\footnote{Another expression for \(B\), valid for finite \(\hb\), can be found by acting once with the Casimir and inverting as a power series in \((1-\zb)\), as in appendix \ref{sec:wSeriesInv}. 
The resulting expression is manifestly expandable in large \(\hb \), but involves derivatives of hypergeometric functions and due to its length we do not present it here. }.

\section{Inversion table} \label{app:inversionTable}

We present here, in tables \ref{tab:log2w} and \ref{tab:nonintPow}, a collection of inversion results, i.e. for a collection of different \(G(\zb)\), we give \(\inv{G(\zb)} \).
For the reader's convenience, we recall the definitions of frequently recurring functions:
\begin{align}
A[p](\hb) &= \frac{2(2\hb-1)\Gamma(\hb+p-1)}{\Gamma(p)^2\Gamma(\hb-p+1)},\\
\mathbf{S_1}[\alpha](\hb) &= 2S_1(\hb-1)-S_1(\hb+\alpha-2)-S_1(\hb-\alpha),\\
R(\hb,\mu) &= \frac{1}{\hb(\hb+2-\mu)}  \,_{3}F_{2}\left(\left. {1,\hb,3-\mu}~\atop~{ \!\!\!\hb+1,\hb+3-\mu}\right|1\right),\\
B(\hb,\mu)   &=    \frac{1}{J^2 \left(J^2-2\right)} \, _4F_3\!\left(\!\left. {1,1,2,\mu+1}~\atop~{\!\!3,3-\hb,\hb+2} \right|1\right)       \nonumber\\
&\qquad   -\frac{2 \pi  \Gamma (\hb) \Gamma (\mu +\hb-1)}{J^2 \Gamma (\mu +1) \sin (\pi  \hb)  \Gamma (2 \hb)} \, _3F_2\!\left(\!\left. {\hb-1,\hb,\hb+\mu -1}~\atop~{\!\!2 \hb,\hb+1} \right|1\right). 
\end{align}
From the inversions in these tables we can deduce many other inversions through the use of the collinear Casimir \(\Dbar\).
An example that is relevant to the first inversion in this paper, giving the first order anomalous dimensions \(\gamma_{T/A,\ell}^{(1)}\), is the following: table \ref{tab:nonintPow} says that \( \inv{\xit^{1-\mu}} = A[\mu-1](\hb)\). 
Since
\begin{equation}
\Dbar \left[ \frac{1 }{(\mu-2)^2} \xit^{1-\mu}\,(1-\zb)\, _2F_1\!\left(\!\left. {1,1}~\atop~{\!\!\!3-\mu} \right|1-\zb\right)  \right] = \xit^{1-\mu} ,
\end{equation}
it follows that
\begin{equation}
\inv{ \frac{1 }{(\mu-2)^2} \xit^{1-\mu}\,(1-\zb)\, _2F_1\!\left(\!\left. {1,1}~\atop~{\!\!\!3-\mu} \right|1-\zb\right) } =  \frac{A[\mu-1](\hb)}{J^2}.
\end{equation}

\begin{table}[H]
\begin{center}
\begin{tabular}{|l|l|}
\hline
$ G(\bar{z}) $ & $\inv{G(\zb)}$\rule[-1.1ex]{0pt}{0pt}  \\\hline
$\log ^2\left(1-\bar{z}\right)$&$\displaystyle 4 \left(2 \bar{h}-1\right) \dfrac{1}{J^2}$
\rule{0pt}{4.5ex} \rule[-4.5ex]{0pt}{0pt} \\
$\log^2(1-\bar z) \, \zb^{3-\mu} $ & $\displaystyle 2 \pi^2 \left(2 \bar{h}-1\right)  R(\hb,\mu)$
 \rule{0pt}{4.5ex} \rule[-4.5ex]{0pt}{0pt} \\
$ \frac{1}{4} \log^2(1-\zb)\, \xit  \,_3F_2\!\left(\! \left. {1,1,\mu +1}~\atop~{\!\!\!\!2,3} \right| -\xit\right)$ & $\displaystyle (2\hb-1)B(\hb,\mu) $
\rule{0pt}{4.5ex} \rule[-4.5ex]{0pt}{0pt} \\
\hline
\end{tabular}
\caption{Inversions for $ G(\bar{z}) $ not containing non-integer powers of \( (1-\zb) \). }\label{tab:log2w}
\end{center}
\end{table}

\begin{table}[H]
\begin{center}
\begin{tabular}{|l|l|}
\hline
$ G(\bar{z}) $ & $\inv{G(\zb)}$\rule[-1.1ex]{0pt}{0pt}  \\\hline
$\left(\frac{\zb}{1-\zb}\right)^p $&$ A[p](\hb)$
\rule{0pt}{4.5ex} \rule[-4.5ex]{0pt}{0pt} \\
$ \frac{1}{(\mu-2)^2}\xit^{1-\mu}  \,  (1-\zb)$ & $\displaystyle    A[\mu-1](\hb) R(\hb,\mu)   $
\rule{0pt}{4.5ex} \rule[-4.5ex]{0pt}{0pt} \\
$  \xit^{1-\mu}  \log \zb $ & $\displaystyle    A[\mu-1](\hb) \mathbf{S_1}[\mu-1](\hb)   $
\rule{0pt}{4.5ex} \rule[-4.5ex]{0pt}{0pt}\\
 \hline

\end{tabular}
\caption{Inversions for $ G(\bar{z}) $ containing non-integer powers of \( (1-\zb) \). }\label{tab:nonintPow}
\end{center}
\end{table}

For completeness, we also include an explicit form for $R(\hb,\mu)$ and $B(\hb,\mu)$ in integer dimensions $2,3,4$:
\begin{align}
R(\hb,1)&=1-J^2\left(2S_{-2}(\hb-1)+\zeta_2  \right),
\\
R(\hb,\tfrac32)&=4+2(2\hb-1)\left( S_1(\hb/2-1)-S_1(\hb/2-1/2)\right),
\\
R(\hb,2)&=2S_{-2}(\hb-1)+\zeta_2 ,
\end{align}
where $S_{-2}(n)=\frac14\left(S_2(n/2)-S_2(n/2-1/2)\right)-\zeta_2/2$,

\begin{align}
B(\hb,1)&=\frac2{J^4}-\frac2{J^2}-4S_3(\hb-1)+4\zeta_3,
\\
B(\hb,\tfrac32)&=\frac1{J^4}+\frac{\left(2\hb-1+2J^2S_1(\hb/2-1)-2J^2S_1(\hb/2-1/2)\right)^2}{3J^4},
\\
B(\hb,2)&=\frac1{J^4}.
\end{align}
To compare with the $\epsilon$-expansion, one also needs
\begin{equation}
\partial_\mu B(\hb,2)=\frac{2S_{-2}(\hb-1)+\zeta_2}{J^2}-2\left(
S_3(\hb-1)-\zeta_3\right)-\frac1{J^2}-\frac{1}{2J^4}.
\end{equation}


\section{Sums of conformal blocks on the unitarity bound}
\label{sec:unitarityboundblocks}

The contributions $H_{T\pm A}^{(2)}$ to the double discontinuity in sections~\ref{sec:S_at_N-1} and~\ref{sec:H_TpA} are sums of conformal blocks of twist $\tau=d-2$, corresponding to operators that saturate the unitarity bound.
To find these contributions, we are interested in computing sums of unitarity bound blocks,
$G^{(d)}_{\tau=d-2,\ell}(z,\zb)
$.
In this appendix, we will find an explicit expression for these blocks and some simple sums of them. 
These results are then used in appendix~\ref{app:HT_pm_A} to find explicit forms for \(H_{T\pm A}^{(2)}\).

\subsection{A solution basis}
Any unitarity bound conformal block is annihilated by the operator \cite{Alday:2016jfr} 
\begin{equation}
\dsat=2(\mu-1) \left(z^2 \frac{\partial }{\partial z}-\zb^2 \frac{\partial
   }{\partial \zb}\right)+2 z \zb (\zb-z) \frac{\partial }{\partial
   z}\frac{\partial }{\partial \zb}.
\end{equation}
The twist conformal blocks $H_{T\pm A}^{(m)}(z,\zb)$, as sums of such blocks, are also annihilated by \(\dsat\):
\begin{equation} \label{eq:dsatEqnforHTA}
\dsat H^{(m)}_{T\pm A}(z,\zb)=0.
\end{equation}
Recall that they also solve a recursion relation using the Casimir in \eqref{eq:CasimirDef}:
\begin{equation}
{\cas}^mH^{(m)}_{T\pm A}(z,\zb)=H^{(0)}_{T\pm A}(z,\zb).
\end{equation}
We shall find the twist conformal blocks by solving the \(\dsat\) equation, and then imposing the Casimir equation on the solution.
To do this, we make the following expansion:
\begin{equation}\label{eq:ansatzfordsat}
f(z,\zb)=(z \zb)^{\mu-1}\sum_kz^k g_k(y),\qquad y=\frac z\zb.
\end{equation}
Acting with \(\dsat\) on \(f(z,\zb)\), the $\dsat$ action respects the powers in $z$ and reduces to a linear ordinary differential equation for each $g_k(y)$:
\begin{align}
\dsat&\left((z \zb)^{\mu-1}z^{k}g_k(y)\right)\\
&=2(z \zb)^{\mu-1}z^{k+1}\left(\frac{k(\mu-1)  g_k(y)}{y}+(\mu-2- k+(\mu+ k) y) g_k'(y)+y (y-1)  g_k''(y)\right)=0. \nonumber
\end{align}
The general solution to this linear ODE is
\begin{equation}
g_k(y)=c_k\, y^{-k} \, _2F_1\left(\mu-1,-k;2-\mu-k;y\right)+\tilde c_k\, y^{\mu-1} \, _2F_1\left(\mu-1,2\mu+k-2;\mu+k;y\right).
\end{equation}
In general, solutions to $\dsat f=0$ will have $k$ ranging over $\mathbb Z$. 
Since conformal blocks behave like $G^{(d)}_{\tau=d-2,\ell}(z,\zb)\sim z^{\mu-1}$ in the limit of small \(z\), we see that we need \(\tilde c_{k}=0\) for all \(k\).
Furthermore we need \(c_k=0\) for all negative \(k\), since the corresponding solutions behave like \(\zb^{2\mu+k-2}\) for small \(\zb\), in contradiction with the behaviour $G^{(d)}_{\tau=d-2,\ell}(z,\zb)\sim \zb^{\mu-1}$ for small \(\zb\).
Therefore sums of unitarity bound blocks can be written as a sum  \begin{equation}\label{eq:UBBexpression}
\sum_{k=0}^\infty c_k(z \zb)^{\mu-1}T_k(z,\zb),\qquad T_k(z,\zb)=\zb^k\,{_2F_1}\left(\mu-1,-k;2-\mu-k;\frac z\zb\right).
\end{equation}
Notice that each $T_k$ is a symmetric polynomial in $z,\zb$ of homogeneous degree $k$.

\subsection{Examples}

\subsubsection*{Scalar block at \(\Delta = d-2\)}
The scalar conformal blocks $g^{(d)}_{\Delta,\ell}(z,\zb)$ are given in \eqref{eq:ScalarBlockAnyD}. 
In the particular case where $\Delta=d-2$, this is
\begin{equation}
G^{(d)}_{d-2,0}=(z \zb)^{\mu-1}g^{(d)}_{d-2,0}(z,\zb)=(z \zb)^{\mu-1}\sum_{k=0}^\infty c_kT_k(z,\zb),
\end{equation}
where
\begin{equation}
c_{k}= \frac{\left(\mu-1\right)_k^2}{ k! \,(2\mu-2)_k} .
\end{equation}
This is readily found by performing the sum over $n$ in \eqref{eq:ScalarBlockAnyD} and then expanding both sides for small $z$.

\subsubsection*{Tree-level correlator $H_{T\pm A}^{(0)}(z,\zb)$}
Also the tree-level twist conformal block \( H^{(0)}_{T-A}(z,\zb)\) is easily decomposed into the basis of functions \(T_k(z,\zb)\)
\begin{equation}\label{eq:HTmAasTsum}
H^{(0)}_{T-A}(z,\zb)=\left(\frac{z\zb}{(1-z)(1-\zb)}\right)^{\mu-1}=(z \zb)^{\mu-1}\sum_{k=0}^\infty  \frac{\left(\mu -1\right)_k}{k!} T_k(z,\zb).
\end{equation}
It is clear that \( H^{(0)}_{T+A}(z,\zb) \) is simply equal to a single such function:
\begin{equation}\label{eq:HTpAasTsum}
H^{(0)}_{T+A}(z,\zb) = (z \zb)^{\mu-1} \sum_{k=0}^\infty  \delta_{k,0} T_k(z,\zb)=(z \zb)^{\mu-1}.
\end{equation}

\subsubsection*{Spinning unitarity bound conformal block}
We provide here the explicit results for the unitarity bound blocks for spinning operators, as an infinite sum. As far as we are aware, this is a new result.
\begin{equation}\label{eq:solSpinningUBBlocks}
G^{(d)}_{\tau=d-2,\ell}(z,\zb)=(z \zb)^{\mu-1}\sum_{k=0}^\infty c_{k,\ell}T_k(z,\zb),
\end{equation}
where
\begin{equation}
c_{k,\ell}=\frac{\Gamma \left(\mu-1+k\right)^2 \Gamma (2\mu+2 \ell-2)}{\Gamma
   \left(\mu-1+\ell\right)^2 \Gamma (k-\ell+1) \Gamma (2\mu+k+\ell-2)}.
\end{equation}
This was found by looking for simultaneous solutions of 
\begin{equation}
\dsat G^{(d)}_{\tau=d-2,\ell}(z,\zb)=0,\qquad \cas G^{(d)}_{\tau=d-2,\ell}(z,\zb)=J_{d-2,\ell}^2\, G^{(d)}_{\tau=d-2,\ell}(z,\zb),
\end{equation}
where $J_{\tau,\ell}^2=\left(\ell+\tau/2\right)\left(\ell+\tau/2-1\right)$ and the Casimir $\cas$ was defined in \eqref{eq:CasimirDef}. 

For consistency we have checked that the solutions \eqref{eq:solSpinningUBBlocks} reduce to the well-known spinning conformal blocks in four dimensions.

\section{Computing $H_{T\pm A}^{(2)}$} \label{app:HT_pm_A}
In this appendix we provide some more details about the computation of the contributions $H_{T\pm A}^{(2)}$ used in sections~\ref{sec:S_at_N-1} and~\ref{sec:H_TpA}. As described in detail in these sections, they are special cases of the following twist conformal blocks at the unitarity bound:
\begin{equation}
H_{T\pm A}^{(m)}(z,\zb) = \sum_{\ell=0,1,2,\dots} \frac{a^{\text{free}}_{\mu,\ell} }{ J^{2m}_{d-2,\ell}} (\mp)^{\ell}G^{(d)}_{\tau=d-2,\ell}(z,\zb),
\end{equation} 
where $a^{\text{free}}_{\mu,\ell}$ is OPE coefficients for a free scalar field in $d=2\mu$ dimensions. 

In section~\ref{sec:HTpmAKampe} we will use the expression \eqref{eq:UBBexpression} for sums of spinning unitarity bound blocks to write down $H_{T\pm A}^{(2)}$ in terms of the so-called Kamp{\'e} de Feriet functions \cite{exton1978handbook}. 
These functions give an expression for the full \(H_{T\pm A}^{(2)}(z,\zb) \) as an infinite sum in powers of \(z,\zb\); however we are interested in the limit \(w = 1- \zb \ll 1\), which is difficult to extract from this expression.
In section \ref{sec:HTpmADiffEqn} we describe a method using coupled differential equations that allows us to find the small \(w\) limit explicitly, with some integration constants fixed by the Kamp{\'e} de F{\'e}riet expansion.

\subsection{As Kamp{\'e} de F{\'e}riet functions} \label{sec:HTpmAKampe}
Notice that the conformal blocks satisfy the following relation
\begin{equation}
G^{(d)}_{\tau,\ell}(z,\zb)=(-1)^\ell G^{(d)}_{\tau,\ell}\left(\tfrac{z}{z-1},\tfrac{\zb}{\zb-1}\right),
\end{equation}
from which it follows that 
\begin{equation}
H^{(m)}_{T-A}(z,\zb)=\left.H^{(m)}_{T+A}(z,\zb)\right|_{z,\zb\to\tfrac{z}{z-1},\tfrac{\zb}{\zb-1}}.
\end{equation}
This can clearly be seen for $m=0$ in equations \eqref{eq:HTmAasTsum} and \eqref{eq:HTpAasTsum}. This means that we can focus our attention on computing for instance $H^{(2)}_{T+A}(z,\zb)$, from which $H^{(2)}_{T-A}(z,\zb)$ will follow immediately. 

$H^{(m)}_{T+A}(z,\zb)$ is a sum of unitarity bound blocks, and we therefore look for an expression of the form \eqref{eq:UBBexpression}
\begin{equation}\label{eq:HTpAansatz}
H^{(m)}_{T+A}(z,\zb)=(z \zb)^{\mu-1}\sum_{k=0}^\infty c^{(m)}_{k}T_k(z,\zb), \qquad c^{(0)}_{k}=\delta_{k,0}.
\end{equation}
The action of the Casimir \eqref{eq:CasimirDef} gives a recursion relation amongst the basis functions:
\begin{equation}
\cas \left((z \zb)^{\mu-1}T_k(z,\zb)\right)=(z \zb)^{\mu-1}(\mu-1+k)(\mu-2+k)T_k(z,\zb)+(z \zb)^{\mu-1}(\mu-1+k)^2T_{k+1}(z,\zb).
\end{equation}
With the ansatz and initial condition \eqref{eq:HTpAansatz}, the recursion relation leads to a difference equation that can be solved case by case in $m$. The first two solutions are
\begin{equation}
c^{(1)}_{k}=\frac{(-1)^k}{(\mu-2)(\mu-1+k)},\qquad c^{(2)}_{k}=\frac{(-1)^k(k+1)}{(\mu-2)^2(\mu-1+k)^2}, \qquad k=0,1,2,\ldots
\end{equation}
Using the finite sum representation for $T_k(z,\zb)$, we can express the $H^{(m)}_{T+A}(z,\zb)$ as double sums over $z,\zb$. 
After some manipulations, the explicit results are
\begin{align}
H^{(1)}_{T+A}(z,\zb)&=\frac{(z\zb)^{\mu-1}}{(\mu-1)(\mu-2)}
 F^{11}_{10}\left(
\begin{matrix}1\\\mu\end{matrix}
\middle|
\begin{matrix}\{\mu-1\},\{\mu-1\}\\\phantom x\end{matrix}
\middle|z,\zb\right) \nonumber\\
& = \frac{(z\zb)^{\mu-1}}{(\mu-1)(\mu-2)} \mathrm F_1(1;\mu-1,\mu-1;\mu;z,\zb),
\label{eq:H1TpAasKdF}
\\
H^{(2)}_{T+A}(z,\zb)&=\frac{(z\zb)^{\mu-1}}{(\mu-1)^2(\mu-2)^2}
 F^{21}_{20}\left(
\begin{matrix}2,&\mu-1\\\mu,&\mu\end{matrix}
\middle|
\begin{matrix}\{\mu-1\},\{\mu-1\}\\\phantom x\end{matrix}
\middle|z,\zb\right),
\label{eq:HTpAasKdF}
\end{align}
where $F^{pq}_{rs}$ is the Kamp\'e de F\'eriet function defined in \eqref{eq:KdFdef}, and $ \mathrm F_1$ is Appell's hypergeometric function \cite{zbMATH02708199,Appell1925}. $H^{(2)}_{T-A}(z,\zb)$ follows from \eqref{eq:HTpAasKdF} upon replacing $z,\zb\to\tfrac{z}{z-1},\tfrac{\zb}{\zb-1}$.

\subsection{Differential equation approach} \label{sec:HTpmADiffEqn}
Here we present a different approach, in which the small \(z\) limit of \(H_{T\pm A}^{(2)}(1-\zb,1-z)\) is evident from the start, and which is based on the observation that both the Casimir \(\cas\) and \(\dsat\) preserve integrality and non-integrality of powers of \(z, (1-\zb) \).
We can therefore expand \(H^{(2)}_{T\pm A}(z,1-w)\) in powers of \(w=1-\zb\) and treat the differential equations for integer-power part completely separately.
That is, we write:
\begin{equation}
H^{(2)}_{T\pm A,\text{int}}(1-\zb,1-z) = \xit^{\mu-1} \sum_{k=0}^\infty z^k \left(f^{(\pm)}_k(\xit)+g^{(\pm)}_k(\xit)\log z\right),
\end{equation}
where \(\xit= \tfrac{1-\zb}{\zb}\), and we impose the \(\dsat\) and Casimir equations order by order in \( z\).

By isolating the \(\log z\) piece of these equations, we can find all the \(g_k^{(\pm)}\) order by order:
\begin{align}\label{eq:g0result}
g_0^{(-)}(\xit) &= -\frac{\pi  \csc (\pi \mu )}{\mu-2},                        & g_1^{(-)}(\xit) &= -\frac{ (\mu -1)^2  \pi \csc (\pi  \mu )}{\mu -2} \xit, & \dots,\\
g_0^{(+)}(\xit) &= -\frac{1}{(\mu-2)^2},                       & g_1^{(+)}(\xit) &= -\frac{1}{\mu-2}-\frac{(\mu-1)^2  }{(\mu-2)^2}\xit, & \dots ,
\end{align}
where the overall constant was fixed from the Kamp{\'e} de Feriet expansions 
\eqref{eq:HTpAasKdF}.

Plugging this back into the differential equations, we get a set of two coupled differential equations for \(f_0(\xit)\) and \(f_1(\xit)\).
We can eliminate \(f_1(\xit)\) from the Casimir equation and put this in the \(\dsat\) equation to get a single, third order differential equation for \( f_0(\xit) \).
This is an inhomogeneous differential equation, with inhomogeneous part determined by {\( \xit^{1-\mu} H_{T\pm A}^{(1)}(1-\zb,1-z)\big|_{z^0} \)}
, which can be extracted from \eqref{eq:H1TpAasKdF}\footnote{They satisfy
\begin{align*}
\left. H_{T-A}^{(1)}(1-\zb,1-z)\right|_{z^0} &= -\frac{\pi  \csc (\pi \mu )}{\mu-2} \xit ^{\mu-1} \, _2F_1\left(2-\mu,\mu-1;1;-\xit \right),\\
\left. H_{T+A}^{(1)}(1-\zb,1-z)\right|_{z^0} &= -\frac{1}{\mu -2} \xit^{\mu-1}\log z -\frac{S_1\left(\mu-2 \right)}{\mu -2} \xit^{\mu-1}   -\frac{\mu -1}{\mu -2} \xit ^{\mu}\,_3F_2\!\left(\!\left. {1,1,\mu}~\atop~{\!\!\!2,2} \right| -\xit\right) .
\end{align*}}.
Specifically, we find the equations:
\begin{align}
\xit  (\xit +1) \partial_{\xit}^3 f^{(-)}_0 (\xit )+2 (2 \xit +1) \partial_{\xit}^2 f^{(-)}_0 (\xit ) - \mu(\mu-3) \partial_{\xit} f^{(-)}_0 (\xit )&= \frac{1-\!\, _2F_1\left(3-\mu,\mu;2;-\xit \right)}{4 \xit }, \label{eq:HTminA2DiffEqnf0}  \\
\xit  (\xit +1) \partial_{\xit}^3 f^{(+)}_0 (\xit )+2 (2 \xit +1) \partial_{\xit}^2 f^{(+)}_0 (\xit ) - \mu(\mu-3)  \partial_{\xit} f^{(+)}_0 (\xit )&= \frac{(1+\xit)^{1-\mu}+(\mu-1) \xit-1}{(\mu-2) \xit ^2}.\label{eq:HTplusA2DiffEqnf0}
\end{align}
The equation \eqref{eq:HTminA2DiffEqnf0} for \(f^{(-)}_0\) is easily solved exactly. 
Demanding that the solution has no logarithmic singularity at \(\xit=0\), and that it is compatible with the Kamp{\'e} de Feriet expansions, fixes the integration constants to give
\begin{align}
 f^{(-)}_0 (\xit ) = \pi  \csc (\pi  \mu )   & \bigg[\frac{\pi  \cot (\pi  \mu )}{\mu -2} -\frac{1}{(\mu -2)^2} +(\mu -1)\xit  \,_4F_3\!\left(\!\left. {1,1,3-\mu,\mu}~\atop~{\!\!\!2,2,2} \right| -\xit\right) \nonumber \\
 & \qquad-\frac{\, _2F_1(2-\mu ,\mu -1;1;-\xit )-1}{(\mu -2)^2}\bigg].
\end{align}
As for \(f^{(+)}_0(\xit)\), we are unable to solve the differential equation directly, but we can find it as a series in \(\xit\) and match this up to known functions.
This way we find that
\begin{align}
 f^{(+)}_0 (\xit ) = \frac{1}{(\mu -2)^2}& \bigg[ \frac{(\mu -2)^2 (S_2(\mu -2)-\zeta_2)+\mu -3}{\mu -2}\, _2F_1(2-\mu ,\mu-1;1;-\xit ) \nonumber \\
& \quad  + \frac{1}{2} (\mu -2) (\mu -1) \mu \left(\overline{D}-K^2\right)^{-1} \left[\xit  \,_3F_2\!\left(\! \left. {1,1,\mu +1}~\atop~{\!\!\!\!2,3} \right| -\xit\right)\right] \nonumber\\
  &  \quad -\frac{\mu -3}{\mu -2} - S_1(\mu -2) \bigg].
\end{align}
Here \(\overline{D}\) is the collinear Casimir operator defined in \eqref{eq:dbardefn}, and \(K^2= (\mu-1)(\mu-2) \). 
Note that \( \left(\overline{D}-K^2\right)^{-1} \left[ g(\xit) \right] \) is the solution to an ODE with kernel equal to \( \, _2F_1(2-\mu ,\mu-1;1;-\xit )\).
In the expression above, the coefficient of the kernel element is chosen such that in a small \(\xit\) expansion \( \left(\overline{D}-K^2\right)^{-1} \left[ g(\xit) \right] \) starts at one order higher than \( g(\xit)\), i.e.
\begin{equation}
\left(\overline{D}-K^2\right)^{-1} \left[\xit  \,_3F_2\!\left(\! \left. {1,1,\mu +1}~\atop~{\!\!\!\!2,3} \right| -\xit\right)\right] = c_2 \xit^2  + c_3 \xit^3 + c_4\xit^4 + \dots.
\end{equation}
This choice of boundary condition guarantees that the inversion integrals of these functions are related by
\begin{equation}
\widehat{\mathrm{INV}}\left[ \left(\overline{D}-K^2\right)^{-1} \big[ g(\xit) \big]\right] = \frac{1}{J^2-K^2} \widehat{\mathrm{INV}} \left[ g(\xit) \right].
\end{equation}
%
\section{Mixed correlator bootstrap for OPE coefficients}\label{sec:mixed}

In this appendix we show how to derive the mixed OPE coefficients 
\begin{equation}
\alpha_{n,\ell}:=c_{\varphi\varphi[\sigma,\sigma]_{n,\ell}}c_{\sigma\sigma[\sigma,\sigma]_{n,\ell}}
\end{equation}
 from the inversion integral by studying crossing for the mixed correlator $\langle\varphi^i\varphi^i\sigma\sigma\rangle$. From the $\alpha_{n,\ell}$ we extract the expression~\eqref{eq:sigsigOPECoeffs} for $c^2_{\varphi\varphi[\sigma,\sigma]_{n,\ell}}$, which is used in the main text to compute the contributions from $[\sigma,\sigma]_{n,\ell}$ in section~\ref{sec:sigmasigma}.

The crossing equation for the mixed correlator reads
\begin{equation}
\G_{
\varphi^i\varphi^i\sigma\sigma
}(u,v)
=\frac{u^{\Delta_\varphi}}{v^{\frac{\Delta_\varphi+\Delta_\sigma}2}}
\G_{\sigma\varphi^i\varphi^i\sigma}(v,u),
\end{equation}
and to leading order we may use $\Delta_\varphi=\mu-1$ and $\Delta_{\sigma}=2$. As in the main text, we consider contributions from operators in the crossed channel with non-vanishing double-discontinuitites at leading order. An important difference is that now the identity operator does not appear, since all crossed channel operators in the $\varphi^i\times\sigma$ OPE transform in the vector representation: $ S\otimes\mathbf N=\mathbf N$. We can express the structure of this OPE as in \eqref{eq:OPEorders} in the introduction:
\begin{equation}
\sigma\times \varphi^i= [\sigma,\varphi]^i_{n,\ell}+\frac1N\varphi^i +O(N^{-2}).
\end{equation}

The operators $[\sigma,\varphi]^i_{n,\ell}$ have double-twist dimensions and they therefore have zero double-discontinuity until order $N^{-2}$. Thus, to leading order in $N^{-1}$ we only need to consider the contribution from the (space-time) scalar $\varphi^i$. Its (squared) OPE coeffiecient correlator in this channel is equal to
$c_{\varphi\varphi\sigma}^2=a_{\sigma}^{(0)}/N+O(N^{-2})$, which was given in \eqref{eq:asigma0res}. This means that we have to compute
\begin{equation}\label{eq:phiblock}
\INV\bigg[
 \frac{a_{\sigma}^{(0)}}N\frac{u^{\Delta_\varphi}}{v}g^{(d)}_{\Delta_\varphi,0}(v,u)
 \bigg]
\end{equation}
where we used $\Delta_\sigma=2$ and where $v^{\Delta_\varphi/2}g^{(d)}_{\Delta_\varphi,0}(v,u)$ is the crossed channel scalar conformal block \emph{in the mixed correlator}. The expression for this block is given by a simple modification compared to the case \eqref{eq:ScalarBlockAnyD} of identical operators \cite{Dolan:2000ut}:
\begin{equation}
g^{(d)}_{\Delta,0}(v,u)=\sum_{m,n=0}^{\infty}v^m(1-u)^n\frac{\big(\frac{\Delta+\Delta_{12}}2\big)_m\big(\frac{\Delta-\Delta_{34}}2\big)_m\big(\frac{\Delta-\Delta_{12}}2\big)_{m+n}\big(\frac{\Delta+\Delta_{34}}2\big)_{m+n}}{m!n!(\Delta)_{2m+n}(\Delta+1-\mu)_m},
\end{equation}
with $\Delta_{ij}=\Delta_i-\Delta_j$. We will evaluate this at leading order: $\Delta_1=\Delta_4=2$ and $\Delta_2=\Delta_3=\mu-1$. From the expression, one can see that the scalar block contribution \eqref{eq:phiblock} only contains integer powers of $(1-\zb)$. Since non-negative integer powers of $(1-\zb)$ have zero double-discontinuity we only need to consider the term proportional to $(1-\zb)^{-1}$. Keeping only this term after performing the $m$ sum, the sum over $n$ can also be computed and gives 
\begin{equation}
\frac{u^{\Delta_\varphi}}{v}g^{(d)}_{\Delta_\varphi,0}(v,u)=\frac{z^{\mu-1}}{(1-z)(1-\zb)}{_2F_1}(\mu-2,\mu-2;\mu-1;1-z)+O((1-\zb)^0).
\end{equation}
In the small $z$ limit, this expression has two types of terms, non-integer powers $z^{\mu-1+m}$, $m\in\mathbb Z$, contributing to $[\varphi,\varphi]_{n,\ell}$ in the direct channel, and integer powers $z^{m+2}$ contributing to $[\sigma,\sigma]_{n,\ell}$. Focusing on the integer powers we get the expansion
\begin{equation}
\frac{u^{\Delta_\varphi}}{v}g^{(d)}_{\Delta,0}(v,u)=\frac{\mu-2}{\mu-3}\left(
z^2+\frac{\mu-5}{\mu-4}z^3+\frac{\mu^2-10\mu+27}{(\mu-4)(\mu-5)}z^4+\ldots.
\right)\frac{1}{1-\zb}+\ldots
\end{equation}
The leading term immediately gives that
\begin{equation}
\alpha_{0,\ell}=\frac{\Gamma(2+\ell)^2}{\Gamma(4+2\ell)}\frac{(\mu-2)a_{\sigma}^{(0)}}{N(\mu-3)}A[1](2+\ell)=\frac{2\Gamma(\ell+2)^2}{\Gamma(2\ell+3)}\frac{\mu-2}{\mu-3}\frac{a_{\sigma}^{(0)}}N.
\end{equation}
Here we used $\hb=2+\ell$ and that the inversion of $(1-\zb)^{-1}$ is simply $A[1](\hb)=2(2\hb-1)$.

To compute the OPE coefficients for $[\sigma,\sigma]_{n,\ell}$ with $n>0$ we should analyse the term proportional to $z^{n+2}$, as well as project away the contribution from operators at lower order. These contributions can be computed from the subleading collinear blocks as given in appendix A of \cite{Henriksson:2018myn} (see also \cite{Simmons-Duffin:2016wlq}). For instance, at $n=1$ we look at the term proportional to $z^3$:
\begin{equation}\label{eq:projontoz3}
\!\sum_{\ell}\!\!\alpha_{0,\ell}\left[
c_{1,-1}k_{\ell+1}(\zb)+c_{1,0}k_{\ell+2}(\zb)+c_{1,1}k_{\ell+3}(\zb)
\right]
+\alpha_{1,\ell}k_{\ell+3}(\zb)=\frac{(\mu-2)(\mu-5)}{(\mu-3)(\mu-4)}\frac{a_{\sigma}^{(0)}/N}{(1-\zb)}+\text{reg.},
\end{equation}
where
\begin{equation}
c_{1,-1}=\frac{\ell(\mu-1)}{\ell+\mu-2},\quad c_{1,0}=1,\quad c_{1,1}=\frac{(\ell+2)^2(\ell+3)(\mu-1)}{4(2\ell+3)(2\ell+5)(5+\ell-\mu)}.
\end{equation}
Using the inversion integral for $\hb=3+\ell$ we see that the right-hand side of \eqref{eq:projontoz3} is produced by a sum over terms $\frac{(\mu-2)(\mu-5)}{(\mu-3)(\mu-4)}\frac{2\Gamma(\ell+3)^2}{\Gamma(2\ell+5)}\frac{a_{\sigma}^{(0)}}{N}k_{\ell+3}(\zb)$. Up to regular terms, we can further make a shift in $\ell$ in the first two terms in the left-hand side so that all terms multiply $k_{\ell+3}(\zb)$. This gives an equation for $\alpha_{1,\ell}$ which is solved by
\begin{equation}
\alpha_{1,\ell}=\frac{(\mu-2)^2(\ell+1)(\ell+4)}{(\ell+\mu)(5+\ell-\mu)(4-\mu)}\frac{\Gamma(\ell+3)^2}{\Gamma(2\ell+5)}\frac{a_{\sigma}^{(0)}}N.
\end{equation}
Proceeding in this fashion for $n=2,3,\ldots$ one finds that
\begin{equation}\label{eq:alphanlres}
\alpha_{n,\ell}=\frac{(\mu-2)^2}{(n+2-\mu)(n+3-\mu)J^2_{2n+4,\ell}}\frac{a_\sigma^{(0)}a^{\mathrm{GFF}}_{n,\ell}|_{\Delta =2}}{N},
\end{equation}
where $J^2_{2n+4,\ell}=(n+\ell+2)(n+\ell+1)$ and $a^{\mathrm{GFF}}_{n,\ell}|_{\Delta=2}$ are the (squared) OPE coefficients for a $\Delta=2$ generalized free scalar.
Using that $a^{\mathrm{GFF}}_{n,\ell}|_{\Delta=2}=c^2_{\sigma\sigma[\sigma,\sigma]_{n,\ell}}$ we eliminate $c_{\sigma\sigma[\sigma,\sigma]_{n,\ell}}$ and finally arrive at the expression 
\begin{equation}
c^2_{\varphi^i\varphi^i[\sigma,\sigma]_{n,\ell}} = \frac{\big(a_\sigma^{(0)}\big)^2}{N^2}\frac{(\mu-2)^2}{(\mu-3)^2} \frac{(\mu-1-n)_n^2}{(\mu-3-n)_n^2}  \frac{1}{J^4_{2n+4,\ell}} \left. a^{\mathrm{GFF}}_{n,\ell}\! \right|_{\Delta=2} \, +\order(N^{-3}),
\end{equation}
quoted in the main text in \eqref{eq:sigsigOPECoeffs}.

Notice that we derived the result \eqref{eq:alphanlres} for $\alpha_{n,\ell}$ purely from large spin perturbation theory. We can compare the result with the literature by looking at the known order $N^{-1}$ expression for the mixed correlator \cite{Lang:1992pp}. Using \(\overline{D}\) functions \cite{Dolan:2000ut}, this result reads, in our conventions\footnote{Notice that this correlator was translated into \(\overline{D}\) functions already in \cite{Alday:2015ewa}. 
The $(u,v)$ there are defined by other conventions, namely $u_{\mathrm{there}}=\frac{x_{13}^2x_{24}^2}{x_{12}^2x_{34}^2}=\frac 1u$, $v_{\mathrm{there}}=\frac{x_{14}^2x_{23}^2}{x_{12}^2x_{34}^2}=\frac{v}{u}$. 
This can be transformed into the form \eqref{eq:phiphisigmasigmacorr} by using identities for $\overline D$ \cite{Dolan:2000ut}.}
\begin{align}
\left.\langle\varphi^i\varphi^j\sigma\sigma\rangle\right|_{\frac1N} &=\frac{\delta^{ij}}{x_{12}^{2\Delta_\varphi}x_{34}^{2\Delta_\sigma}} \gamma_\varphi^{(1)}   \bigg[   \frac{2\mu(\mu-1)(2\mu-3)}{(\mu-2)^2\Gamma(\mu-2)}   u^{\mu-1}\overline D_{\mu-1,\mu-1,1,1}(u,v) \nonumber\\  \label{eq:phiphisigmasigmacorr}
& \qquad + 
\frac{\Gamma(\mu+1)\,\,\,}{\Gamma(\mu-1)^2} \Big(u^{\mu-1}  \overline D_{\mu-2,\mu-1,1,2}(u,v)     +     u^2 \overline D_{1,2,\mu-1,\mu-2}(u,v) \Big)    \bigg].
\end{align}
A conformal block decomposition of this, again using the subleading corrections to the conformal blocks, gives exactly \eqref{eq:alphanlres}.

\section{Infinite sum for \(S(v,u)\) from Mellin amplitude} \label{app:MellinForSigSig}
From the Mellin representation of this contribution, we can deduce a closed form expression: 
\begin{equation}
S(v,u)= K(\mu)^2\sum_{j=2}^\infty \sum_{k=2}^\infty \frac{2 \Gamma (j+k-2)^2}{\Gamma (k-1)^2 \Gamma(j-1)^2} c_{j,k} v^j u^{k-2},                                     
\end{equation}
where
\begin{align} \label{eq:genSigsigTerm}
c_{j,k} =& \quad\,\frac18 \log^2 v~  R_{k,j} \nonumber\\
&+ \frac12 \log v \left[ \left(S_1(k+j-3)-S_1(k-2) \right)R_{k,j} + \Sigma_1^{(0)}(k,j)+ \Sigma_2^{(0)}(k,j) \right] \nonumber\\
& +  \left(S_1(k+j-3)-S_1(k-2) \right) \left(\Sigma_1^{(0)}(k,j) + \Sigma_2^{(0)}(k,j) \right) + \Sigma_1^{(1)}(k,j) + \Sigma_2^{(1)}(k,j)  \nonumber\\
& +\frac{1}{4} \left[2 \zeta _2+2 \left(S_1(j+k-3)-S_1(k-2)\right)^2  -  \left(S_2(j+k-3)-S_2(k-2)\right)\right]R_{k,j},
\end{align}
where
\begin{align}
R_{k,j}& = 2\Gamma(j-1)^2 \res_{s=2j} \left[ \frac{1}{4} \Gamma\left(2-\frac{s}{2}\right) \tilde{R}(k,s) \right] \nonumber \\
&= \frac{2 (\mu -3)\Gamma(j-1)^2 \Gamma (\mu-j -1)^2 }{\Gamma (\mu -3)} \sum_{i=0}^{j-2} \frac{(-1)^i}{ (-i+j+k-3) \Gamma (i+1) \Gamma (\mu-i -2)}, \\  
\Sigma_1^{(0)}(k,j) &= \frac12 \sum_{\substack{\tilde{k} \geqslant 2\\ \tilde{k}\neq k}} \frac{R_{\tilde{k},j}}{k-\tilde{k}}, \qquad\qquad\qquad\,
\Sigma_1^{(1)}(k,j) = - \frac14 \sum_{\substack{\tilde{k} \geqslant 2\\ \tilde{k}\neq k}} \frac{R_{\tilde{k},j}}{(k-\tilde{k})^2}, \\
\Sigma_2^{(0)}(k,j) &= \frac12 \sum_{\substack{\tilde{k} \geqslant 2
}} \frac{R_{\tilde{k},j}}{4-(j+\tilde{k}+k)}, \qquad
\Sigma_2^{(1)}(k,j) = \frac14 \sum_{\substack{\tilde{k} \geqslant 2
}} \frac{R_{\tilde{k},j}}{(4-(j+\tilde{k}+k))^2}.
\end{align}
We have been able to evaluate some, but unfortunately not all, of these sums by suitably swapping orders of summation.
Recall that in the end, we only need the small \(u\) limit of \(S(v,u)\), so that we only need to know the sum \( \sum_{j\geqslant2} \frac{2 \Gamma(j)^2}{\Gamma(j-1)^2} c_{j,2} v^j \). 

The simplest one is simply \(R_{2,j}\), which equals
\begin{equation}
R_{2,j} = 2 (\mu -3) \frac{ \Gamma (j-1) }{(j-1) (4-\mu )_{j-2}} \left(S_1(2-\mu ) - S_1(j-\mu +1)\right) .
\end{equation}
This implies a contribution to \(S(v,u)\large|_{u^0}\) of the form
\begin{equation}
\sum_{j\geqslant2} \frac{2 \Gamma(j)^2}{\Gamma(j-1)^2} R_{2,j} v^j    =     4 v^2 \bigg[  \,_2F_1\!\left(\!\left. {\,\,1,2}~\atop~{\!\!4-\mu}\right| v \right) +     (\mu -3) ~ \dda  \,_2F_1\!\left(\!\left. {\,\,1,2}~\atop~{\!\!4-\mu+a}\right| v\right) \!    \bigg],
\end{equation}
where, as in the main text, $\dda f = \left.\frac{\partial f}{\partial a}\right|_{a=0}$.
Up to an overall factor, this is essentially the \(\log^2 v\) piece, and completely determines \(U^{(2)}_{T/A,\hb}\) in \eqref{eq:fullU2_TA_Nmin2}.

The next sum that we have been able to find, is \(\Sigma_2^{(0)}\):
\begin{equation}
\Sigma_2^{(0)}(2,j) = -\frac{(\mu -3) }{(\mu -2)}  \frac{\Gamma (j-1) }{(j-1)  (4-\mu )_{j-2}} \left(S_1(2-\mu )-S_1(j-\mu +1)+S_1(j-1)\right).
\end{equation}
Therefore this contributes with a sum
\begin{align}
\sum_{j\geqslant2} \frac{2 \Gamma(j)^2}{\Gamma(j-1)^2} \Sigma_2^{(0)}(2,j) v^j 
&= -2 v^2 \bigg[         \,_2F_1\!\left(\!\left. {\,\,1,2}~\atop~{\!\!4-\mu}\right| v\right)   +  \frac{\mu -3}{\mu -2}  \dda  \,_2F_1\!\left(\!\left. {\,\,1,2+a}~\atop~{\!\!4-\mu+a}\right| v\right)          \bigg].
\end{align}

We have partial results for the other sums: specifically, we have been able to evaluate the \(\zeta_2\) pieces of the sums \(\Sigma_2^{(1)}(1,j)\) and \( \Sigma_2^{(1)}(2,j) \).
These are, respectively:
\begin{align}
\sum_{j\geqslant2} \frac{2 \Gamma(j)^2}{\Gamma(j-1)^2} \Sigma_1^{(0)}(2,j) v^j\Big|_{\zeta_2} &= - v^2 \bigg[  \,_2F_1\!\left(\!\left. {\,\,1,2}~\atop~{\!\!4-\mu}\right| v\right)    + (\mu -3) \dda \,_2F_1\!\left(\!\left. {\,\,1,2}~\atop~{\!\!4-\mu+a}\right| v\right)     \bigg], \\
\sum_{j\geqslant2} \frac{2 \Gamma(j)^2}{\Gamma(j-1)^2} \Sigma_2^{(0)}(2,j) v^j\Big|_{\zeta_2} &=-\frac{v^2}{\mu -2} \bigg[ \,_3F_2\!\left(\!\left. {\,\,1,2,2}~\atop~{\!\!4-\mu,4-\mu}\right| v\right)   +(\mu -3) \,_2F_1\!\left(\!\left. {\,\,1,2}~\atop~{\!\!4-\mu}\right| v\right) \! \bigg].
\end{align}
Together with the explicit \(\zeta_2\) term in \eqref{eq:genSigsigTerm}, this gives the full \(\zeta_2\) piece of the small \(u\) limit of \(S(v,u)\) as
\begin{equation}\label{eq:Svuz2partAppendix}
 S(v,u)|_{u^0\zeta_2} = K(\mu)^2 v^2\Bigg[\frac{\,_2F_1\!\left(\!\left. {\,\,1,2}~\atop~{\!\!4-\mu}\right| v\right)}{\mu -2}-\frac{\,_3F_2\!\left(\!\left. {\,1,2,2}~\atop~{4-\mu ,4-\mu}\right| v\right)}{\mu -2}+(\mu -3) \dda \,_2F_1\!\left(\!\left. {\,\,1,2}~\atop~{\!\!4-\mu+a}\right| v\right)\!\!\Bigg] \! .
\end{equation}
\subsection{Series expansion for $s_\mu(\zb)$}\label{sec:taylorseriessmu}
The Taylor coefficients for the function $s_\mu(1-w):=\left.S(V,U)\right|_{z^0,\log^0z}$ around small $w=1-\zb$ have the large $n$ dependence given by \eqref{eq:smuzbcoefs}, where
\begin{align}
a_2(\mu)&=\frac12\left(S_1(2-\mu)+S_1(\mu-2)\right),
\\
a_1(\mu)&=-S_1(1-\mu )^2+S_1(\mu -2) S_1(1-\mu )+\frac{2 S_1(2-\mu )}{\mu -2}-S_2(1-\mu )-2 S_2(\mu
-2)-\zeta_2,
\\
a_0(\mu)&=
\frac{31}{2} \zeta _2 S_1(1-\mu )-\frac{11 \zeta _2}{2 (\mu -2)}+\frac{3}{2} \zeta _2 S_1(\mu
   -2)+\frac{25}{6} S_1(1-\mu )^3
   \nonumber
   \\&\quad
   -\frac{3 S_1(1-\mu
   ){}^2}{2 (\mu -2)}+\frac{7}{2} S_1(\mu -2){}^2 S_1(1-\mu )+\frac{5 S_1(\mu -2) S_1(1-\mu )}{\mu
   -2}
   \nonumber
   \\&\quad
   +5 S_2(\mu -2) S_1(1-\mu )+\frac{1}{12} S_1(\mu
   -2)^3-\frac{S_1(\mu -2)^2}{\mu -2}-\frac{2 S_1(\mu -2)}{(\mu -2)^2}
   \nonumber
   \\&\quad
   -\frac{3}{4} S_1(\mu -2)
   S_2(1-\mu )-\frac{3 S_2(1-\mu )}{2 (\mu -2)}-\frac{5}{4} S_1(\mu -2) S_2(\mu -2)-\frac{2 S_2(\mu
   -2)}{\mu -2}
   \nonumber
   \\&\quad+
   \frac{1}{3} S_3(1-\mu )+\frac{5}{3} S_3(\mu -2)-\frac{35}{4} S_1(\mu -2) S_1(1-\mu )^2+\frac{9}{2} S_2(1-\mu ) S_1(1-\mu ).
\end{align}

\section{Results for T/A OPE coefficients}\label{app:completeresults}
Below are the results to order \(1/N^2\) for the OPE coefficients in the traceless symmetric and antisymmetric representations.
The functions \(R(\hb,\mu)\) and \(B(\hb,\mu)\) are given in appendix \ref{app:inversionTable}.
\begin{align}
&a_{T/A,\ell}=\pm a_{\mu,\ell}^{\text{free}}
\pm 2a_{\mu,\ell}^{\text{free}}\left(\frac{\gamma_\varphi^{(1)}}{N}+\frac{\gamma_\varphi^{(2)}}{N^2}\right)
\bigg[
S_1(\mu +\ell-2)-S_1(\mu -2)
\nonumber
\\&
\quad+
\frac{(\ell-1) (2 \mu +\ell-2) \left(S_1(2 \mu +\ell-4)-S_1(2 \mu +2 \ell-3)\right)}{J_0^2}
+
\frac{2 \mu ^2+\mu  (2 \ell-3)+(\ell-1)^2}{(\mu +\ell-1)^2 (2 \mu +2 \ell-3)}
\bigg]
\nonumber
\\&
+
\frac{\left(\gamma_\varphi^{(1)}\right)^2}{N^2}\frac{ (\mu -1)^2 \mu ^2 \Gamma (\ell+\mu -1)^2}{(\mu -2)^3\Gamma (2 \ell+2\mu -2)}
\Bigg[
\frac{2 (\mu -2) \left(2 (\mu +\ell-1)^2-2 (\mu +\ell-1)+1\right)}{J_0^4}
\nonumber
\\&
\quad+
\frac{4 (\mu -2) (2 \mu +2 \ell-3) \left(S_1(2 \mu +2 \ell-3)-S_1(\mu +\ell-2)\right)}{J_0^2}
\nonumber
\\&
\quad+
(2 \mu +2 \ell-3)\bigg(
\frac{(\mu -2)^2 (\mu -1) \mu B(\mu +\ell-1,\mu )}{\ell (2 \mu +\ell-3)}
-
\frac{2 \left(\mu +(\mu -2) S_1(\mu -2)-3\right)}{J_0^2}
\nonumber
\\&
\quad+
\frac{2 \left(\mu +(\mu -2)^2 S_2(\mu -2)-(\mu -2)^2 \zeta_2-3\right)}{\ell (2 \mu +\ell-3)}
\bigg)
\Bigg]
\nonumber
\\&
\pm 
a_{\mu,\ell}^{\text{free}}\frac{\left(\gamma_\varphi^{(1)}\right)^2}{N^2}\left(A_3+A_2+A_1+A_0\right)+O(N^{-3}),
\label{eq:aTANm2Final}
\end{align}
where the upper sign corresponds to $T$ and the lower to $A$. 
Here
\begin{equation}
a^{\text{free}}_{\mu,\ell} =\frac{2 \Gamma (\ell+\mu -1)^2 \Gamma (\ell+2 \mu -3)}{\Gamma (\ell+1) \Gamma (\mu -1)^2 \Gamma (2 \ell+2 \mu -3)} , \quad J_0^2=(\ell+\mu-1)(\ell+\mu-2),
\end{equation}
and the expressions $A_i$ are given by
\begin{align}
&A_3=-\frac{(\mu -1)^2 \mu ^2 \left(\mu ^2-3 \mu +\ell^2+(2 \mu -3) \ell+3\right) }{J_0^6}\left(S_1(\ell)+S_1(2 \mu +2 \ell-3)\right)
\nonumber
\\&
+
\frac{S_1(2 \mu +2 \ell-3)-S_1(2 \mu +\ell-4)}{J_0^6 (\mu -2)^2 (2 \mu +2 \ell-3)}
\Big[
4 \left(8 \mu ^4-35 \mu ^3+67 \mu ^2-76 \mu +40\right) \ell^5
\nonumber
\\&
\quad-
4 (\mu -2)^2 \ell^6
+2 \left(80 \mu ^5-444 \mu
   ^4+1005 \mu ^3-1248 \mu ^2+920 \mu -324\right) \ell^4
\nonumber
\\&   
\quad+
2 \left(159 \mu ^6-1110 \mu ^5+3203 \mu ^4-5016 \mu ^3+4690 \mu ^2-2600 \mu +680\right) \ell^3
\nonumber
\\&
\quad+
\left(314 \mu ^7-2651 \mu ^6+9472 \mu ^5-18763 \mu
   ^4+22642 \mu ^3-17040 \mu ^2+7584 \mu -1560\right) \ell^2
\nonumber
\\&
\quad+
(\mu -1)^2 \left(154 \mu ^6-1210 \mu ^5+3897 \mu ^4-6680 \mu ^3+6596 \mu ^2-3680 \mu +928\right) \ell
\nonumber
\\&
\quad+
(\mu -2)^2 (\mu -1)^2 \left(30 \mu ^5-155 \mu ^4+327 \mu ^3-371 \mu ^2+232 \mu -56\right)
\Big]
\nonumber
\\&
-
\frac{2S_1(\mu +\ell-2)}{J_0^6 (\mu -2) (2 \mu +2 \ell-3)}\Big[
-2 (\mu -2) \ell^6+2 \left(4 \mu ^4-26 \mu ^3+86 \mu ^2-127 \mu +66\right)
   \ell^4
\nonumber
   \\&
\quad   -
   6 \left(2 \mu ^2-7 \mu +6\right) \ell^5
+
   2 \left(15 \mu ^5-88 \mu ^4+246 \mu ^3-403 \mu ^2+355 \mu -126\right) \ell^3
\nonumber
   \\&
\quad+
   3 (\mu -1)^2 \left(14 \mu ^4-71 \mu ^3+148 \mu ^2-168 \mu +88\right) \ell^2
\nonumber
   \\&
\quad+
\left(26 \mu ^7-226 \mu ^6+835 \mu
   ^5-1728 \mu ^4+2195 \mu ^3-1718 \mu ^2+760 \mu -144\right) \ell
\nonumber
   \\&
\quad   +
   (\mu -1)^2 \left(6 \mu ^6-51 \mu ^5+173 \mu ^4-305 \mu ^3+302 \mu ^2-160 \mu +32\right)
\Big],
\end{align}
\begin{align}
&A_2=
\frac{(\mu -1)^2 \mu ^2 \left(2 \mu ^2-6 \mu +2 \ell^2+4 \mu  \ell-6 \ell+5\right) }{J_0^4 (2 \mu +2 \ell-3)}
R(\mu +\ell-1,\mu)
\nonumber
\\&
-
\frac{(\mu -1)^2 \mu ^2 }{2 J_0^4}\left(2 J_0^2 \partial_\hb R(\mu +\ell-1,\mu)+S_2(\ell)\right)
-
\left(S_2(2 \mu +\ell-4)-S_1(2 \mu
   +\ell-4)^2\right)\times
\nonumber
\\&
\quad\times\frac{\left(-\mu ^2-3 \mu +4 \mu  \ell+2 (\ell-3) \ell+4\right) \left(\mu ^2-5 \mu +4 \mu  \ell+2 (\ell-3) \ell+4\right) }{2 J_0^4}
\nonumber
\\&
+
\frac{2 (\ell-1)^2 (2 \mu +\ell-2)^2 \left(S_1(2 \mu +2 \ell-3)^2+S_2(2 \mu +2 \ell-3)\right)}{J_0^4}
\nonumber
\\&
-\frac{(\mu -1)^2 \mu ^2 \left(S_1(\ell)-2 J_0^2R(\mu +\ell-1,\mu)-4 S_1(\mu +\ell-2)+2 S_1(2 \mu +\ell-4)\right) S_1(\ell)}{2 J_0^4}
\nonumber
\\&
+
\frac{2 (\ell-1) (2 \mu +\ell-2) \left(2 (\mu -1)^2+\ell^2+(2 \mu -3) \ell\right)}{J_0^4} S_1(\mu +\ell-2)^2
\nonumber
\\&
-\frac{4 (\ell-1)^2 (2 \mu +\ell-2)^2 }{J_0^4}S_1(2 \mu +\ell-4) S_1(2 \mu +2 \ell-3)
\nonumber
\\&
+
S_1(\mu +\ell-2) S_1(2 \mu +\ell-4)
\Big[
\frac{2  (\mu -1)^2 \left(8 \mu^3 -11 \mu ^2+8\right)+4 \ell^4+8  (2 \mu -3) \ell^3}{J_0^4 }
\nonumber
\\&
\quad+
\frac{4 \left(4 \mu ^4-5 \mu ^3-19 \mu ^2+45 \mu -26\right) \ell^2+4 (\mu -1) (2 \mu -3)
   \left(4 \mu ^3-5 \mu ^2-4 \mu +8\right) \ell
   }{J_0^4 (\mu -2)}
   \Big],
\end{align}
\begin{align}
&A_1=
-\frac{(\mu -1)^2 \mu ^2}{J_0^2} R(\mu +\ell-1,\mu) \left(2 S_1(\mu +\ell-2)+S_1(2 \mu +\ell-4)-2 S_1(2 \mu +2 \ell-3)\right)
\nonumber
\\&
-
\frac{ (\mu -1)^3 \mu ^2\zeta _2}{J_0^2 \ell (2 \mu +\ell-3)}
+
\frac{8 (\mu -1) \left((2 \mu -3) \mu ^2+2\right)+4 (\mu -2) \ell^2+4 (\mu -2) (2 \mu -3) \ell }{J_0^2 (\mu -2)}\times
\nonumber
\\&
\quad\times
\Big[
S_1(\mu -2) \left(S_1(2 \mu +2 \ell-3)-S_1(2 \mu +\ell-4)\right)
-S_1(\mu +\ell-2) S_1(2 \mu +2 \ell-3)
\Big]
\end{align}
and
\begin{align}
&A_0=\frac{ (\mu -2) (\mu -1)^2 \mu ^2\zeta _2}{\ell (2 \mu +\ell-3)} R(\mu +\ell-1,\mu)
+
\frac{(\mu -1)^2 \mu ^2\rho_1(\mu +\ell-1,\mu )}{(\mu -3)^2 (\mu -2)^2}
\nonumber
\\&
-
\frac{4 \left((\mu -2) \ell^2+2 \left(\mu ^2-3 \mu +2\right) \ell-4 \mu ^4+12 \mu ^3-15 \mu ^2+9 \mu-2\right) }{(\mu -2) (\mu +\ell-1)^2 (2 \mu +2 \ell-3)}S_1(\mu -2)
\nonumber
\\&
+
\frac{2(\mu -1)^2 ((\mu -2) \mu  (\mu  (5 \mu -7)+8)+4)+2\left((\mu -2) \mu  \left(8 \mu ^2-13 \mu +8\right)+4\right) \ell^2
   }{(\mu -2)^2 (\mu +\ell-1)^4 (2 \mu +2 \ell-3)}
  \nonumber 
  \\&
   +
\frac{4 (\mu -1) \left(8 \mu ^4-29 \mu ^3+34 \mu ^2-16 \mu +4\right)
   \ell
  }{(\mu -2)^2 (\mu +\ell-1)^4 (2 \mu +2 \ell-3)}
  -4 S_1(\mu -2) S_1(\mu +\ell-2)+2 S_1(\mu -2){}^2
\nonumber
  \\&
  +
\left(\frac{\mu ^2 (\mu -1)^2}{J_0^4}-\frac{4 \mu  (2 \mu -1) (\mu -1)^2}{J_0^2 (\mu -2)}-1\right) S_2(\mu +\ell-2)+S_2(\mu -2).
\end{align}


\bibliographystyle{JHEP}
\bibliography{ONmodel}

\end{document}